\documentclass[twocolumn]{aastex701}

\newcommand{\vect}[1]{\boldsymbol{#1}}

\shorttitle{SMC Structure \& Kinematics}
\shortauthors{Rathore et al.}

\usepackage{multirow}
\usepackage{csquotes}
\usepackage{amsmath}
\begin{document}
\title{A Galactic Transformation---Understanding the SMC's Structural and Kinematic Disequilibrium}

\author[orcid=0009-0009-0158-585X]{Himansh Rathore}
\affiliation{Department of Astronomy and Steward Observatory, University of Arizona, 933 North Cherry Avenue, Tucson, AZ 85721, USA}
\email[show]{himansh@arizona.edu}

\author[0000-0003-0715-2173]{Gurtina Besla}
\affiliation{Department of Astronomy and Steward Observatory, University of Arizona, 933 North Cherry Avenue, Tucson, AZ 85721, USA}
\email[]{gbesla@arizona.edu}

\author[0000-0001-7827-7825]{Roeland P. van der Marel}
\affiliation{Space Telescope Science Institute, 3700 San Martin Drive, Baltimore, MD 21218, USA}
\affiliation{Center for Astrophysical Sciences, The William H. Miller III Department of Physics \& Astronomy, Johns Hopkins University, Baltimore, MD 21218, USA}
\email[]{marel@stsci.edu}

\author[0000-0002-3204-1742]{Nitya Kallivayalil}
\affiliation{Department of Astronomy, University of Virginia, 530 McCormick Road, Charlottesville, VA 22904, USA}
\email[]{njk3r@virginia.edu}

\begin{abstract}
The SMC is in disequilibrium. Gas line-of-sight (LoS) velocity maps show a gradient of $60-100$~km~s$^{-1}$, generally interpreted as a rotating gas disk consistent with the Tully-Fisher relation. Yet, the stars don't show rotation. Despite a small on-sky extent ($\sim4$~kpc), the SMC exhibits a large ($\sim10$~kpc) LoS depth, and the stellar photometric center is offset from the HI kinematic center by $\sim$1~kpc. With N-body hydrodynamical simulations, we show that a recent ($\sim$100~Myr ago) SMC-LMC collision (impact parameter $\sim2$~kpc) explains the observed SMC's internal structure and kinematics. The simulated SMC is initialized with rotating stellar and gaseous disks. Post-collision, the SMC's tidal tail accounts for the large LoS depth. The SMC's stellar kinematics become dispersion dominated ($v/\sigma\approx0.2$), with radially outward motions at $R>2$~kpc, and a small ($<10$~km~s$^{-1}$) remnant rotation at $R<2$~kpc, consistent with observations. Post-collision gas kinematics are also dominated by radially outward motions, without remnant rotation. Hence, the observed SMC's gas LoS velocity gradient is due to radial motions as opposed to disk rotation. Ram pressure from the LMC's gas disk during the collision imparts $\approx30$~km~s$^{-1}$ kick to the SMC's gas, sufficient to destroy gas rotation and offset the SMC's stellar and gas centers. Our work highlights the critical role of group processing through galaxy collisions in driving dIrr to dE/dSph transformation, including the removal of gas. Consequently, frameworks that treat the SMC as a galaxy in transformation are required to effectively use its observational data to constrain interstellar medium and dark matter physics.
\end{abstract}

\keywords{\href{https://astrothesaurus.org/uat/1468}{SMC (1468)}; \href{https://astrothesaurus.org/uat/903}{LMC (903)}; \href{https://astrothesaurus.org/uat/602}{Galaxy kinematics (602)}; \href{https://astrothesaurus.org/uat/582}{Galaxy morphology (582)}; \href{http://astrothesaurus.org/uat/767}{Hydrodynamical simulations (767)}; \href{http://astrothesaurus.org/uat/600}{Galaxy interactions (600)}; \href{http://astrothesaurus.org/uat/416}{Dwarf galaxies (416)}}

\section{Introduction} \label{sec:intro}
\begin{figure*}
    \centering
    \includegraphics[width = 0.49\textwidth, height = 0.40\textwidth]{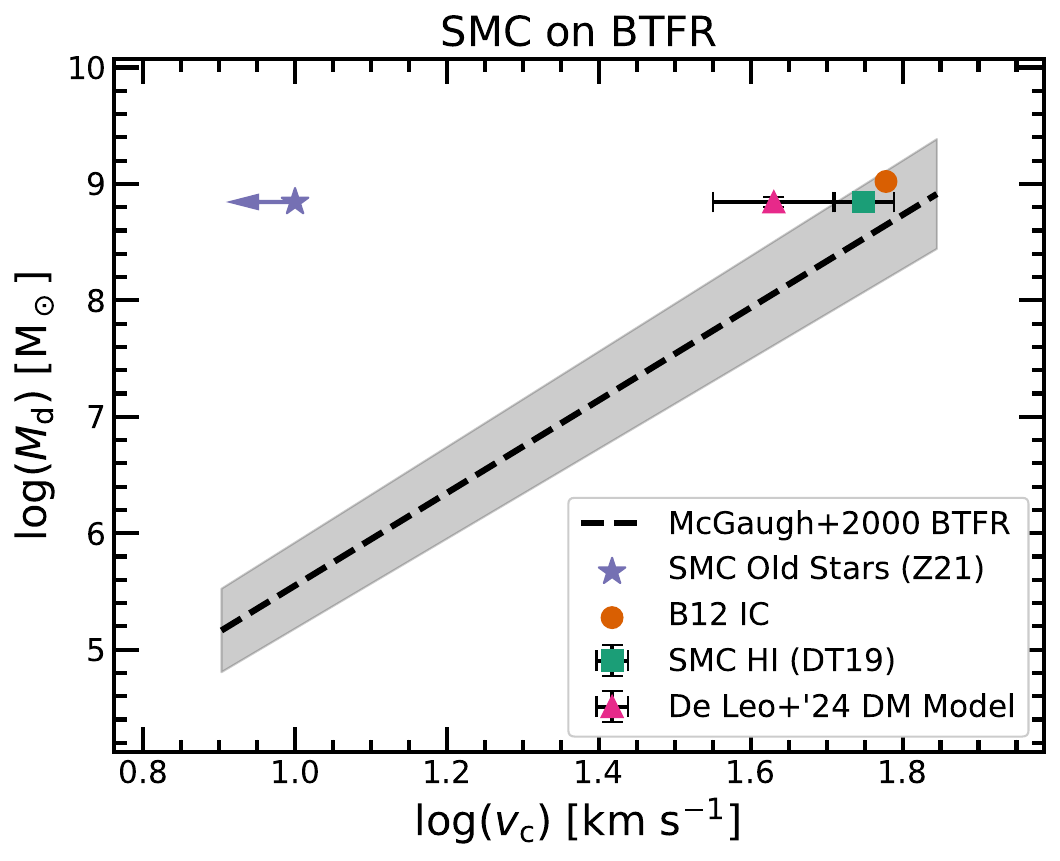}
    \includegraphics[width = 0.4\textwidth, height = 0.4\textwidth]{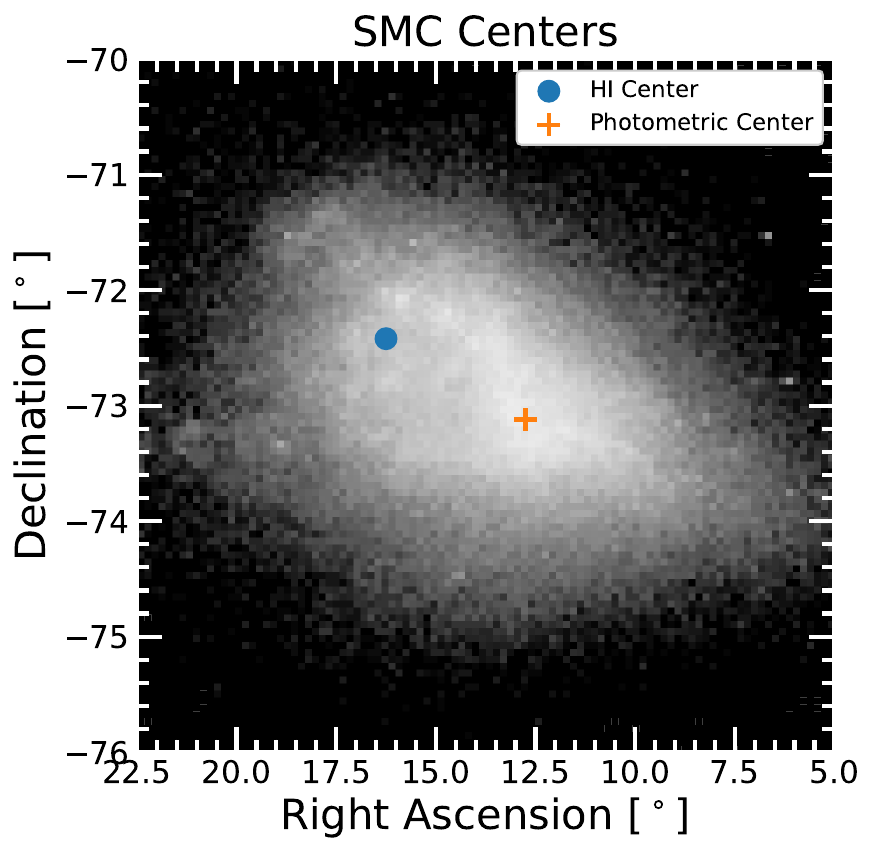}
    \caption{{\em Left panel:} Placing the SMC on the Baryonic Tully-Fisher Relation (BTFR). The black dashed line denotes the BTFR fit taken from \cite{McGaugh2000}, with the grey shaded region denoting the $1-\sigma$ uncertainty. The SMC's total baryonic mass (stars $+$ HI) is $(7.0 \pm 0.6) \times 10^8$ M$_\odot$ \citep{Stanimirovic1999, Harris2004}. Symbols denote the SMC's inferred peak HI rotation velocity of $56 \pm 5$ km s$^{-1}$ \citep[][DT19]{DiTeodoro2019}, peak old star rotation velocity of $< 10$ km s$^{-1}$ \citep[][Z21]{Zivick2021}, and the expected SMC peak velocity of $42.70 \pm 8.06$ km s$^{-1}$ obtained by assuming a DM profile motivated by $\Lambda$CDM \citep{DeLeo2024}, which is consistent with abundance matching expectations \citep{Busha2011}. The kinematics of the SMC's HI and old stars are strongly discrepant. The orange circle denotes the SMC's initial disk in the B12 simulation, which is consistent with BTFR by design. {\em Right panel:} the SMC's photometric center $(12.80^\circ, -72.83^\circ)$ \citep{Gonidakis2009} and the HI kinematic center $(16.26^\circ, -72.42^\circ)$ \citep{Stanimirovic2004} are plotted over a background of SMC stars selected from the NN Optimal Gaia DR3 sample of \cite{Arranz2023}. The two centers are separated by $1-2$ kpc on-sky, which is a significant fraction of the SMC's on-sky extent of $\approx 4-5$ kpc. Explaining the kinematic discrepancies between the old stars and HI (rotation peaks and centers) is the goal of this work.
    }
    \label{fig:btfr}
\end{figure*}

The SMC is the most massive satellite galaxy of the LMC, and is the second most massive satellite of the Milky Way (MW). Given the SMC's proximity \citep[$\sim 60$~kpc,][]{Cioni2000}, a wealth of astrometric, photometric, and spectroscopic data is available for this galaxy from both ground- and space-based observatories. As such, the SMC is a benchmark for studies of star formation, the interstellar medium (ISM), and dust physics in the low metallicity regime \citep[e.g.][]{RusselDopita1992, Gordon1998, Murray2024Scylla}. Hence, the SMC is a standard calibrator used to study high-redshift galaxies \citep[e.g.][]{Kulkarni2021, Schouws2022, Joseph2023, Markov2023, Roman-Duval2025, Ocvirk2025}. Further, with the advent of Gaia, the internal stellar kinematics of the SMC are now well constrained \citep{Luri2021, Zivick2021, Dhanush2025}, making the SMC a potential unique laboratory for dark matter (DM) physics. 

However, Gaia proper motions (PM) reveal that the SMC is in a high state of disequilibrium \citep{DeLeo2020, Zivick2021, Luri2021}. Hence, equilibrium assumptions \citep[e.g.][]{Ostriker2010} cannot be used to describe the SMC's ISM structure and kinematics, or the SMC's DM distribution. Theoretical models of the SMC's disequilibrium are urgently required---not only to interpret the SMC's observational data, but also to assesss the reliability of the SMC as a calibrator for low-metallicity galaxies. In this work, we present a new theoretical framework to understand the origin of the SMC's disequilibrium in structure and kinematics, in the context of its interaction history with the LMC. 

The SMC's disequilibrium is clearly illustrated by placing it on the Baryonic Tully-Fisher relation \citep[BTFR, ][]{TF1977, McGaugh2000}. The SMC has a gas-to-stellar mass ratio close to 1 \citep{Stanimirovic1999, Harris2004, Bruns2005}. In the local universe, dwarf galaxies with such high gas fractions generally possess rotation supported disks \citep[e.g. The LITTLE THINGS survey -- ][]{Hunter2012, Johnson2012, Johnson2015}, see also \cite{Geha2006}. Indeed, local gas-rich dwarfs satisfy the BTFR \citep[e.g.][]{Geha2006, Begum2008, McGaugh2021}. This implies that the SMC should also satisfy the BTFR. In isolated disk galaxies where stellar kinematics are measurable, both stars and gas serve as tracers of the same underlying rotational field \citep{Hinz2003, Williams2010}. In Figure \ref{fig:btfr}, the SMC is placed on the BTFR using the rotation amplitude derived from both stars and gas. The SMC's old stars (age $>$ 1 Gyr) and gas (HI) are strongly discrepant in the BTFR \citep[see also][]{BekkiChiba2008, BekkiChiba2009}---illustrating the SMC's gas and stars are not in equilibrium. The old stellar kinematics is highly dispersion dominated \citep[$v_{\rm rot}/\sigma < 0.6$,][]{Harris2006}, with a small rotation amplitude \citep[$< 20$ km s$^{-1}$, ][]{Hardy1989, Harris2006, Dobbie2014, Zivick2018, Niederhofer2021, DeLeo2020, Zivick2021, Dhanush2025}. However, HI surveys have revealed a velocity gradient of 60-100 km s$^{-1}$ across the SMC's $\sim 5^\circ$ extent. This has been interpreted as the signature of a rotating gas disk with a peak velocity of $\approx$ 50 km s$^{-1}$ \citep[e.g.][]{Hindman1967, Stanimirovic2004, DiTeodoro2019, Pingel2022}. Further, the observed stellar kinematics are highly discrepant with the peak velocity expected from a SMC DM profile motivated by the Lambda Cold Dark Matter ($\Lambda$CDM) cosmology \citep{DeLeo2024} and abundance matching \citep{Busha2011}.

Addressing the discrepancy between the SMC's stellar and gas kinematics is key to understanding the dynamical state of this galaxy, which is the main goal of this work. Significantly, the SMC's high gas fraction conflicts with its dispersion-dominated stellar kinematics, potentially signaling a new pathway for how \enquote{dwarf irregular} (dIrr) galaxies can lose rotational support.

The spatial distribution of the SMC's old stellar populations and gas also points to the system being in a state of disequilibrium. The SMC has a large stellar line-of-sight (LoS) depth, which ranges from 5 kpc to 20 kpc depending on the measurement technique \citep{Mathewson1986, Hatzidimitriou1993, Crowl2001, Subramanian2012, Nidever2013, Ripepi2017, Subramanian2017, Muraveva2018, Tatton2021, Mercia-Jones2021, Almeida2024, Dhanush2025, Oden2025}. HI surveys show a bi-modal gas LoS velocity profile \citep{Johnson1961, Hindman1967, Torres1987, Pingel2022}, with two brightness temperature peaks separated by $\approx$50 km s$^{-1}$ due to gas clumps 5–10 kpc apart along the LoS \citep{Murray2024}. Understanding the SMC's complex gas distribution is crucial to properly interpret maps of dust extinction \citep{Murray2024} and HI column densities \citep{Pingel2022}, which are ultimately used to derive star formation rates \citep{Bolatto2011, Jameson2016}. Further, understanding the large stellar LoS depth is important for converting PMs to 3D velocities. This work investigates the origin of the SMC's large stellar LoS depth and the bi-modal LoS gas distribution. 

Unsurprisingly, the SMC's center is ill-defined given the discrepant stellar and gas kinematics, and the complex 3-D structure. The photometric center derived from near Infra-Red (IR) star counts \citep{Gonidakis2009} is separated from the HI kinematic center \citep{Stanimirovic2004} by $1-2$ kpc on-sky, as illustrated in {\em right panel} of Figure \ref{fig:btfr} (see also \citealt{DeLeo2020}). For reference, the SMC is $\sim$4 kpc across on the sky, meaning a spatial difference of 1 - 2 kpc in stellar- and gas-derived centers must point to a high degree of disequilibrium. This study aims to provide a physical explanation for the separation between the SMC's different centers, enabling us to identify the correct center for analyzing stellar and gas kinematics.

The coincident star formation histories of the SMC and LMC over the last $\sim 4$ Gyr \citep{Noel2007, Noel2009, Harris2009, Weisz2013, Massana2022, Burhenne2026} suggest that the Clouds have been bound for at least that amount of time. Further, the formation of the $150^\circ$ long gas stream (MS) \citep{Mathewson1974, Braun2004, Nidever2010} and the SMC-LMC gas bridge \citep{Kerr1957, Putman2003, Bruns2005} require the Clouds to complete multiple pericenters about each other \citep[e.g.][]{Murai1980, Besla2010, Pardy2018, Lucchini2021}. There is increasing observational evidence that the most recent ($100 - 200$ Myr ago) SMC-LMC pericenter was a direct collision between the Clouds with an impact parameter of $\sim 2$ kpc \citep{Olsen2011, Besla2016, Zivick2018, Zivick2019, Choi2022, Dhanush2024, Arranz2025, Rathore2025a, Rathore2025b}. 

In this work, we use hydrodynamic simulations of the SMC-LMC-MW interaction history from \citet[hereafter B12,][]{Besla2012} to: (i) understand how a recent ($\sim 100$ Myr ago) SMC-LMC direct collision (impact parameter $\approx 2$ kpc) affects the SMC's internal structure and kinematics; (ii) provide a means to interpret observations of the SMC's stars and gas; and (iii) assess the validity of equilibrium assumptions for measuring the SMC's DM content.

This manuscript is organized as follows. Section \ref{sec:methods} gives an overview of the B12 simulations and presents a coordinate system for analyzing the simulated SMC. Then, we study the effects of the SMC-LMC collision on the SMC's stellar morphology (section \ref{sec:stellar_struc}), stellar kinematics (section \ref{sec:stellar_kine}), and gas kinematics (section \ref{sec:gas_kine}). In section \ref{ref:discussion}, we investigate the LoS distribution of the simulated SMC's gas, understand the effect of the LMC ISM's ram pressure on the SMC's gas during the collision, discuss ideas to determine the SMC's DM content without relying on equilibrium assumptions, compare our results to previous studies and assess the limitations of our analysis and prospects for the future. We conclude in section \ref{sec:conc}.   

\section{Methods} \label{sec:methods}

\subsection{Hydrodynamic Simulations of the LMC-SMC-MW Interactions} \label{sec:sims}

\begin{figure}
    \centering
    \includegraphics[width=\columnwidth]{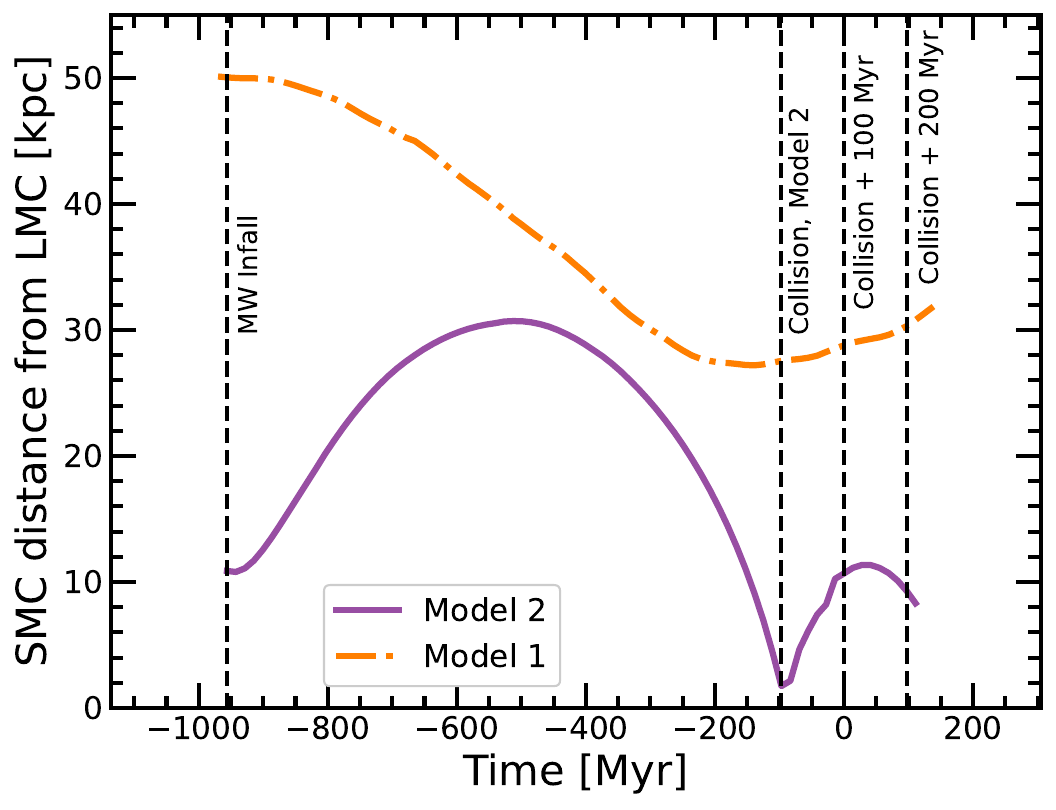}
    \caption{The SMC's orbit about the LMC in B12 Model 1 (orange dash-dot line) and Model 2 (purple solid line) simulations, after their MW infall. Three additional key epochs corresponding to Model 2 are marked with vertical dashed lines: SMC-SMC collision (impact parameter $\approx 2$ kpc); 100 Myr and 200 Myr post-collision. The fiducial present day is denoted as time $= 0$, and in Model 2, this corresponds to 100 Myr post-collision. The Model 2 SMC is expected to be highly morphologically and kinematically disturbed post-collision, and will be the primary subject of investigation in this work. Model 1, without a collision, serves as a control.}
    \label{fig:orbit}
\end{figure}

In this section, we give an overview of the B12 simulations. For more details, we refer the reader to the B12 paper. B12 modeled the SMC-LMC interaction history over the past $\approx 7$ Gyr, in which the last $\approx 1$ Gyr was the first infall orbit of the Clouds in the MW halo.

The initial LMC and SMC are modeled with live Hernquist DM halos \citep{Hernquist1990}, live exponential stellar disks and Smooth Particle Hydrodynamic (SPH) gas disks. The DM, stellar and gas masses of the LMC (SMC) were chosen to be $1.76 \times 10^{11} \: (2 \times 10^{10})$ M$_\odot$, $2.5 \times 10^{9} \: (2.6 \times 10^{8})$ M$_\odot$ and $1.1 \times 10^{9} \: (7.9 \times 10^{8})$ M$_\odot$ respectively. The scale radii of the LMC (SMC) stellar and gas disks were chosen to be 1.7 (1.1) kpc and 1.7 (3.3) kpc. The scale heights of the LMC (SMC) stellar and gas disks were chosen to be 0.34 (0.11) kpc and 0.34 (0.33) kpc respectively. These chosen values of the scale radii and scale height are consistent with isolated dIrr galaxies in the local universe \citep{Swaters2002, Connors2004, Kreckel2011}. The gravitational softening length for the gas and stars was 0.1 kpc.

The DM, stellar and gas mass resolution for the LMC (SMC) was $1.76 \times 10^6 \: (1.4 \times 10^6)$ M$_\odot$, $2500 \: (2600)$ M$_\odot$ and $3667 \: (2633)$ M$_\odot$ respectively. The MW was modeled as a static NFW \citep{NFW1997} potential, with a total mass of $1.5 \times 10^{12}$ M$_\odot$, Virial concentration of 12 and Viral radius of 300 kpc. 

For the full table of simulation parameter choices and their justification, we refer the reader to Table 1 of B12.

The SMC's initial halo and disk properties were chosen in B12 such that the resulting rotation curve (B12 Figure 1) is consistent with the BTFR (Figure \ref{fig:btfr} {\em left panel}), i.e. the SMC is initialized with a baryonic mass of $1.05 \times 10^9$ M$_\odot$ and a peak rotation velocity of $\approx 60$ km s$^{-1}$. The simulated SMC is initially in equilibrium; the stellar density and gas kinematic centers are initially coincident.

The B12 simulation was performed using the Gadget-3 N-body SPH code \citep{Springel2005}. A sub-resolution multiphase model with an effective equation of state was used for the Interstellar medium (ISM) \citep{Springel2003}. Radiative cooling and star formation were implemented according to the \cite{Springel2003} prescription. 

Note that ram pressure from the Clouds' ISM might play an important role in shaping the gas morphology and kinematics of the Clouds. However, ram pressure is known to be underestimated in standard SPH \citep[e.g.][]{Agertz2007}. We will discuss the role of ram pressure in some detail in section \ref{sec:ram_pressure}, and provide an analytical means to understand the effect of ram pressure on the Clouds.

B12 present two scenarios - Model 1 and Model 2. In Model 1, the SMC and LMC remain far from each other with their closest separation being $\approx 30$ kpc. In Model 2, after three pericentric passages about each other, the SMC and LMC collide, with an impact parameter of $\approx 2$ kpc. The collision happened at a lookback time of $\approx100$ Myr. Figure \ref{fig:orbit} illustrates the separation of the SMC with respect to the LMC as a function of time since MW infall, where the \enquote{present day} is denoted at time 0. For Model 2, the present day is $\sim$100 Myr after the SMC-LMC collision epoch (see vertical dashed lines in Figure \ref{fig:orbit}). The B12 simulation runs $\approx 100$ Myr beyond the inferred present day.

Both Model 1 and Model 2 are in reasonable agreement with several external features associated with the Clouds. These features include: the observed SMC's trailing gas stream (MS) \citep{Mathewson1974, Braun2004, Nidever2010} and its stellar counterpart \citep{Chandra2023, Zaritsky2025}; the SMC-LMC gas bridge \citep{Kerr1957, Putman2003, Bruns2005}; and the LMC's galactocentric position and velocity at present day as determined by \citet{Kallivayalil2006}. We refer the reader to the B12 paper and \cite{Besla2013} for a detailed comparison between the aforementioned observed features and the simulation. 

The SMC-LMC collision scenario in the Model 2 simulation is particularly successful at explaining the LMC's observed internal features. These features include - the LMC's outer disk morphology \citep{Besla2016}, disturbed kinematics of the LMC's old (age $>$ 1 Gyr) stellar populations \citep{Choi2022}, the disturbed morphology and kinematics of the LMC's bar \citep{Rathore2025b, Rathore2025a}, and the LMC's gas distribution \citep[][hereafter R25]{Rathore2025b}. Further, the observed stellar PM vectors in the LMC-SMC bridge are also in reasonable agreement with Model 2 \citep{Zivick2019}. None of the above observations are explainable in Model 1, which does not invoke a collision.

Given the success of the LMC-SMC collision scenario in explaining several of the LMC's disturbed features, we investigate whether the same scenario can simultaneously help us to understand the SMC's disequilibrium.

In this work, we will primarily use Model 2 to understand the internal structure and kinematics of the simulated SMC. We will treat Model 1 (where the SMC and LMC do not collide) as a control simulation. The results of Model 2 will be compared with Model 1 to assess whether a SMC-LMC collision is necessary to explain the SMC's disturbed structure and kinematics.

As with any simulation, B12 Model 2 is just one possibility of the SMC - LMC - MW interaction history, and the model has several limitations. We will discuss these limitations and how they affect our main conclusions in detail in section \ref{sec:future}. We emphasize that our aim is not to exactly represent the observed SMC's structure and kinematics with the B12 simulation. Rather, we want to use the B12 simulation to understand the physical mechanisms that operate when the SMC collides with the LMC, and provide a framework to interpret the SMC's observations. 

One limitation of Model 2, which is clear from Figure \ref{fig:orbit}, is that the simulated SMC-LMC separation at present day ($\approx 10$ kpc) is inconsistent with the observed separation between the Clouds \citep[20 kpc, ][]{Kallivayalil2006}. Given this limitation, we will refrain from labeling the epochs in the simulation with respect to the present day. Rather, we will analyze the simulated SMC's structure and kinematics as a function of time elapsed since the SMC-LMC collision (key times are marked in Figure \ref{fig:orbit}). This approach will help us understand the rate at which the SMC's disequilibrium grows post-collision. 

The B12 simulation does qualitatively match the relative locations of the LMC, SMC and MW with respect to each other. More specifically, the simulated SMC is located south of the LMC at present day, and is at a larger Galactocentric distance as compared to the Galactocentric distance of the LMC. Quantitatively, the SMC's Galactocentric $x$ and $z$ positions are off by $\approx 10$ and $\approx 5$ kpc respectively when compared to observations. The simulated SMC's Galactocentric velocities at present day are in reasonable agreement with observations from \cite{Kallivayalil2006, Kallivayalil2006b}. As B12 mention, a more accurate orbital solution for the SMC can be achieved by fine tuning the simulation initial conditions. However, such fine tuning is unlikely to change the physical mechanisms that govern the LMC-SMC collision. For a more detailed comparison of the simulated phase space of the Clouds to the observations, we refer the reader to Table 2 of B12.

Although star formation is included in the B12 simulations, in this work, we only follow the star particles (hereafter referred to as stars) that were used to create the initial SMC disk. These stars are suitable for comparison with the structure and kinematics of the observed SMC's old stellar populations (age $>$ 1 Gyr). We defer study of younger stellar populations (age $<$ 1 Gyr) to later work. 

The B12 simulation does not explicitly follow different gas phases; however, the gas mass of the SMC is dominated by HI. As such, we follow the SPH gas particles (hereafter referred to as gas), to compare with the observed SMC's HI distribution and kinematics.

The fiducial coordinate system in the simulation is Galactocentric. In this system, we will refer to the coordinate axes as ($x, y, z$) and velocities as ($v_{x}, v_{y}, v_{z}$).

\subsection{Centering and Alignment of the Simulated SMC} \label{sec:centering}

\begin{figure*}
    \centering
    \includegraphics[width=\textwidth]{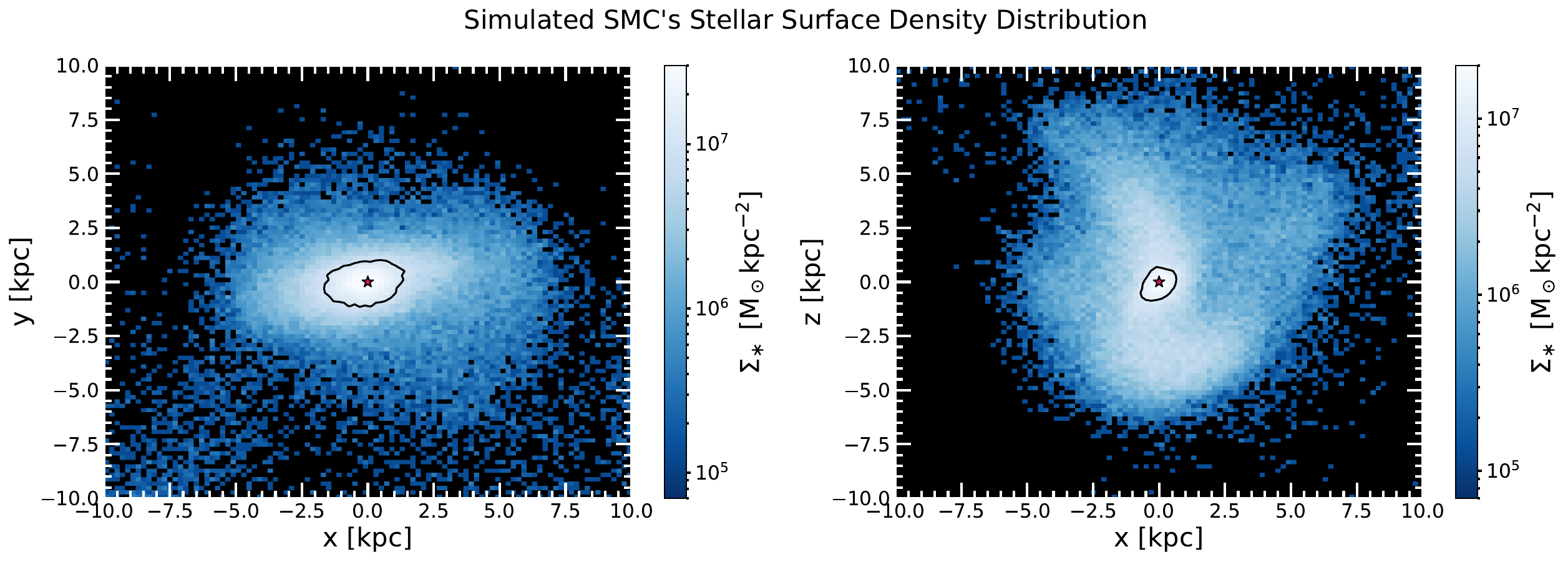}
    \includegraphics[width=\textwidth]{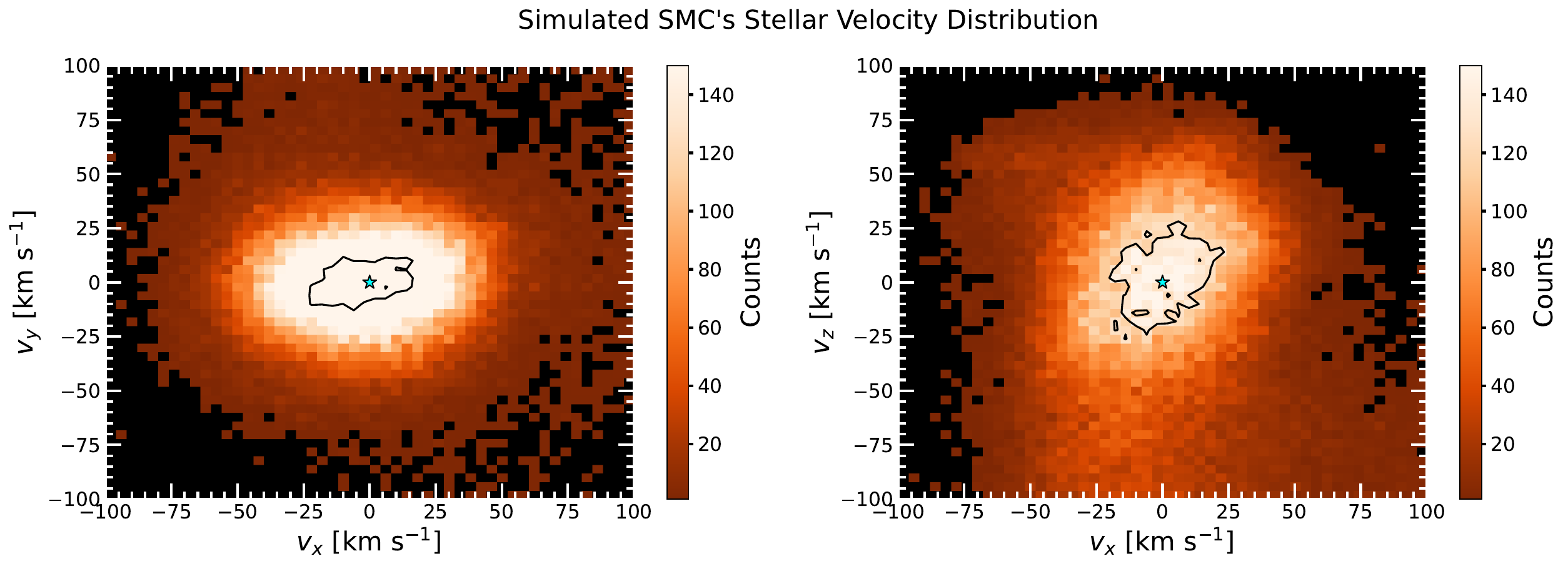}
    \caption{{\em Top row:} Identifying the stellar density center of the simulated SMC. The {\em left (right)} panel shows the x-y (x-z) projection of the SMC's stellar surface density distribution 100 Myr after the SMC-LMC collision in B12 Model 2. The Galactocentric axes are translated to the inferred density center, depicted by the red star at (0, 0). The red star is a reasonable representation of the SMC's stellar density peak (a $10^7$ M$_\odot$ kpc$^{-2}$ contour is included to guide the eye). The iterative shrinking sphere method succeeds in identifying the SMC's stellar density center despite its highly disturbed morphology. {\em Bottom row:} Computing the systemic velocity of the simulated SMC. The {\em left (right) panel} shows the $v_{x}$ - $v_{y}$ ($v_{x}$ - $v_{z}$) projection of the SMC's stellar velocity distribution for the same epoch as the {\em top row}. The Galactocentric velocities are translated to the calculated stellar systemic velocity, depicted by the cyan star at (0, 0). The colorscale depicts star particle counts in 4 km~s$^{-1}$ by 4 km~s$^{-1}$ bins. The cyan star is a reasonable representation of the center of the SMC's velocity field (a contour corresponding to 75\% of the peak value of the distribution is shown to guide the eye). The iterative shrinking sphere method succeeds in identifying the SMC's systemic velocity despite its disturbed kinematics.}
    \label{fig:dens_center_check}
\end{figure*}

\begin{figure*}
    \centering
    \includegraphics[width = 0.49\textwidth]{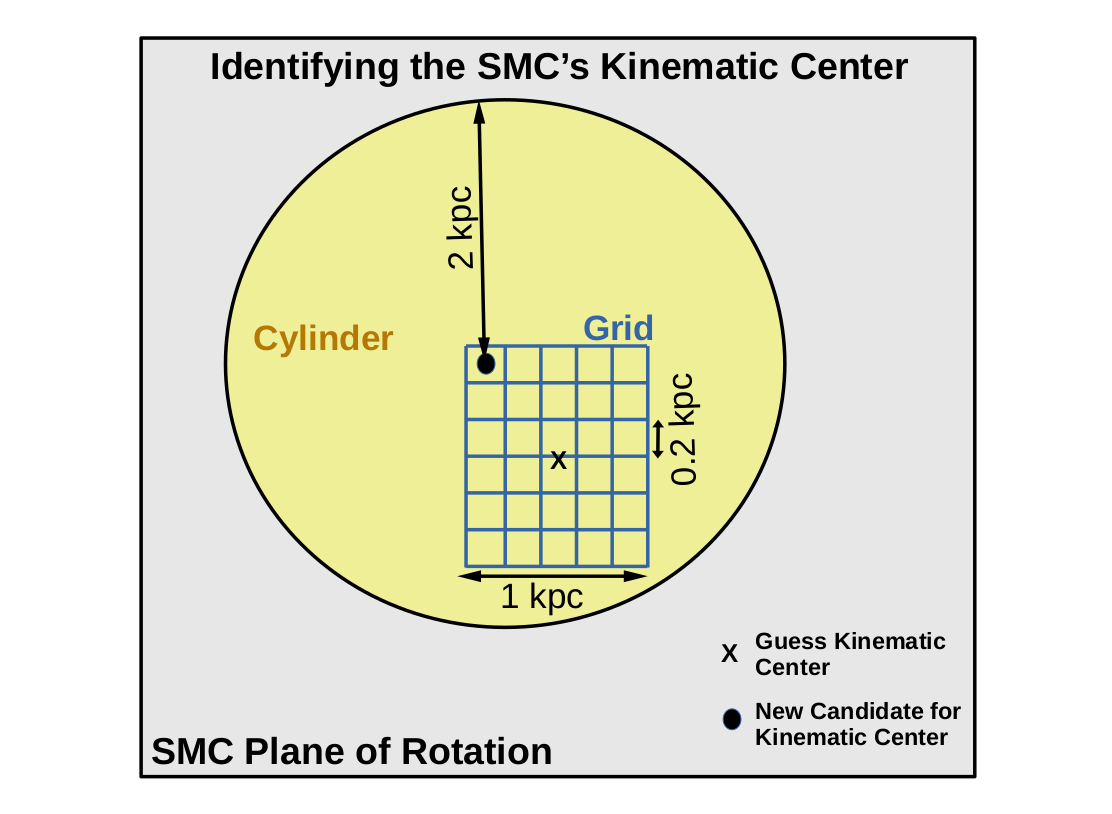}
    \includegraphics[width = 0.49 \textwidth]{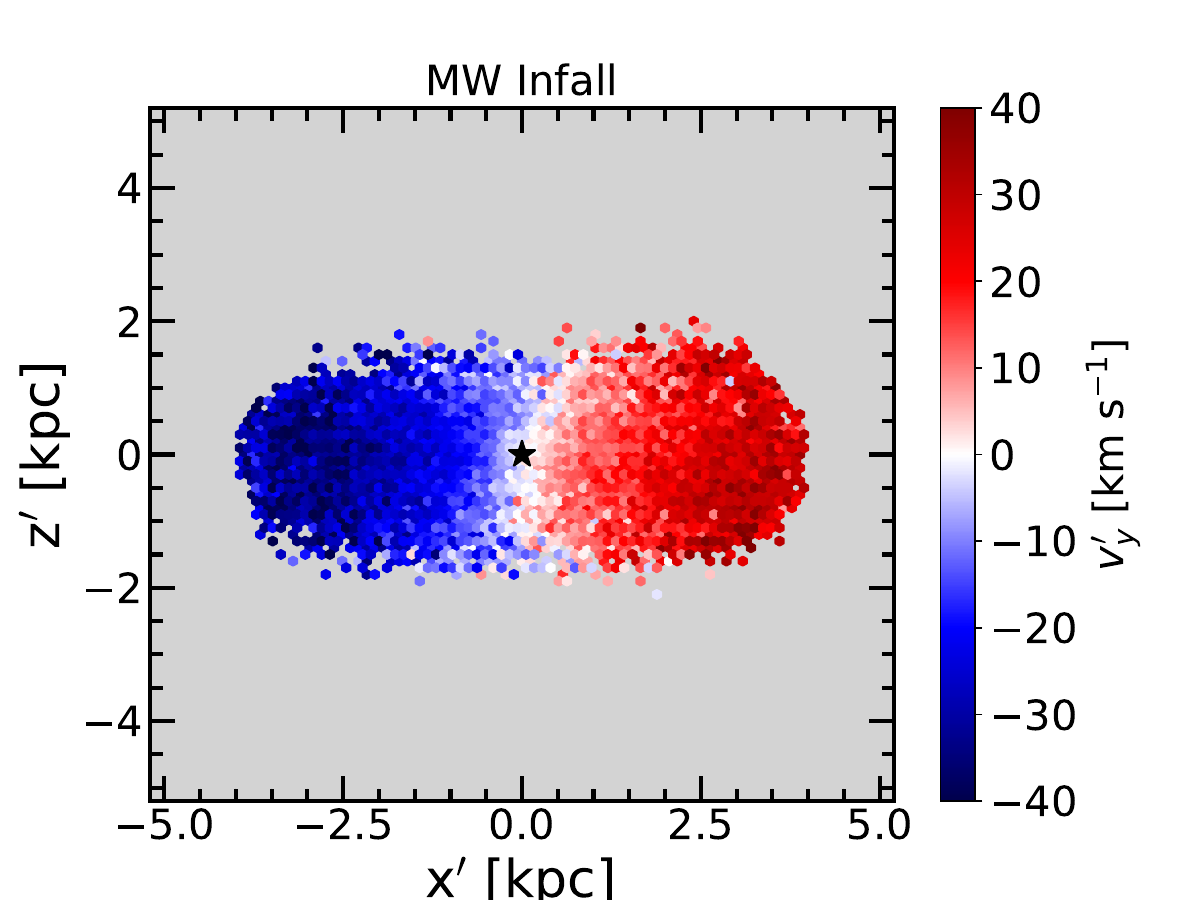}
    \includegraphics[width = 0.49 \textwidth]{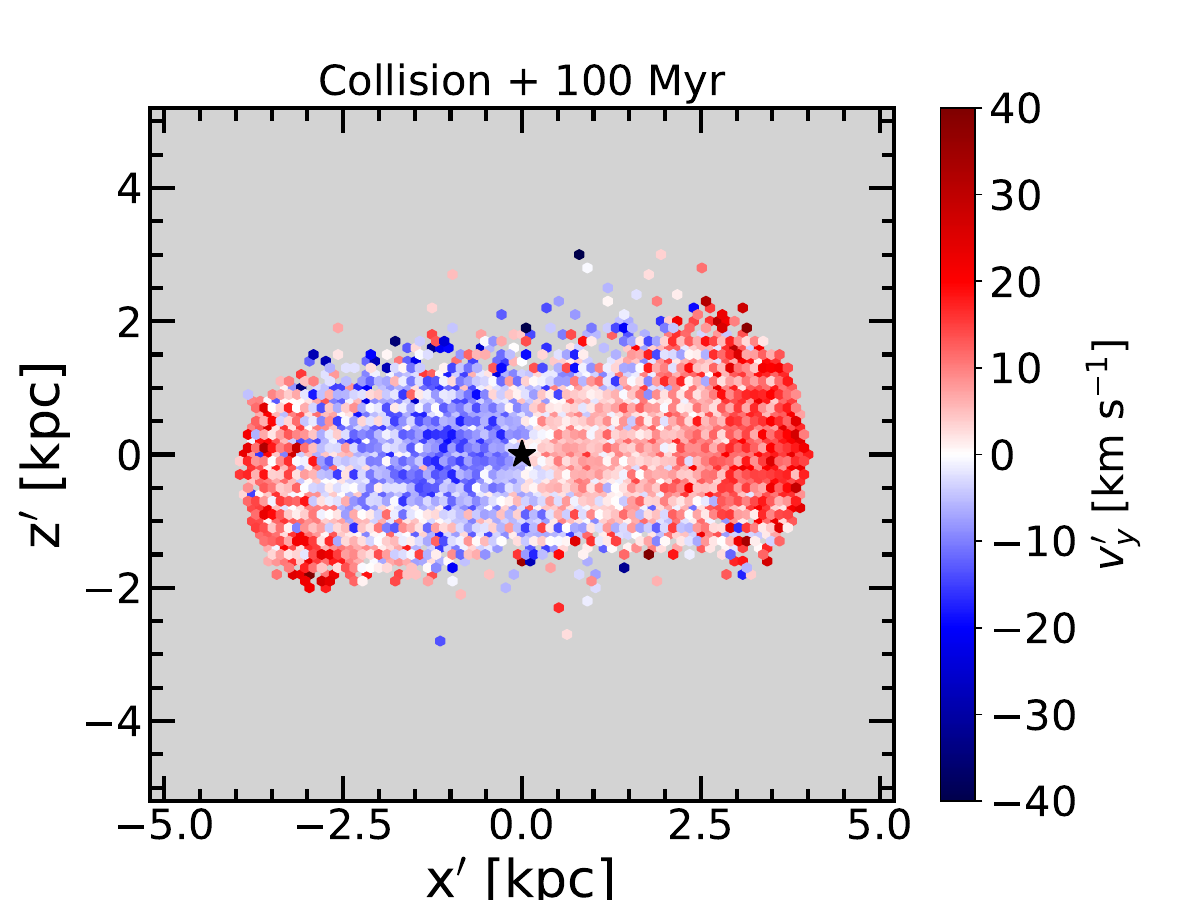}
    \includegraphics[width = 0.49 \textwidth]{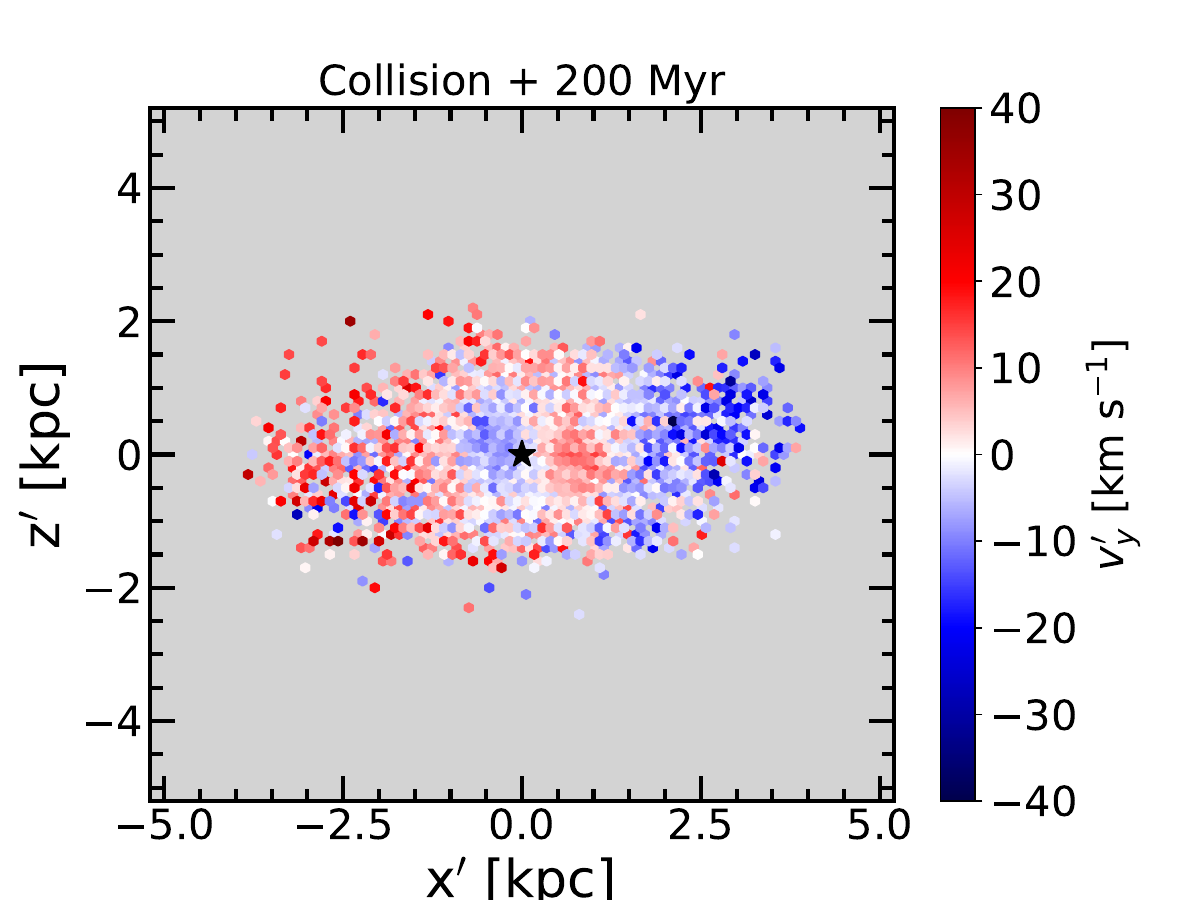}
    \caption{Identifying the stellar kinematic center of the Model 2 simulated SMC. The SMC is oriented edge-on ($x' - z'$ projection) and the in-plane velocities are mapped ($v_y'$ in this projection) through the color bar. The red (blue) colors depict the values of $v_y'$ for stars going into (coming out of) the plane of the paper. Three epochs are shown - MW infall ({\em left panel}), 100 Myr, and 200 Myr post SMC-LMC collision ({\em middle panel} and {\em right panel} respectively). The SMC's stellar kinematic center (translated to the origin) is marked by the black star. The black star is a reasonable representation of the center of the SMC's stellar rotation field, even when the internal kinematics are significantly disturbed and the amplitude of rotation is small ($< 20$ km s$^{-1}$).}
    \label{fig:kine_cen_check}
\end{figure*}

The simulated SMC's structure and kinematics will be highly disturbed post LMC collision. Hence, it is important to correctly define the SMC's \enquote{center} and plane of rotation, in order to establish an appropriate coordinate system in which we can study the SMC's kinematics and structure. We adopt two different definitions for the SMC stellar center:

\begin{itemize}
    \item {\it Stellar density center:} for highly disturbed galaxies, like the SMC, it is challenging to define a center of mass for the stellar distribution. A more appropriate center to use is the stellar density center, which corresponds to the point in 3D position space where the stellar density distribution achieves a global maximum.
    \item {\it Stellar kinematic center:} an internal rotation field of SMC stars can be constructed after subtracting the stellar systemic velocity from the velocities of the individual stars. The stellar kinematic center refers to the point in 3D position space about which the SMC can be considered to be rotating. Mathematically, this is the point about which the average of the azimuthal velocities in the plane of rotation is maximized. 
\end{itemize}
\noindent Here, the stellar systemic velocity is the bulk velocity at which the SMC is moving. We take the systemic velocity as the point with the highest density in the 3D velocity space. To compute the kinematic center, we must determine the systemic velocity. Next, we detail the computation of: (i) the stellar density center, (ii) the stellar systemic velocity, and (iii) the stellar kinematic center. 

We compute the SMC's stellar density center with an iterative shrinking sphere algorithm. We start with a guess for the density center, which is the mass weighted average of the positions of all SMC stars. Then, we consider a spherical volume of radius 5 kpc centered at the guess density center. The mass weighted average position of stars that reside inside that spherical volume is evaluated. This gives a new estimate for the density center. We re-center the sphere at the new density center, shrink its radius by 30\%, and re-compute the mass weighted average position of stars that reside inside the shrunken sphere. Convergence is established when the density center estimates between two successive iterations differ by less than 0.1 kpc, which corresponds to the softening length of the simulation. We experiment with different values of the hyper-parameters (like the starting radius of the sphere, shrinking fraction, convergence threshold), finding that the above adopted values give the best estimate of the density center (see Figure \ref{fig:dens_center_check}). Such an iterative approach is commonly used to calculate density centers for disturbed stellar and dark matter distributions in simulations and observations \citep{Power2003, GC2019, Rathore2025a, Rathore2025b}. 

Figure \ref{fig:dens_center_check} {\em top row} shows projections of the SMC's stellar distribution and the calculated stellar density center 100 Myr after the SMC-LMC collision in Model 2. The shrinking sphere algorithm reasonably identifies the peak of the highly disturbed stellar distribution, as indicated by the $10^7$ M$_\odot$ kpc$^{-2}$ density contour. We have verified this agreement for all simulation snapshots.

We compute the SMC's stellar systemic velocity by applying the iterative shrinking sphere algorithm in velocity space. We again start with a guess for the stellar systemic velocity by computing the mass weighted average velocity of stars located within 5 kpc of the stellar density center. Then, we construct a sphere of radius 30~km s$^{-1}$ in velocity space, centered at the guess systemic velocity. The mass weighted average velocity of stars that reside inside this sphere is computed, which gives a new estimate for the stellar systemic velocity. We recenter the sphere at the new systemic velocity and shrink its radius by 50\%, and re-compute the mass weighted average velocity of the stars that reside inside the shrunken sphere. Convergence is established when the difference between the systemic velocity obtained with two successive iterations is less than 1 km s$^{-1}$. We experiment with different values of the hyper-parameters (like the starting radius of the sphere, shrinking fraction, convergence threshold), finding that the above adopted values give the best estimate of the systemic velocity.

Figure \ref{fig:dens_center_check} {\em bottom row} shows projections of the SMC's stellar velocity distribution and the computed systemic velocity at a snapshot 100 Myr post collision. The shrinking sphere algorithm reasonably identifies the peak of the highly disturbed stellar velocity distribution, as indicated by the 75\% contour level. 
We have verified this agreement for all simulation snapshots.

The identification of the SMC's stellar kinematic center post-collision is especially tricky. Computing the stellar kinematic center requires the azimuthal velocities, consequently requiring a plane of rotation. A standard practice in simulations is to use the total angular momentum vector of the stellar velocity field to identify the plane of rotation \citep[e.g.][]{Gomez2016, Gomez2017, Rathore2025b}. But, computing the total angular momentum vector requires us to already know the kinematic center. For isolated disks, the stellar density center can be used as a good measure of the kinematic center. However, for highly disturbed kinematics, i.e. a post-collision system, the kinematic center cannot be apriori assumed to coincide with the density center. Hence, the kinematic center and the total angular momentum vector (or equivalently the plane of rotation) of the post-collision SMC cannot be determined independently of each other. Needed is an iterative approach that can jointly constrain both the kinematic center and the total angular momentum vector.

We subtract the stellar systemic velocity from the SMC's total stellar velocity field to obtain the internal stellar velocity field, at each epoch in time. We start with a guess for the kinematic center, which we set as the stellar density center. We calculate the angular momentum vector of stars that reside in a spherical volume of radius 2 kpc centered at the guess kinematic center. We define a new z'-axis in the direction of the total angular momentum vector, thereby defining an initial plane of rotation. 

Next, in the plane of rotation, we define a square grid (dimension of 1 kpc, grid size of 0.2 kpc), centered at the guess kinematic center. At the center of each grid cell, we place a cylinder of cross section radius 2 kpc, and height of 1 kpc. The cross section of the cylinder is parallel to the plane of rotation, and the cylinder extends 0.5 kpc above and below this plane. By design, the cross section of the cylinder is larger than the area of the grid. We compute the average $v_\phi$ of stars enclosed within the cylinder. The grid cell with the largest value of average $v_\phi$ is chosen as the guess kinematic center for the next iteration. Convergence is established when the stellar kinematic centers of two successive iterations differ by less than the grid size. We experiment with different values of the hyper-parameters, finding that the above adopted values give the best estimate of the kinematic center.

We hereby define a new internal coordinate system for the SMC ($x', y', z'$ and $v_x', v_y', v_z'$), where the $x'-y'$ plane is aligned with the plane of rotation, $z'$ is aligned with the angular momentum vector, and the origin is the SMC's stellar kinematic center. Hereafter, the $x'-z'$ plane will be referred to as the \enquote{vertical plane}.

Figure \ref{fig:kine_cen_check} shows the SMC's stellar velocity field and kinematic center, in the vertical plane, for three epochs in Model 2: the MW infall epoch, 100 Myr and 200 Myr post SMC-LMC collision. The iterative algorithm reasonably identifies the kinematic center for both un-perturbed and significantly perturbed stellar velocity fields, as indicated by the spatial coincidence of the black star (kinematic center) with the center of a residual inner rotation pattern (red to blue). The algorithm is sensitive enough to identify the kinematic center even if the rotation field has a very small amplitude ($< 20$ km s$^{-1}$). 

\section{Results} \label{sec:results}

\subsection{The Simulated SMC's Stellar Structure} \label{sec:stellar_struc}
\begin{figure*}
    \centering
    \includegraphics[width = 0.49\textwidth]{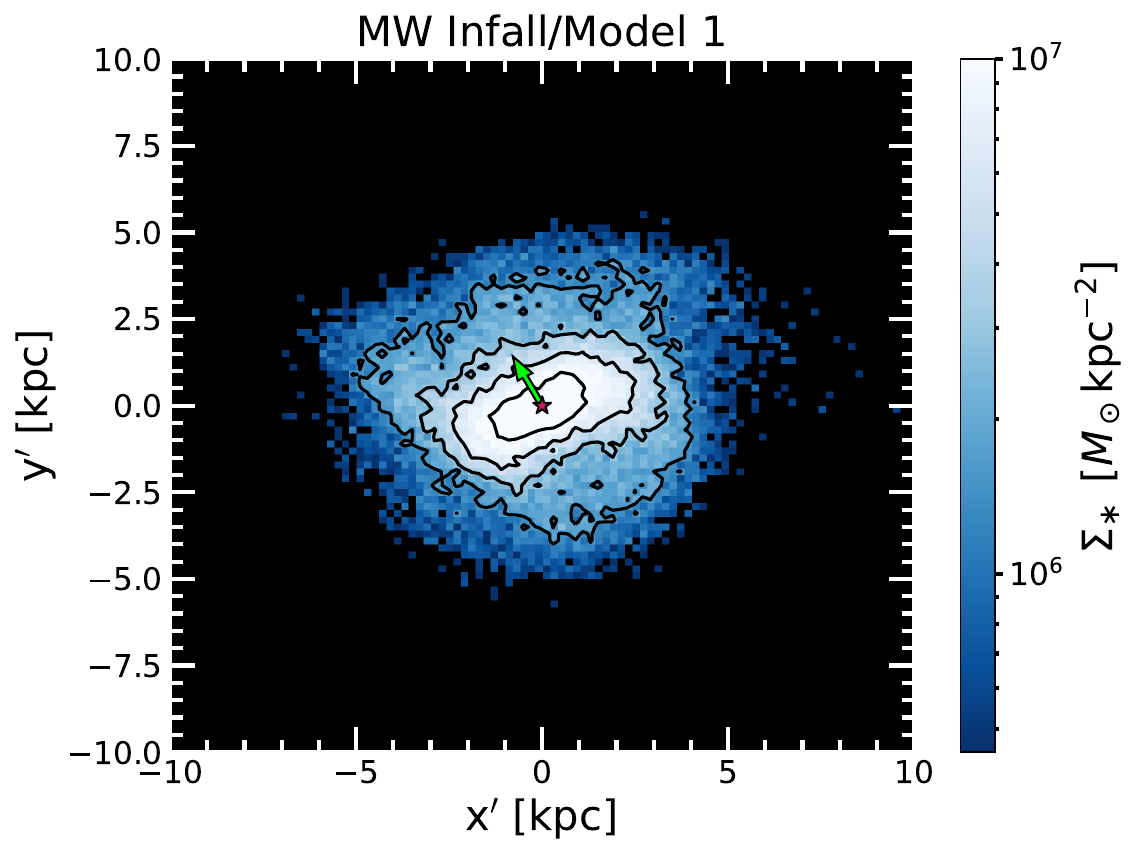}
    \includegraphics[width = 0.49\textwidth]{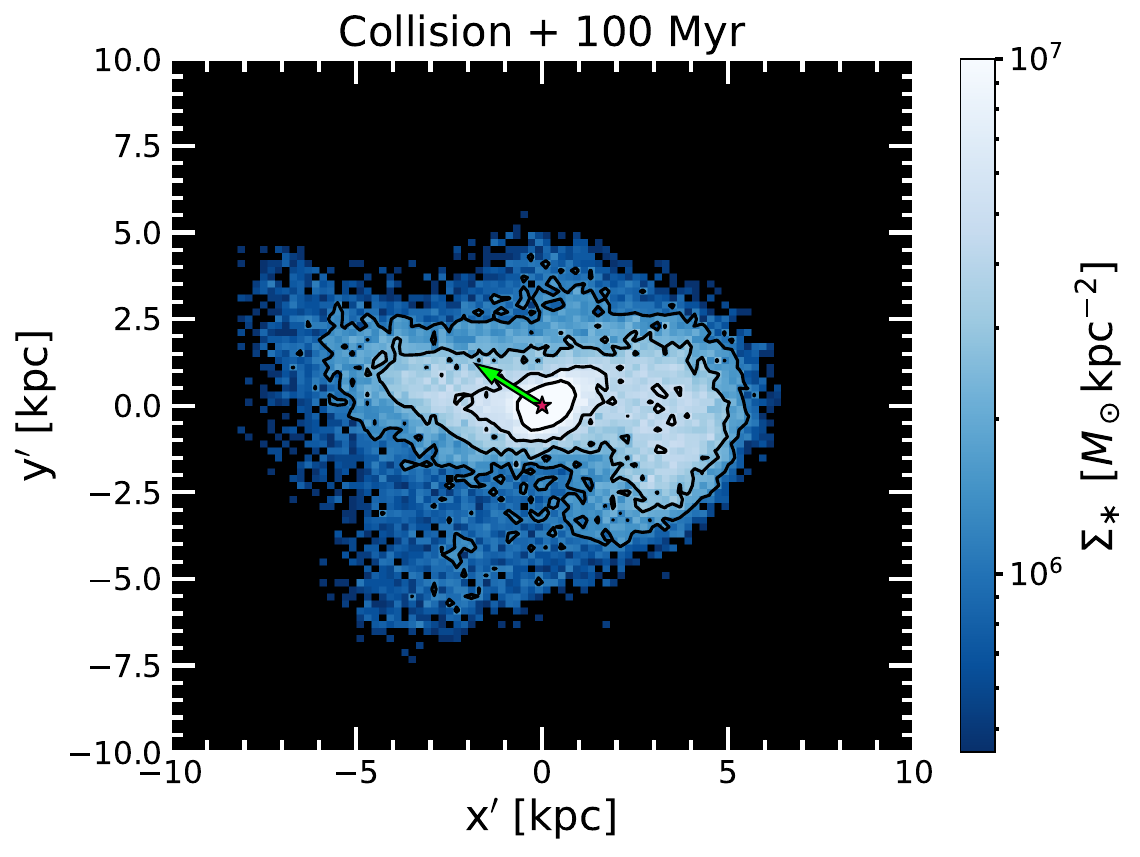}
    \includegraphics[width = 0.49\textwidth]{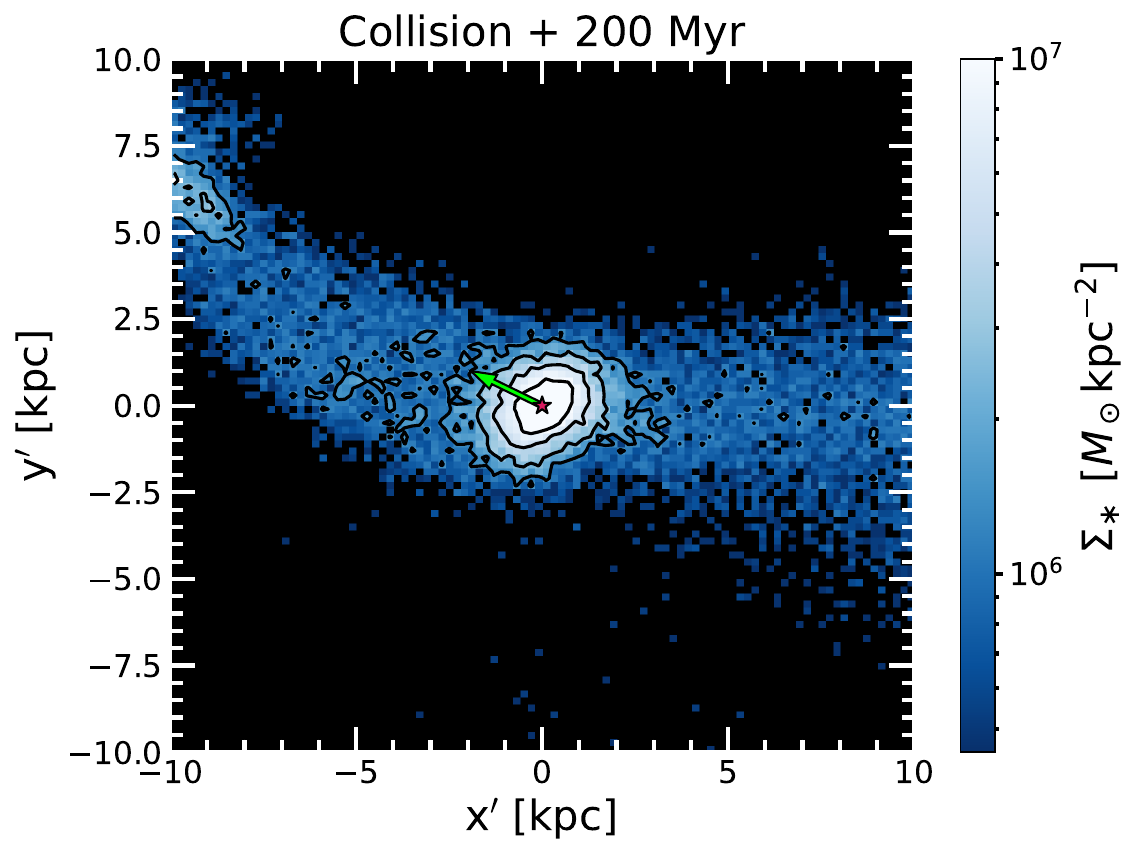}
    \includegraphics[width=0.49\textwidth]{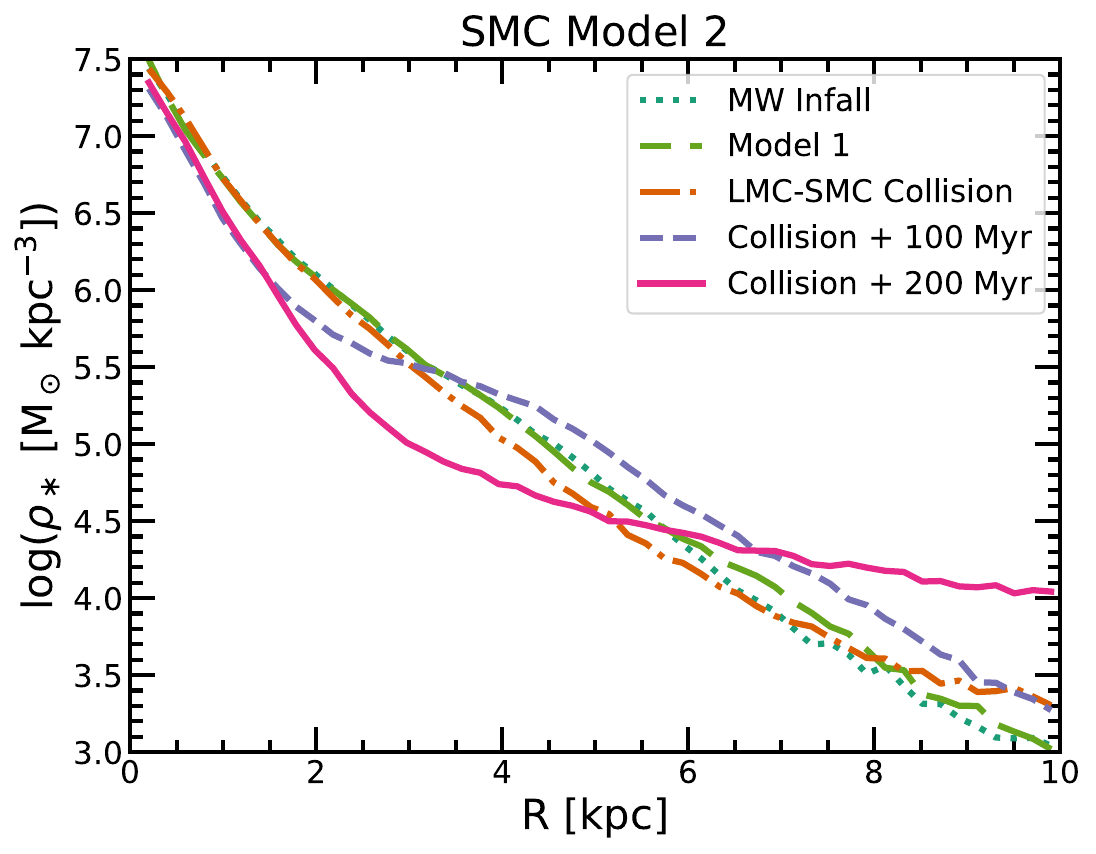}
    \caption{Time evolution of the Model 2 SMC's stellar surface density distribution in the plane of rotation. MW infall epoch, 100 Myr and 200 Myr post-collision are shown. The red star marks the stellar density center. Contour levels represent 5\%, 10\%, 20\%, 40\% of the peak surface density. The arrow points towards the LMC. Post-collision, the SMC's stellar distribution becomes significantly elongated along tidal structures. {\em Bottom right:} the spherically averaged stellar density profile for the SMC at different epochs. At MW Infall/Model 1 control, the SMC stellar distribution is roughly consistent with an exponential disk. Post-collision, the stellar density in the SMC's interior (R $\lesssim 3$ kpc) decreases by a factor of 2 - 3, and that in the SMC's outskirts (R $\gtrsim 3$ kpc) increases by a factor of 2 - 3, relative to the infall epoch. Post-collision stellar density profile is a combination of at least two power laws, with a break around 2 - 4 kpc. This means that the LMC's tidal influence is significant within 3-4 scale lengths of the initial SMC disk.}
    \label{fig:smc_m2_dens_profile}
\end{figure*}

In this section, we investigate the effect of the SMC-LMC collision on the simulated SMC's stellar morphology. In Figure \ref{fig:smc_m2_dens_profile}, the SMC's stellar surface density distribution is plotted at different simulation epochs. At MW Infall, the SMC's stellar density profile is roughly consistent with an exponential disk. Post-collision, the disk gets disrupted. The SMC's stellar distribution becomes significantly extended due to the LMC's tides. 

To quantify the tidal extension, we compute the SMC's spherically averaged stellar volume density profiles at different simulation epochs ({\em bottom right} panel of Figure \ref{fig:smc_m2_dens_profile}). At MW Infall, the SMC's stellar density profile is well described by an exponential disk. However, post-collision, the stellar density profile is instead a combination of at least two power laws. The SMC's stellar density at $R \lesssim 3$ kpc decreases by a factor of 2-3. Correspondingly, the stellar densities in the outer regions, $R \gtrsim 3$ kpc, increase by a factor of 2-3. The power-law break radius of the post-collision profiles is between $2-4$ kpc, depending on the time elapsed since the collision. The small break radii indicate that the internal structure of the SMC is significantly affected by the LMC's tides.
Indeed, there is observational evidence that the LMC's tides are significantly influencing the SMC's stellar distribution at $R < 4$ kpc (Z21), which we discuss in more detail in section \ref{sec:stellar_kine}. 

As mentioned in section \ref{sec:sims}, we utilize the B12 Model~1 SMC as a control, where the SMC is still subjected to MW and LMC tides, but does not collide with the LMC. Here, the Model 2 MW Infall stellar density profile is representative of the SMC profile in Model 1 at {\it present day}. Hence, neither weak LMC tides nor MW tides can significantly affect the SMC's stellar density profile at $R < 4$ kpc. A collision is required to significantly disrupt the internal stellar morphology of the SMC.

In the next section, we study the consequences of the SMC's post-collision extended stellar distribution on the LoS depth. 

\subsubsection{The SMC's LoS Elongation} \label{sec:los_elongation}

\begin{figure*}
    \centering
    \includegraphics[width=\textwidth]{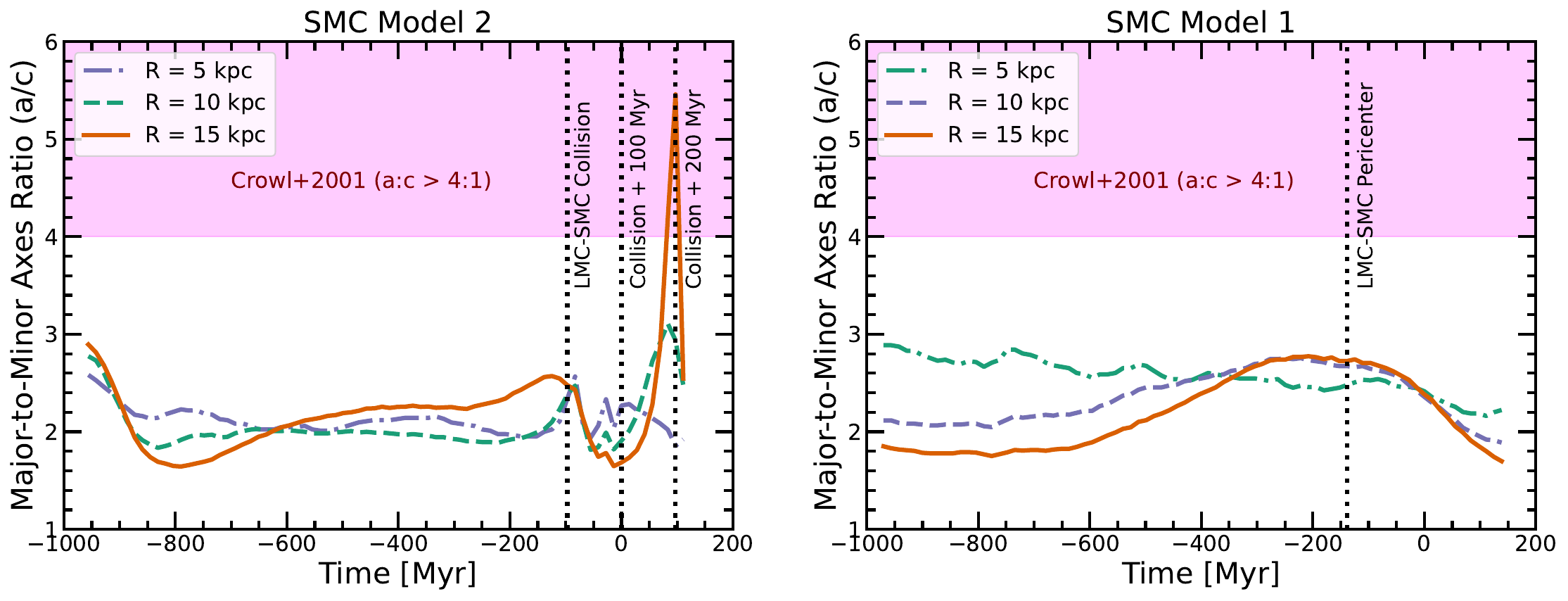}
    \includegraphics[width = \textwidth]{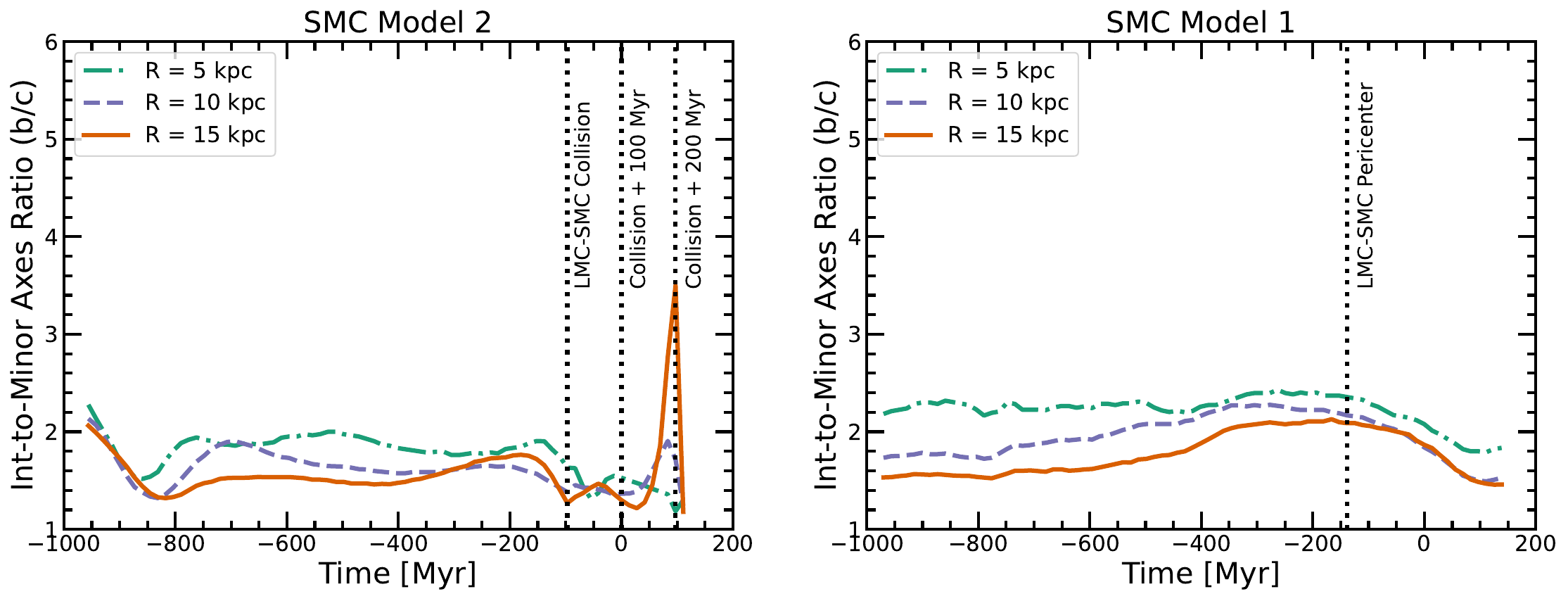}
    \caption{The simulated SMC's principal axis ratios computed at varying radii as a function of time. Top row shows the major-to-minor axis ratio (a/c) and the bottom row shows the intermediate-to-minor axis ratio. Columns show the results for Model 2 (left) and the Model 1 control (right). For Model 2, different epochs (SMC-LMC collision, 100 Myr and 200 Myr post-collision) are marked by dotted vertical lines. Post-collision, for $R<5$ kpc the SMC stellar structure is a triaxial ellipsoid. At larger spatial scales ($R < 15$ kpc), $a:c$ becomes as high as $5.5:1$, indicating that an axis has become significantly more elongated, consistent with observations, which find $a:c > 4:1$ \citep{Crowl2001}. Without a collision (Model 1, right), axis ratios remain $< 2.5:1$ for all spatial scales, which is inconsistent with observations.
}
    \label{fig:ax_ratio_time}
\end{figure*}

Measuring the SMC's LoS depth is challenging in both observations and simulations. With observations, the inferred values depend significantly on the criteria adopted for quantifying the LoS extent, photometric depth of the survey, as well as the dust extinction prescription and the stellar population used (see references in section \ref{sec:intro}). With simulations, it is challenging to define a LoS because the observed SMC's on-sky inclination is not well constrained. 

Given the aforementioned hurdles, we choose to study the simulated SMC's LoS extent by computing the axis ratios of the stellar distribution, since the axis ratios are independent of the assumed orientation. Several observational studies \citep{Crowl2001, Subramanian2012} have measured the axis ratios of the SMC's stellar distribution, finding the major-to-minor axis ratio ($a:c$) to be large ($> 4:1$), which is consistent with a large LoS depth. 

A common way to compute the axis ratios of a stellar distribution is by constructing the moment of inertia tensor ($\mathbf{\tilde{I}}$) \citep{Paz2006, Pejcha2009, Subramanian2012}. We center the simulated SMC at the stellar density center and, following \citet{Subramanian2012}, we write the cartesian representation of ($\mathbf{\tilde{I}}$) as:
\begin{equation} \label{eq:inertia_tensor}
    \mathbf{\tilde{I}_{ij}} = \sum_k m_k \left(||\vect{x_k}||^2 \delta_{ij} + (1 - 2\delta_{ij}) x^{(k)}_i x^{(k)}_j\right)
\end{equation}
\noindent where the indices $i, j = 1, 2, 3$ denote the three cartesian components of the $k^{th}$ star's position vector $\vect{x_k}$, $m_k$ is the mass of the $k^{th}$ star and $\delta_{ij}$ is the Kronecker delta function.

Diagonalizing $\mathbf{\tilde{I}_{ij}}$ yields three eigenvalues $\lambda_1, \lambda_2, \lambda_3$ (in descending order) which correspond to the moment of inertia about the three principal axes of the stellar distribution:
\begin{equation} \label{eq:l1}
    \lambda_1 = M(a^2 + b^2)
\end{equation}
\begin{equation} \label{eq:l2}
    \lambda_2 = M(a^2 + c^2)
\end{equation}
\begin{equation} \label{eq:l3}
    \lambda_3 = M(b^2 + c^2)
\end{equation}
\noindent where $a, b, c$ are the principal axes lengths (in descending order) and $M = \sum_k m_k$. Equations (\ref{eq:l1}), (\ref{eq:l2}) and (\ref{eq:l3}) are solved to obtain $a$, $b$ and $c$.

Further, let the eigenvectors corresponding to $\lambda_1, \lambda_2, \lambda_3$ be $\vect{e_1}$, $\vect{e_2}$, $\vect{e_3}$ respectively. Then, define a transformation matrix $\mathbf{T}$ with columns being the eigenvectors.
\begin{equation} \label{eq:transformation}
    \mathbf{T} = [\vect{e_1}|\vect{e_2}|\vect{e_3}]
\end{equation}
The matrix $\mathbf{T}$ can be used to rotate the stellar distribution such that the principal axes become aligned with the coordinate axes.

We compute $\mathbf{\tilde{I}}$ (and the resulting axis ratios) for the simulated SMC by considering stars within an enclosed radius of 5 kpc, 10 kpc and 15 kpc about the stellar density center. The minimum radius of 5 kpc is chosen to be the boundary of the SMC's main body, reflecting findings from Figure \ref{fig:smc_m2_dens_profile} that the break radius of the SMC's stellar density distribution is $< 4$ kpc. The maximum radius of 15 kpc is chosen to encompass the maximum extent probed by observational studies of the SMC's axis ratios. 

Figure \ref{fig:ax_ratio_time} shows the time evolution of the axis ratios for the SMC in the Model 2 and Model 1 (control) simulations. Different epochs in these simulations are marked with dotted vertical lines. 

The axis ratios at all times in Model 1 and prior to the collision in Model 2 are $(2 - 2.5) : (1.5 - 2): 1$. This indicates that the SMC has undergone significant vertical heating caused by previous distant ($> 10$ kpc separation at pericenter) encounters between the Clouds that have already happened before they infall into the MW (see section \ref{sec:stellar_kine} for more details).

In Model 2, post-collision, axis ratios are similar to the pre-collision values for $R < 5$ kpc, indicating a triaxial main body. This finding is qualitatively consistent with observations; for e.g. \cite{Subramanian2012} used distances to SMC's RR Lyrae stars in the SMC from the OGLE III survey \citep{Soszynski2010} and concluded that the stellar distribution is triaxial, but the exact axis ratios are sensitive to the size of the spatial region used to select the stars.

However, for Model 2 at large spatial scales ($R < 15$ kpc), the major-to-minor axis ratio achieves values as high as $5.5:1$. This indicates that the post-collision SMC has developed a long axis at large radii that was non-existent pre-collision. The simulated post-collision axis ratios at $R < 15$ kpc are qualitatively consistent with observations; e.g. \cite{Crowl2001} used distances to SMC's star clusters and measured axis ratios of $1:2:4$ for $R < 12$ kpc. In Model 1, the major to minor axis ratio remains smaller than $2.5:1$ at all times and for all spatial scales, which is discrepant with the observed SMC's LoS depth.

Note that the Model 2 post-collision major-to-minor axis ratio for $R < 15$ kpc decreases to $\approx 2.5:1$ after attaining a peak of $\approx 5.5:1$. However, we only have access to one more Model 2 simulation snapshot beyond this peak epoch. Hence, at this stage, we are unable to interpret the evolution of the SMC's axis ratio beyond the peak epoch. 

\subsubsection{Orientation of the SMC's Major Axis} \label{sec:orientation}

\begin{figure*}
    \centering
    \includegraphics[height = 0.3\textwidth, width=0.4\textwidth]{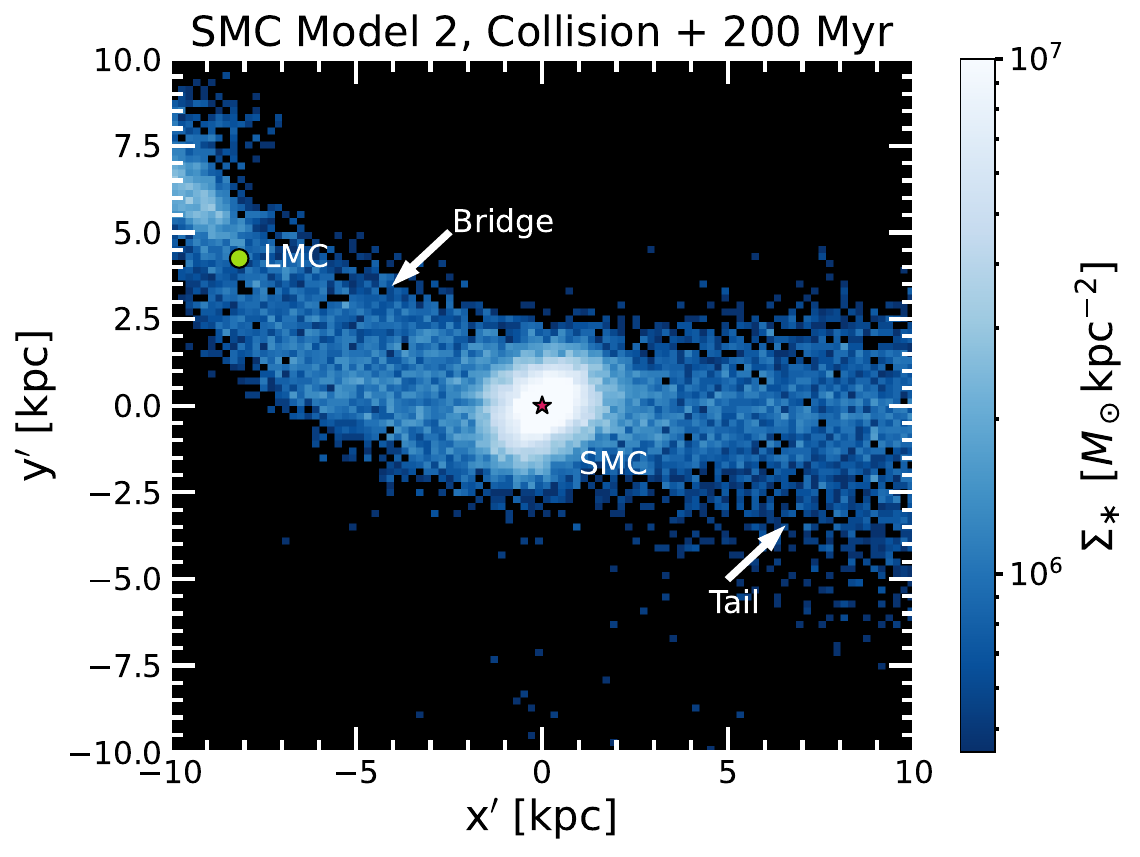}
    \includegraphics[height = 0.3\textwidth, width = 0.5\textwidth]{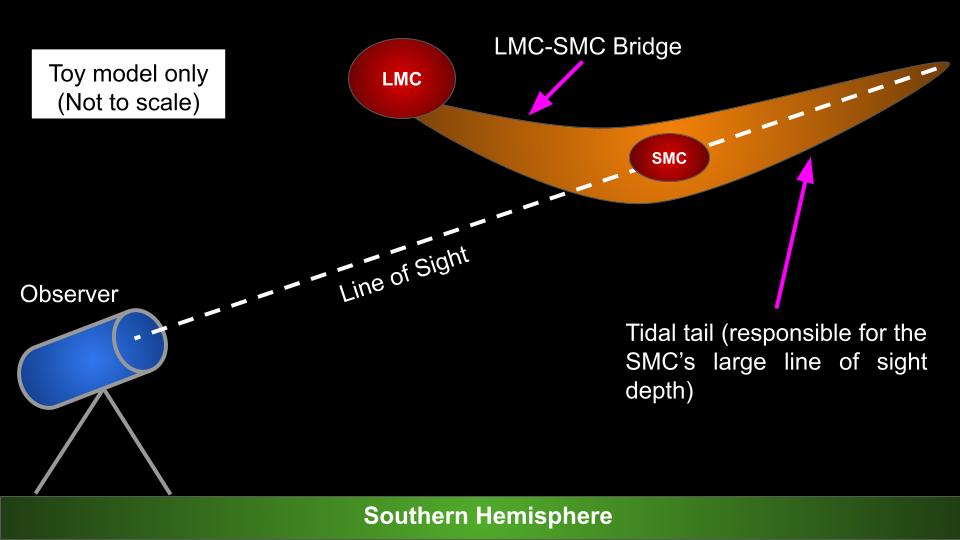}
    \caption{The {\em left panel} shows a projection of the Model 2 simulated SMC's stellar density distribution 200 Myr after the SMC-LMC collision. The x' and y' axis correspond to the SMC plane of rotation. The stellar density centers of the SMC and LMC are marked by the red star and green circle, respectively. The LMC-SMC bridge and the SMC's tidal tail can be clearly seen. The latter is likely responsible for the SMC's large LoS depth \citep{DB2012}. The {\em right panel} shows a cartoon illustration (not to scale) of how we might be observing the SMC. Our LoS likely passes through the SMC's tidal tail, leading to a large observed LoS depth. The LMC-SMC bridge, being misaligned with the tidal tail, does not significantly affect the LoS depth. 
    }
    \label{fig:los_vis}
\end{figure*}

\begin{figure*}
    \centering
    \includegraphics[width=\textwidth]{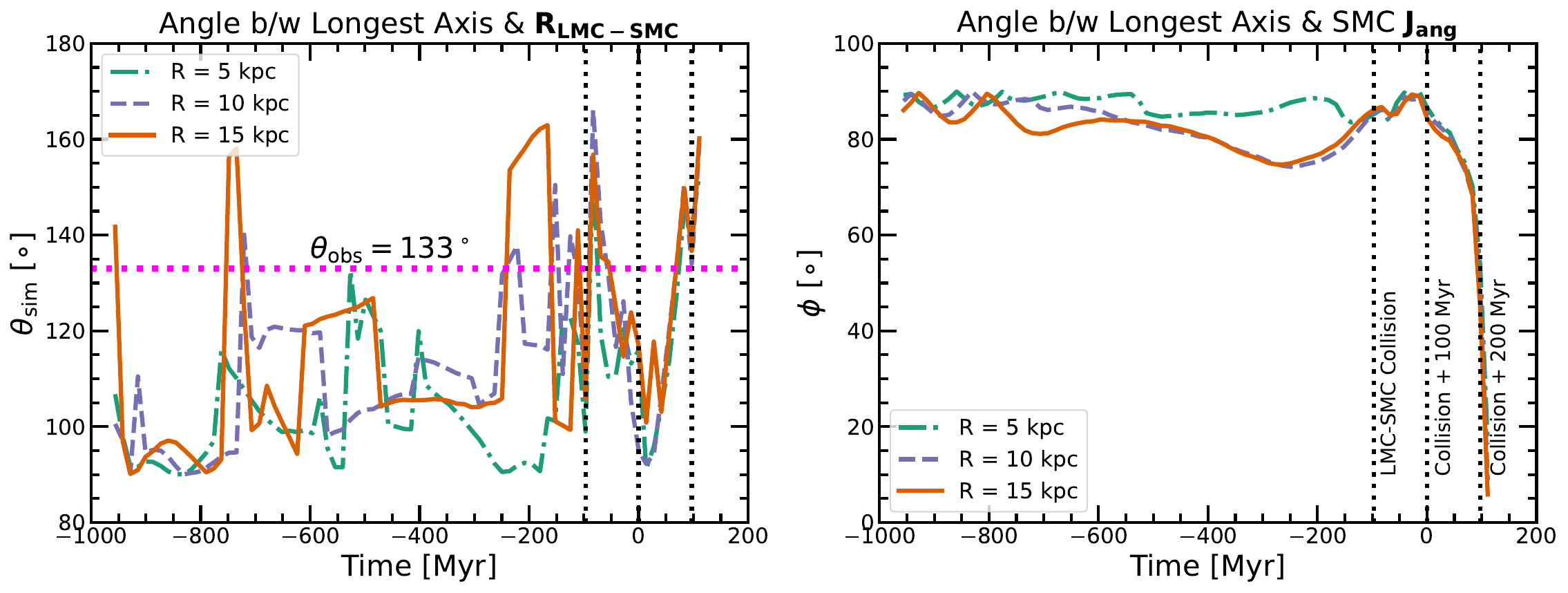}
    \caption{The angle between the Model 2 SMC's longest axis and (1) the LMC-SMC position vector ($\vect{R_{\rm LMC-SMC}}$) ($\theta_{\rm sim}$, {\em left panel}), (2) the SMC's angular momentum vector ($\vect{J_{\rm ang}}$) ($\phi$, {\em right panel}). {\it Left panel:} The longest axis is computed for varying radial extents (5 kpc, 10 kpc, 15 kpc), as described in Figure \ref{fig:ax_ratio_time}. Three epochs are marked with dotted vertical lines: LMC-SMC collision; 100 Myr and 200 Myr post-collision. Post-collision, $\theta_{\rm sim}$ ranges between 100$^\circ$ and 160$^\circ$, representing the angle between the simulated SMC's tidal bridge and tail. Assuming the observed SMC's tidal tail is oriented along the LoS (Figure \ref{fig:los_vis}), the angle between the observed bridge and tail is $\theta_{\rm obs} = 133^\circ$. $\theta_{\rm obs}$ resides within the range of post-collision $\theta_{\rm sim}$, implying that the viewing perspective suggested in Figure \ref{fig:los_vis} is reasonable. {\it Right panel:} Pre-collision, the SMC's longest axis is oriented $\approx 90^\circ$ with $\vect{J_{\rm ang}}$, meaning the longest axis is in the plane of rotation. Post-collision, the longest axis is no longer in the plane of rotation, and is more aligned with $\vect{J_{\rm ang}}$. The SMC is being tidally extended, with a significant extension perpendicular to its plane of rotation. Hence, the observed SMC will have a significant LoS depth even if the plane of rotation is not oriented edge-on.}
    \label{fig:ax_angle}
\end{figure*}

Because the long axis appears only at large spatial scales, it likely corresponds to the SMC's tidal structures born out of the SMC-LMC collision. In Figure \ref{fig:los_vis} {\em left panel}, we show the Model-2 SMC's stellar surface density distribution 200 Myr post-collision, in the SMC's plane of rotation. Two tidal structures can be clearly seen: a bridge of stars connecting the SMC and LMC; and a tidal tail of stars along the positive x'-axis. These are classic examples of tidal bridges and tails that form out of binary galaxy interactions \citep{Toomre1972}.

If the SMC's tidal tail is oriented close to the LoS, the extended stellar distribution would naturally lead to a large LoS depth. Figure \ref{fig:los_vis} {\em right panel} illustrates the proposed viewing perspective, where the LoS is aligned along the the SMC's tidal tail. 

Previous theoretical studies \citep{Yoshizawa2003, Connors2006, DB2012} have also suggested that the SMC's tidal tail is the origin of the large LoS depth; this tidal tail is referred to as the \enquote{counter bridge} in \cite{DB2012}. However, the aforementioned studies did not consider a direct collision between the Clouds, and instead considered more distant encounters (closest separation $\gtrsim 6$ kpc). In this work, we have shown that a direct SMC-LMC collision also leads to a similar tidal tail, thereby explaining the large axis ratios (or equivalently a large LoS depth). 

The viewing perspective suggested in Figure \ref{fig:los_vis} requires the bridge to be misaligned with the tidal tail. Observationally, the angle between the bridge and the LoS ($\theta_{\rm obs}$) is computed using the Galactocentric coordinates of the SMC and the Sun \citep{Kallivayalil2013}:
\begin{equation}
    \vect{R_{\rm SMC}} = [14.83, -38.08, -44.16] \:\: \rm{kpc}
\end{equation}
\begin{equation}
    \vect{R_{\rm LMC}} = [-1.28, -41.05, -27.83] \:\: \rm{kpc}
\end{equation}
\begin{equation}
    \vect{R_{\odot}} = [-8.5, 0, 0] \:\: \rm{kpc}
\end{equation}
\noindent The observed LoS vector is given by $\vect{R_{\rm SMC}} - \vect{R_{\odot}}$, and the vector corresponding to the observed bridge is $\vect{R_{\rm SMC}} - \vect{R_{\rm LMC}}$. Taking a dot product, we find that $\theta_{\rm obs} = 133^\circ$. 

To check for consistency between the proposed theoretical scenario and the observed viewing geometry of the system, we compute the angle between the SMC bridge and tidal tail in Model 2 ($\theta_{\rm sim}$) and compare it with observations ($\theta_{\rm obs}$). 

In the simulation, the orientation of the tidal tail is along the eigenvector $\vect{e_1}$ (eq. \ref{eq:transformation}), and the bridge resides along the SMC-LMC position vector ($\vect{R_{\rm SMC-LMC}}$). Figure \ref{fig:ax_angle} {\em left panel} shows $\theta_{\rm sim}$ as a function of time. Post-collision, the angle between the simulated bridge and tail ranges between 100$^\circ$ and 160$^\circ$. $\theta_{\rm obs}$ resides within the range of $\theta_{\rm sim}$. Hence, the viewing scenario proposed in Figure \ref{fig:los_vis} is reasonable.

Note that the tidal tail discussed in the context of SMC's LoS depth is different from the MS and its stellar counterpart \citep{Chandra2023, Zaritsky2025}. A tidal stream analogous to the MS forms in the B12 simulation at previous SMC-LMC pericentric passages \citep[see also][]{Yoshizawa2003, Connors2006, DB2012, Pardy2018}. Whereas, the tidal tail we discuss here is formed out of the recent SMC-LMC collision, $< 200$ Myr ago.

In principle, the idea that the tidal tail must be along our LoS may also inform the geometry of the SMC's main body. For example, if the SMC is currently an extended disk galaxy seen edge-on, thereby explaining the large LoS depth, the plane of rotation and the longest axis would be roughly aligned. This is counter to observations, where the SMC exhibits a small remnant rotation in the plane of the sky, i.e. 90$^\circ$ to the LoS. Here we demonstrate that, although the SMC was initialized with a stellar disk, the post-collision tidal tail can be significantly misaligned with the plane of rotation.  

Figure \ref{fig:ax_angle} {\em right panel} shows the angle ($\phi$) between the simulated SMC's longest axis and the SMC angular momentum vector $\vect{J_{\rm ang}}$, which defines the plane of rotation. If $\phi = 90^\circ$, it means the longest axis resides in the plane of rotation. If $\phi = 0^\circ$, it means the longest axis is perpendicular to the plane of rotation. Pre-collision, the simulated SMC's longest axis resides in the plane of rotation. This is in accordance with expectations, since the pre-collision SMC has a well defined disk (Figure \ref{fig:smc_m2_dens_profile}), where the longest axis will be the diameter of the disk. However, post-collision, the longest axis becomes increasingly misaligned with the plane of rotation, and becomes more and more aligned with $\vect{J_{\rm ang}}$. This indicates that the SMC's tidal tail has a significant component perpendicular to the plane of rotation. Hence, a SMC-LMC collision scenario can explain the LoS depth of the SMC without requiring the plane of rotation to be seen edge-on along the LoS.

Unfortunately, this also means that the inclination of the SMC's plane of rotation cannot be identified using the SMC's longest axis. Constraining the inclination of the observed SMC's plane of rotation is required to accurately de-project the stellar and gas velocities, which is needed to accurately constrain the SMC's mass profile.

\subsection{The Simulated SMC's Stellar Kinematics} \label{sec:stellar_kine}

\begin{figure*}
    \centering
    \includegraphics[width=0.49\textwidth]{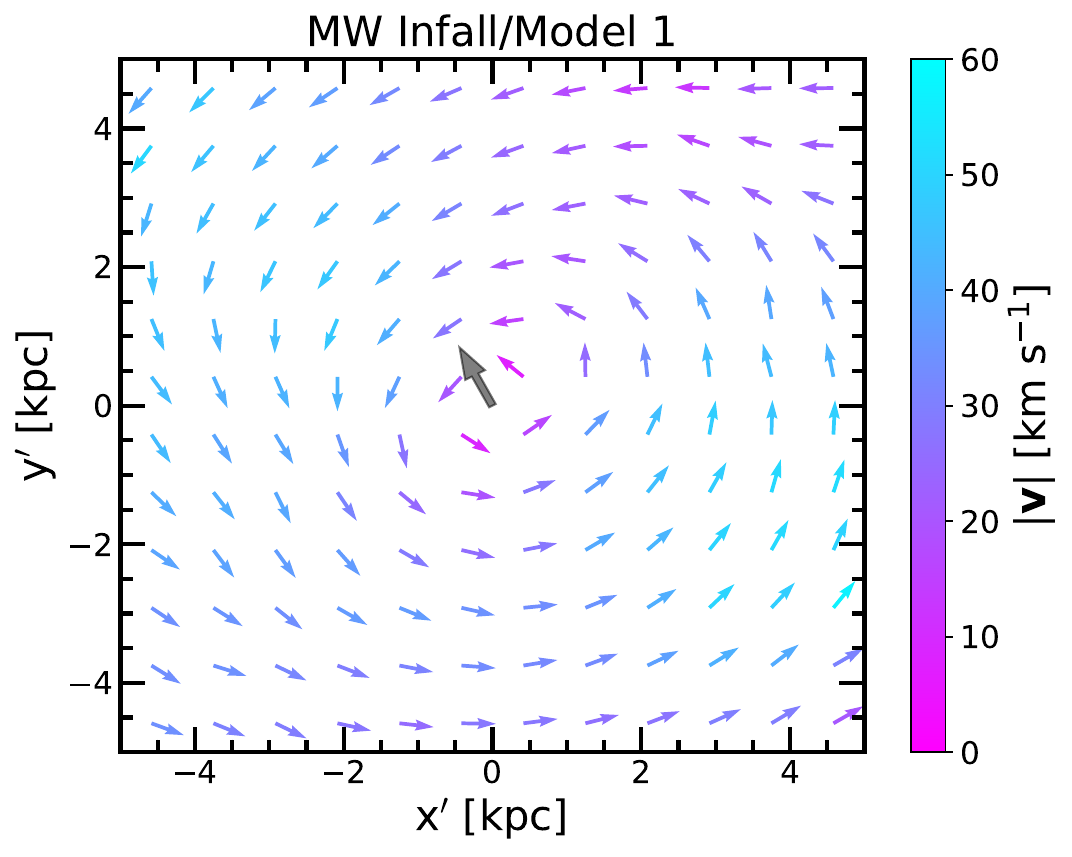}
    \includegraphics[width=0.49\textwidth]{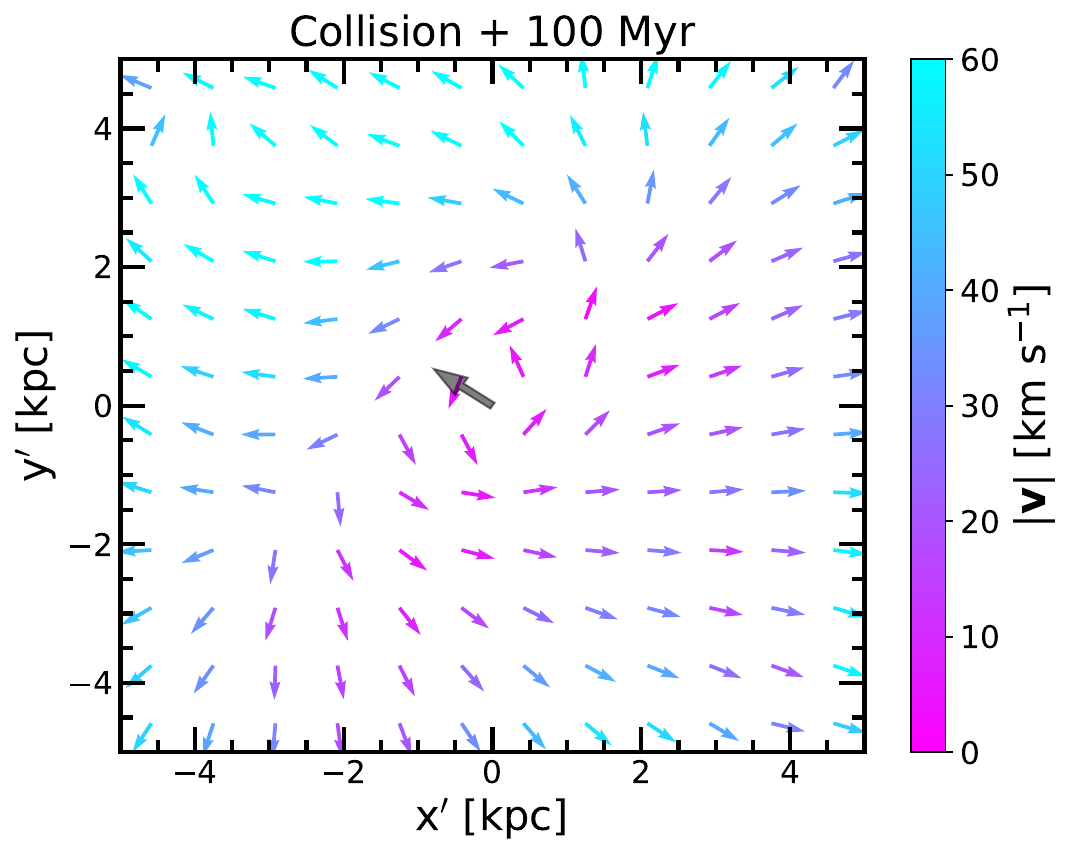}
    \includegraphics[width=0.49\textwidth]{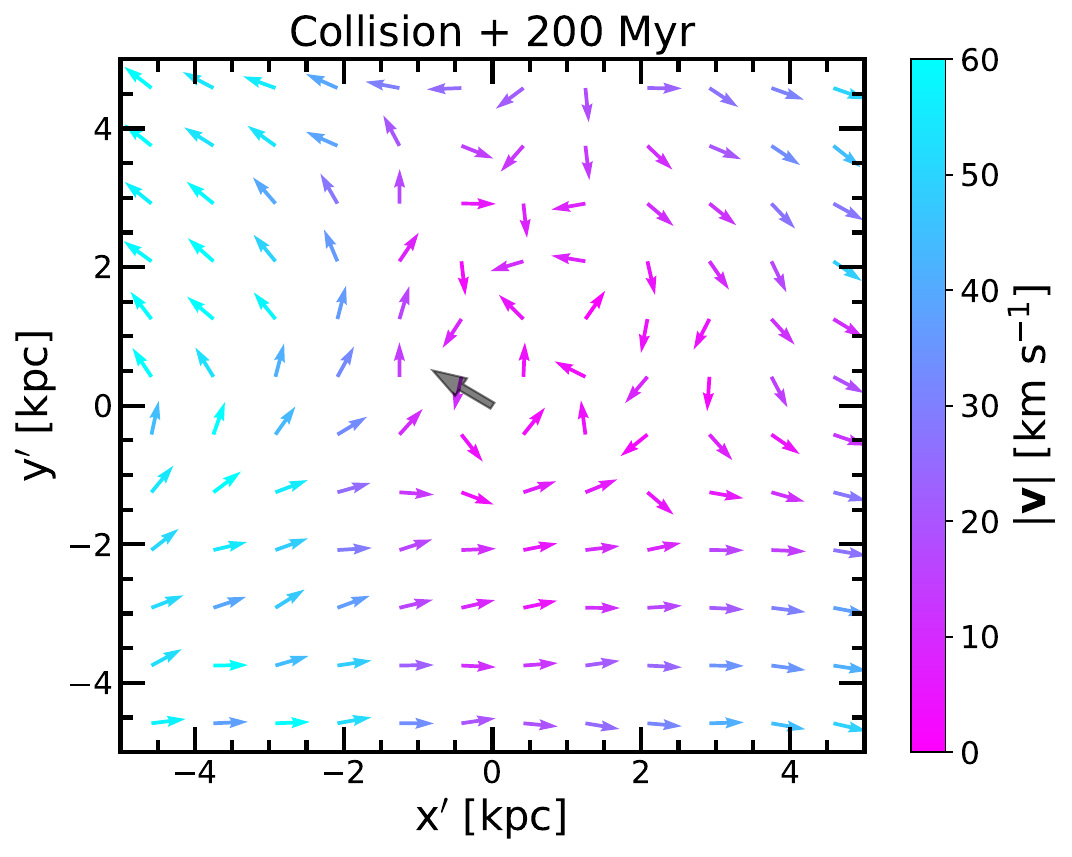}
    \includegraphics[width=0.49\textwidth]{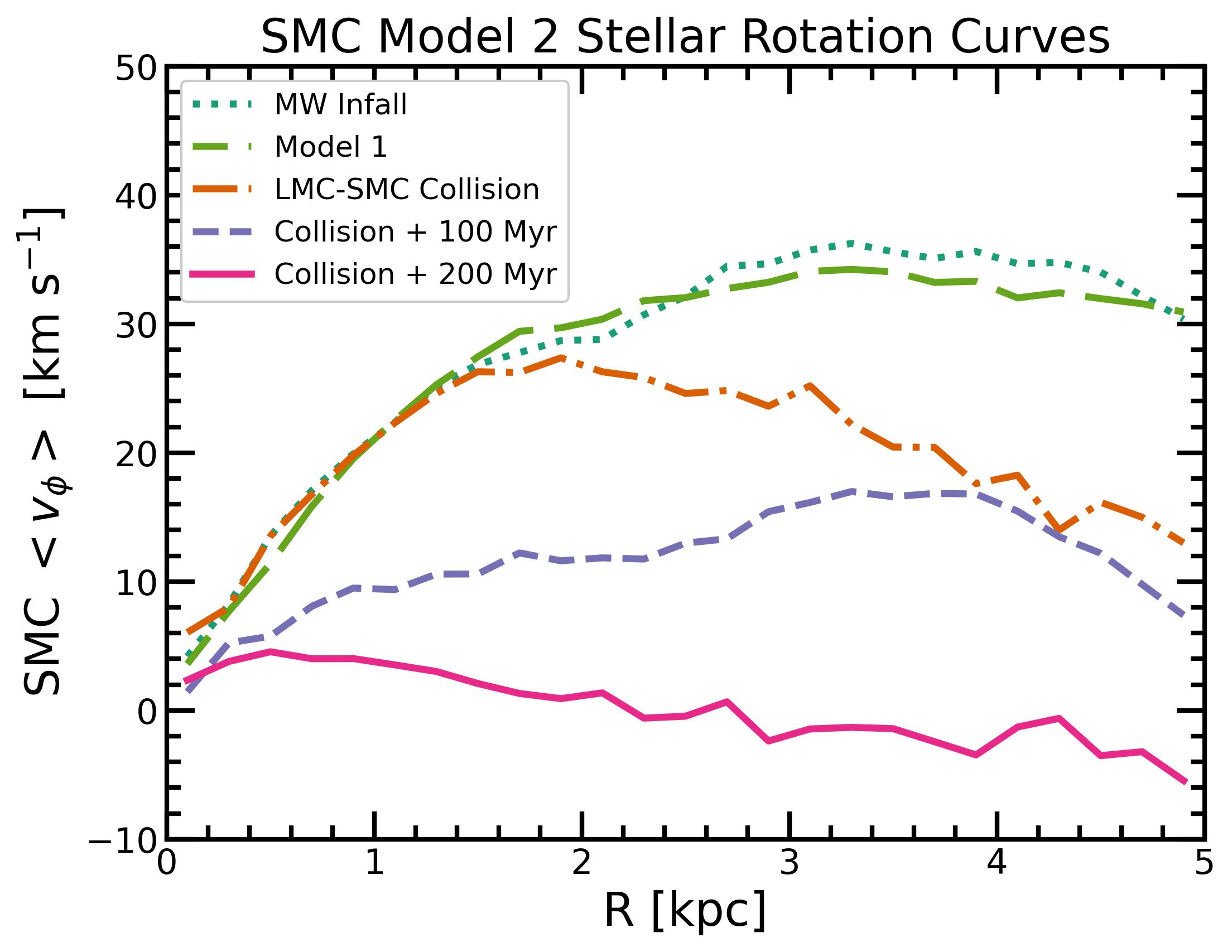}
    \caption{The stellar velocity field and rotation curve of the simulated Model 2 SMC at three epochs: MW infall ({\em top left}); 100 Myr post LMC-SMC collision ({\em top right}); 200 Myr post-collision ({\em bottom left}). Velocity fields are computed in the SMC's plane of rotation. The arrows denote the direction of the average velocity vector in a 0.8 kpc $\times$ 0.8 kpc spatial bin, with the color scale representing the magnitude. The grey arrow is centered at the stellar kinematic center, and points in the direction of the LMC. Pre-collision (MW Infall), the SMC exhibits a coherent rotation field. Post-collision, the SMC's rotation field is disrupted. Most of the SMC's stars at $R > 1$ kpc are moving radially outwards. The direction of maximum radial motion is consistent with the LMC-SMC position vector. This is in agreement with observations (Z21). Depending on the time elapsed since the collision, there is remnant rotation ($<\sim 10$ km s$^{-1}$) at $R < 1$ kpc, as observed (Z21). {\em Bottom right panel} shows the azimuthally averaged stellar rotation ($v_\phi$) curve for different epochs. The peak rotation speed evolves significantly post-collision, decreasing from $
    \sim 40$ km s$^{-1}$ to $< 10$ km s$^{-1}$. The Model 2 rotation field at the MW infall epoch is representative of the Model 1 control at present day. Hence, neither weak LMC tides nor MW tides are sufficient to disrupt the SMC's stellar rotation to the level seen in observations. An SMC-LMC collision is necessary.}
    \label{fig:vel_field}
\end{figure*}

\begin{figure}
    \centering
    \includegraphics[width=\columnwidth]{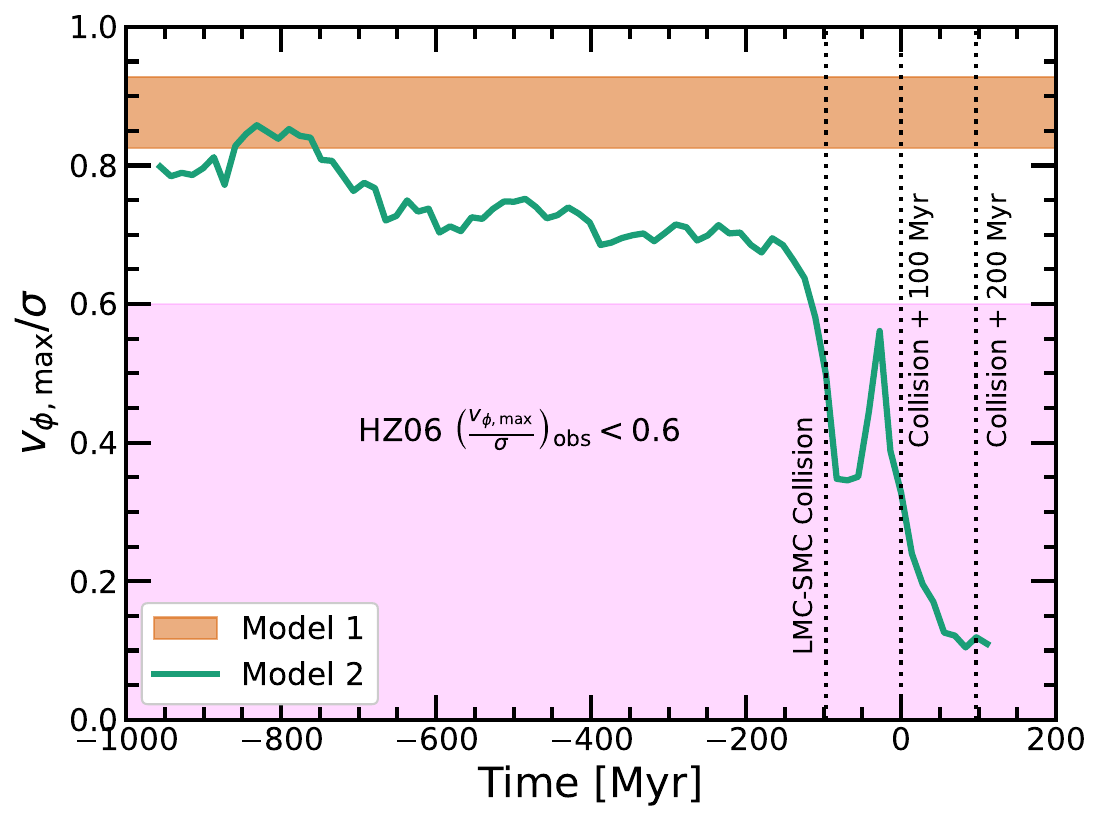}
    \caption{The ratio of the peak stellar rotation speed to the stellar velocity dispersion in the Model 2 SMC, as a function of time. The orange shaded band depicts the time-averaged mean ratio (0.87) and the standard deviation (0.05) for the Model 1 simulation (where the SMC and LMC do not collide). The green solid line depicts the Model 2 simulation (where the SMC and LMC collide). Three epochs are marked with vertical dotted lines: the LMC-SMC collision; 100 Myr; and 200 Myr post-collision. Post-collision, the Model 2 SMC becomes highly dispersion dominated, which is consistent with observations where $v_{\phi, \rm{max}}/\sigma < 0.6$ (HZ06, denoted by the pink shaded region). In Model 1, the SMC has significant rotational support at all times, which is inconsistent with observations. Hence, an LMC-SMC collision can convert an initially rotating SMC disk to a highly dispersion dominated galaxy, offering a new pathway to transform a dIrr type galaxy to dE/dSph type.}
    \label{fig:v_sigma}
\end{figure}

In the previous section we showed that the tidal tail born out of the SMC-LMC collision can explain the origin of the SMC's large LoS depth. In this section, we investigate whether the same collision scenario can explain the SMC's lack of stellar rotation. 

Z21 analyzed Gaia DR2 \citep{Brown2018} PMs of the SMC's red giant stars combined with LoS velocities from \cite{Dobbie2014} and the APOGEE survey \citep{Majewski2017}. They found that the internal rotation field of the SMC is very complex. Stars in the inner $\sim1$ kpc show small rotation ($\approx 10$ km s$^{-1}$), whereas the kinematics beyond $\sim1$ kpc are dominated by radially outward motions \citep[see also][]{Niederhofer2021, Dias2021, Dias2022, Parisi2024}, likely due to the LMC's tides. This suggests that the LMC's tides must be dominant over the SMC's own gravitational force well inside the SMC's main body (R $<4$ kpc).\cite{DeLeo2020} also arrived at a similar conclusion after analyzing the PMs and LoS velocities of a sample of 3000 SMC red giant stars.

Recall that in Figure \ref{fig:smc_m2_dens_profile}, the SMC's post-collision stellar density profile exhibited significant deviations from the initial profile, starting at 2-4 kpc. This suggests that the LMC's tides are significant inside the SMC's main body in Model 2. Here we analyze the corresponding internal stellar kinematics.  

In Figure \ref{fig:vel_field}, the simulated Model 2 SMC's stellar velocity field is visualized in the plane of rotation at different epochs. Pre-collision ({\em top left panel}), the SMC has a well defined stellar rotation field. Post-collision ({\em top right and bottom left panels}), the stellar rotation field is significantly disrupted. Most of the stars at R $>2$ kpc are moving radially outwards. Further, the direction of maximal outward motion is consistent with $\vect{R_{\rm LMC-SMC}}$, indicating that the motions are driven by the LMC's tides. There is some remnant rotation with a small amplitude ($< 10$ km s$^{-1}$) in the inner $\sim1$ kpc. This remnant rotation suggests that stars in the inner $1-2$ kpc are bound to the SMC. At 100 Myr post collision, the simulated SMC's total mass enclosed within a radius of 2 kpc is $6.8 \times 10^8$ M$_\odot$. A rough estimate of the circular velocity at 2 kpc can be obtained by assuming spherical symmetry of the mass enclosed. This circular velocity turns out to be $\approx 38$ km s$^{-1}$, reaffirming that stars at $R \lesssim 2$ kpc are likely bound to the SMC.

The simulated post-collision SMC kinematics are consistent with the remnant rotation and the tidal expansion inferred by Z21. Hence, the SMC-LMC collision can explain the observed SMC's stellar kinematics. 

The {\em bottom right} panel of Figure \ref{fig:vel_field} shows the Model~2 SMC's azimuthally averaged stellar rotation curve at different epochs, which reinforces our findings from visualization of the stellar velocity fields. Pre-collision (MW Infall), the SMC has a well defined rotation curve peaking at $\approx 40$ km s$^{-1}$. The rotation curve starts to disrupt at the SMC-LMC collision epoch, and the rotation amplitude in the inner 2 kpc becomes small ($< 10$ km s$^{-1}$) post-collision.

The Model 2 stellar velocity field and rotation curve at the MW infall epoch are representative of the simulated SMC at present day in the Model 1 control. Hence, in the Model 1 scenario, where the SMC and LMC do not collide, the stellar kinematics exhibit a coherent rotation field out to $\leq$4 kpc with a large rotation amplitude ($\approx$ 40 km s$^{-1}$), which is inconsistent with observations. Thus, neither weak LMC tides nor MW tides are sufficient to disrupt the SMC's stellar rotation field at the level seen in observations. Instead, a strong SMC-LMC interaction, like a recent collision, is required.

Observational studies \citep[e.g.][hereafter HZ06]{Harris2006} have found that the SMC is a highly dispersion dominated galaxy, where the ratio of the peak rotation velocity to the total velocity dispersion ($\frac{v_{\phi, \rm{max}}}{\sigma}$) is significantly less than 0.6. Thus, not only are the SMC's stars moving radially outwards (Z21), their kinematics are also significantly heated. Next, we investigate whether the SMC-LMC collision can explain the heating of the SMC's stellar kinematics. 

We show the time evolution of the simulated Model~2 SMC's $\frac{v_{\phi, \rm{max}}}{\sigma}$ in Figure \ref{fig:v_sigma}. The total velocity dispersion is calculated within the radius of the peak of the azimuthally averaged rotation curve. 

At MW infall, the Model 2 SMC $\frac{v_{\phi, \rm{max}}}{\sigma} \approx 0.8$, indicating a comparable rotation and dispersion support. This is also consistent with the SMC's morphology  prior to the collision (see Figure \ref{fig:ax_ratio_time}). In Figure \ref{fig:v_sigma}, the Model 1 control SMC's time-averaged mean $\frac{v_{\phi, \rm{max}}}{\sigma} = 0.88$ and $1-\sigma$ spread (0.05), are indicated by the orange band. The Model 1 scenario, where the SMC and LMC do not collide, is thus inconsistent with observations. 

It has been shown that multiple distant encounters can significantly heat a disk \citep[e.g.][]{Mayer2001, Kazantzidis2011, Kazantzidis2013, Kazantzidis2017, Lokas2012, Lokas2014, Lokas2015, Semczuk2018}. However, distant encounters with the MW and LMC (Model 1) or just the LMC (Model 2, MW Infall) are not sufficient to make the SMC dispersion dominated at the level seen in observations.

Post-collision, the SMC's $\frac{v_{\phi, \rm{max}}}{\sigma}$ becomes less than 0.2, indicating an almost completely dispersion dominated galaxy with negligible rotation support, which is consistent with observations. A strong SMC-LMC interaction, such as a direct collision, is required to make the SMC's internal kinematics strongly dispersion dominated.

The SMC's dispersion dominated stellar kinematics offer new insights into dramatic transformations of galaxy morphology. Model 2 illustrates that a direct collision can almost completely disrupt the disk of a dIrr type galaxy, converting it into a dE/dSph type. We argue that the SMC has recently undergone this morphological transformation.

Referring back to the BTFR (Figure \ref{fig:btfr} {\em left panel}), the simulated SMC's stellar disk was initialized to be consistent with BTFR. We have shown that the SMC-LMC collision can significantly disrupt the SMC's stellar rotation, which will make the SMC's stellar kinematics strongly discrepant with the BTFR.

\subsection{The Simulated SMC's Gas Kinematics} \label{sec:gas_kine}

\begin{figure*}
    \centering
    \includegraphics[width=0.49\linewidth]{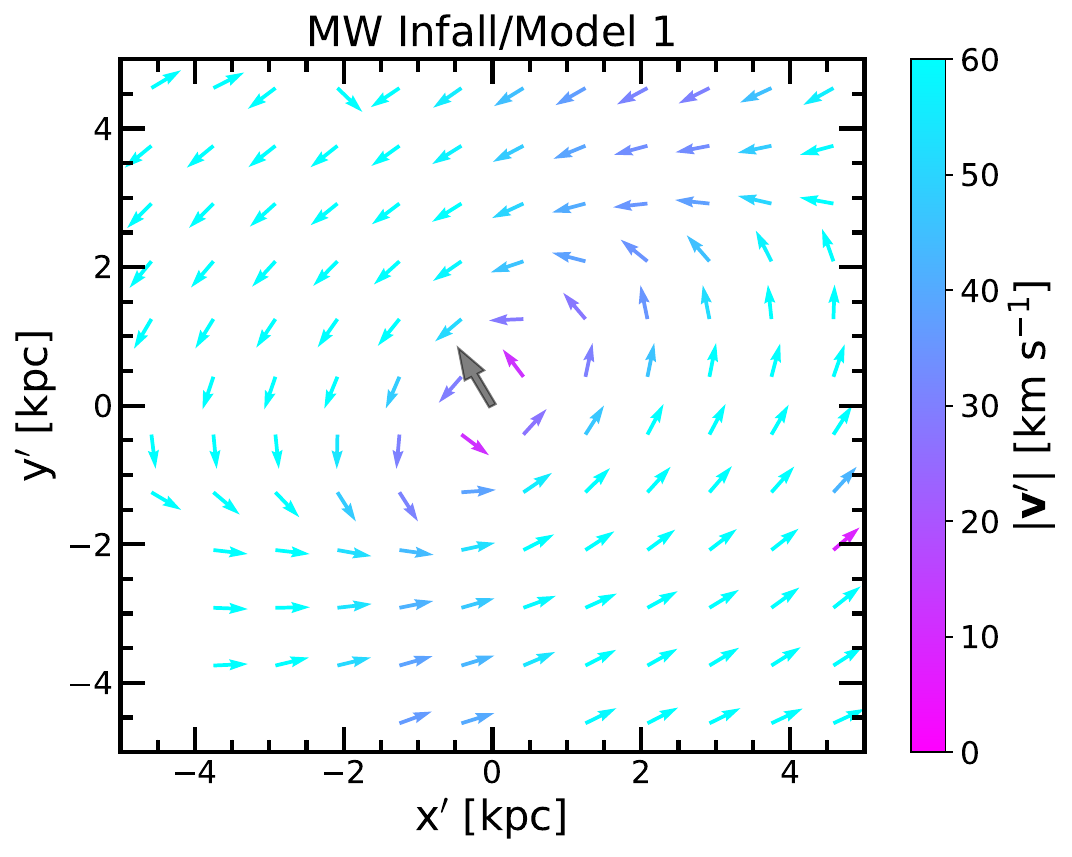}
    \includegraphics[width=0.49\linewidth]{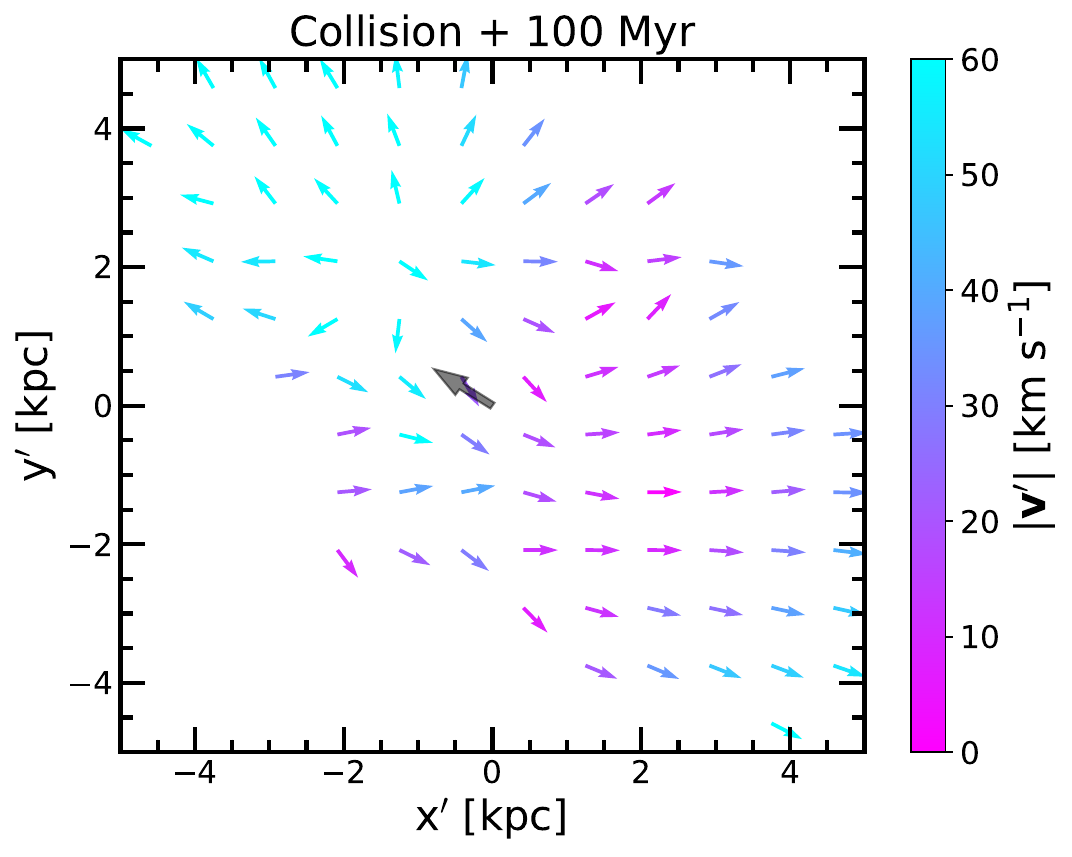}
    \includegraphics[width=0.49\linewidth]{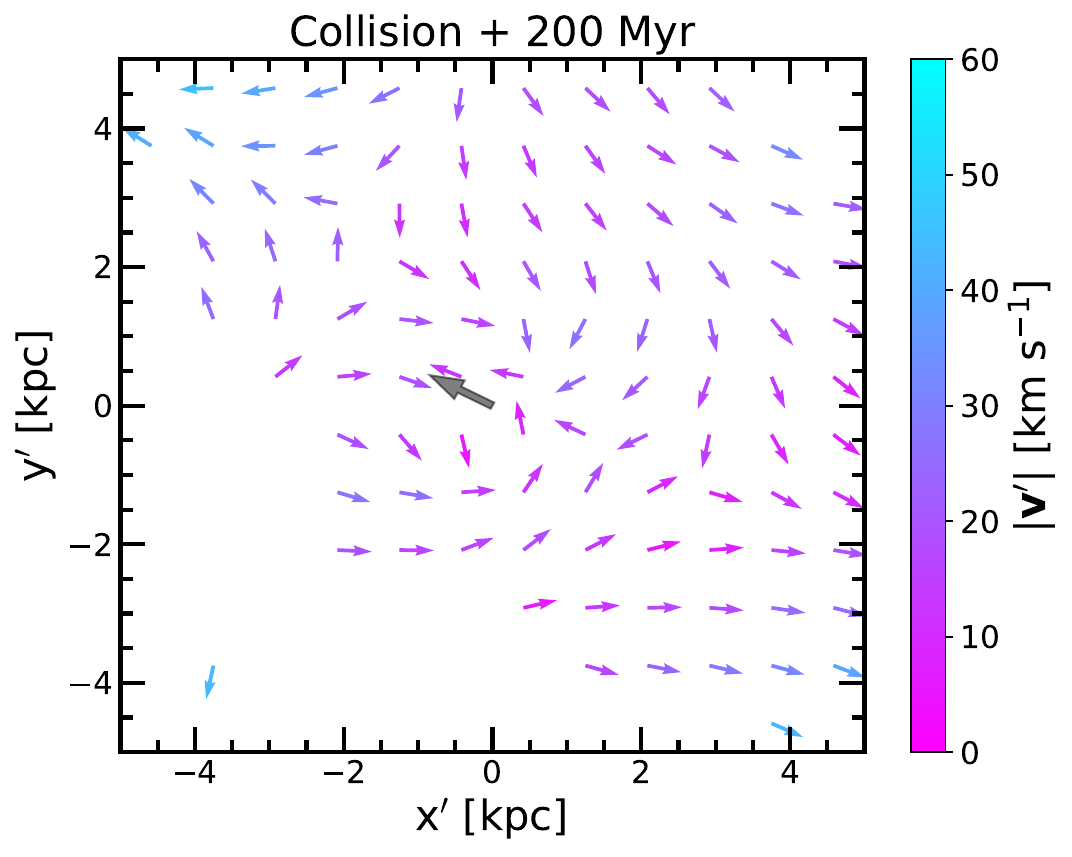}
    \includegraphics[width=0.49\linewidth]{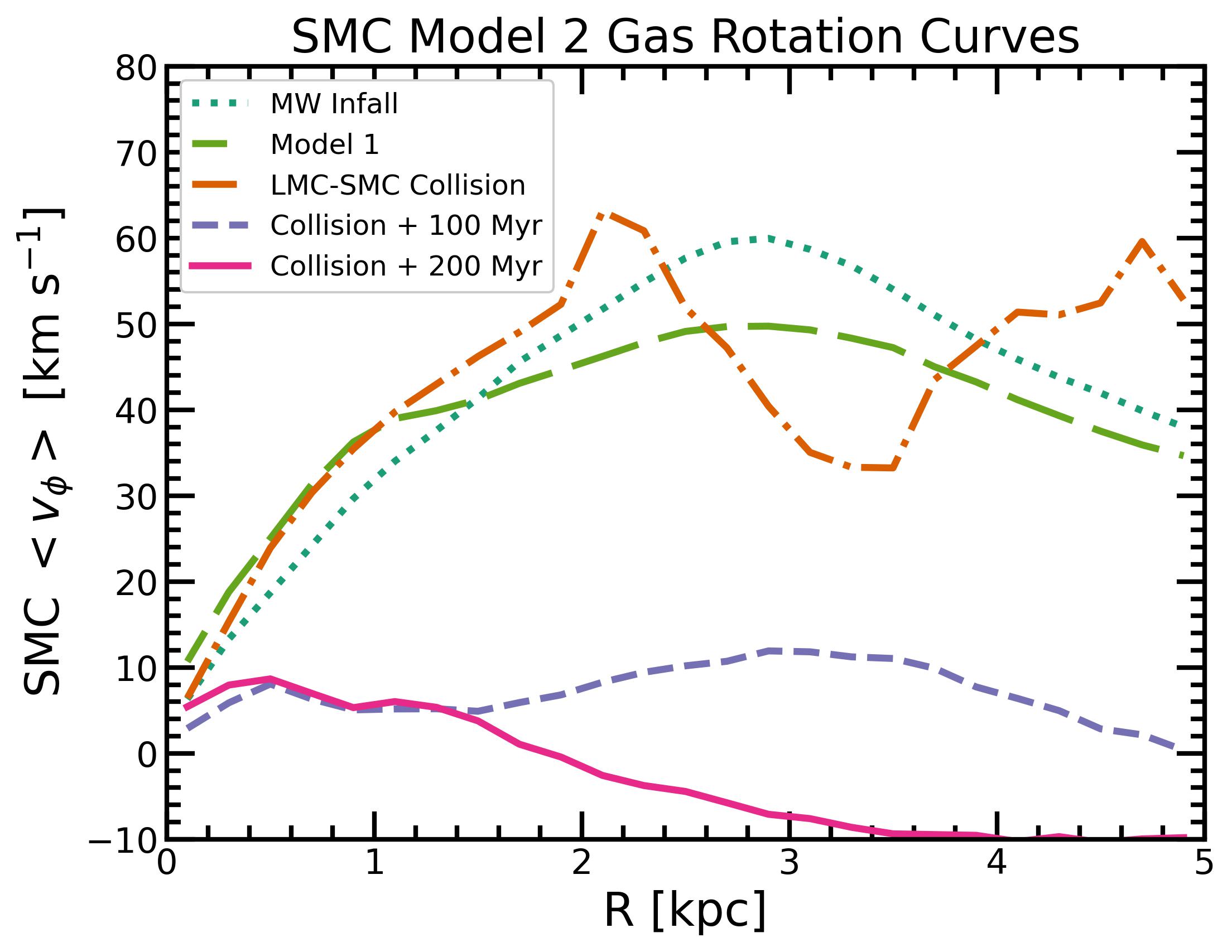}
    \caption{The Model 2 simulated SMC's gas velocity fields, computed in the stellar plane of rotation , and the gas rotation curve. The velocity field is shown for three epochs: MW infall ({\em top left}); 100 Myr post LMC-SMC collision ({\em top right}); and 200 Myr post collision ({\em bottom left}). The grey arrow is centered at the SMC's stellar kinematic center, and points in the direction of the LMC. The smaller arrows denote the average direction of gas velocity in 0.4 by 0.4 kpc spatial bins. The color scale denotes the magnitude of the average in-plane gas velocity vector in the spatial bin. Pre-collision (MW Infall), the gas shows coherent rotation. Post-collision, the gas velocity field is complex, with no evidence of rotation and clear divergence at radii $> 2$ kpc. The direction of gas velocity divergence is qualitatively consistent with the LMC-SMC position vector, consistent with tidally driven motions. {\em Bottom right panel} shows the azimuthally averaged gas $v_\phi$ curve for different epochs in Model 2. At MW Infall, and in the Model 1 control, there is a clear rotation curve. Post-collision, there is significant evolution, with no evidence of a rising rotation curve in the inner 2 kpc.}
    \label{fig:gas_vf_rot_curve}
\end{figure*}

\begin{figure*}
    \centering
    \includegraphics[width=0.32\textwidth]{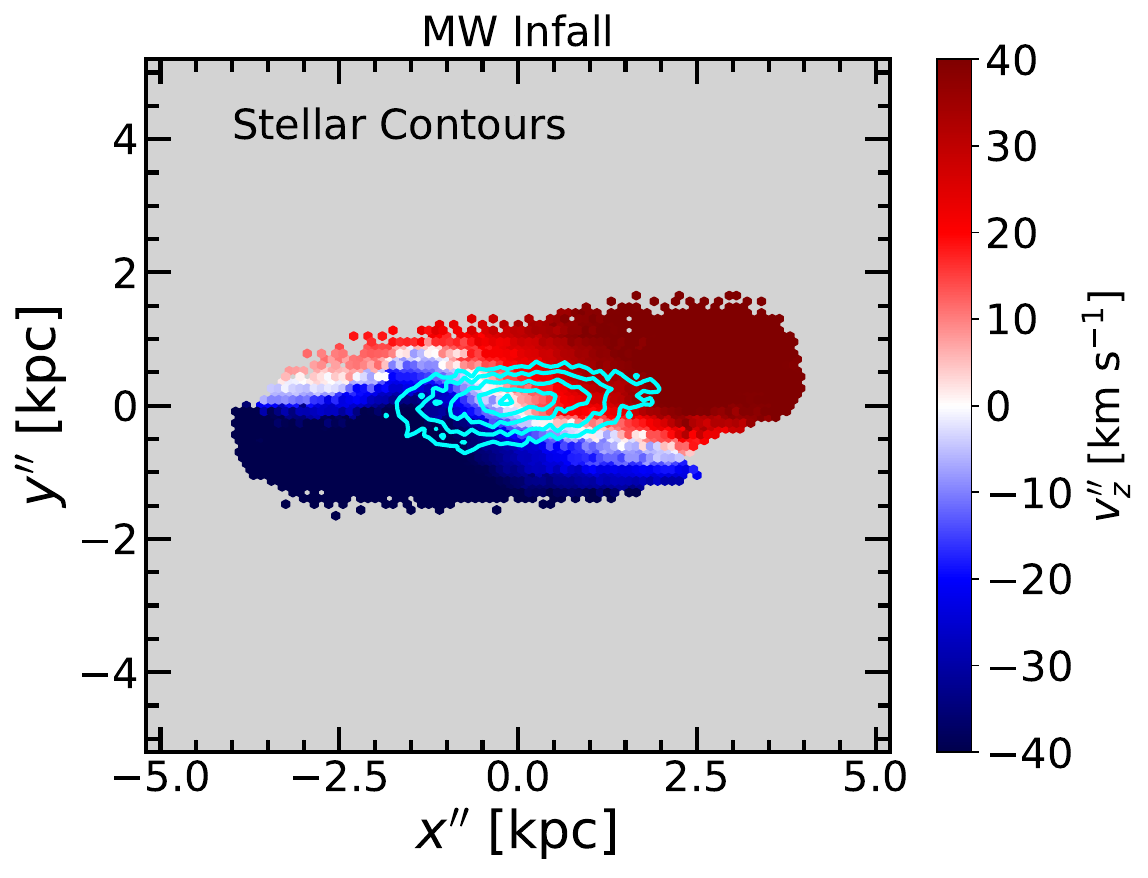}
    \includegraphics[width=0.32\textwidth]{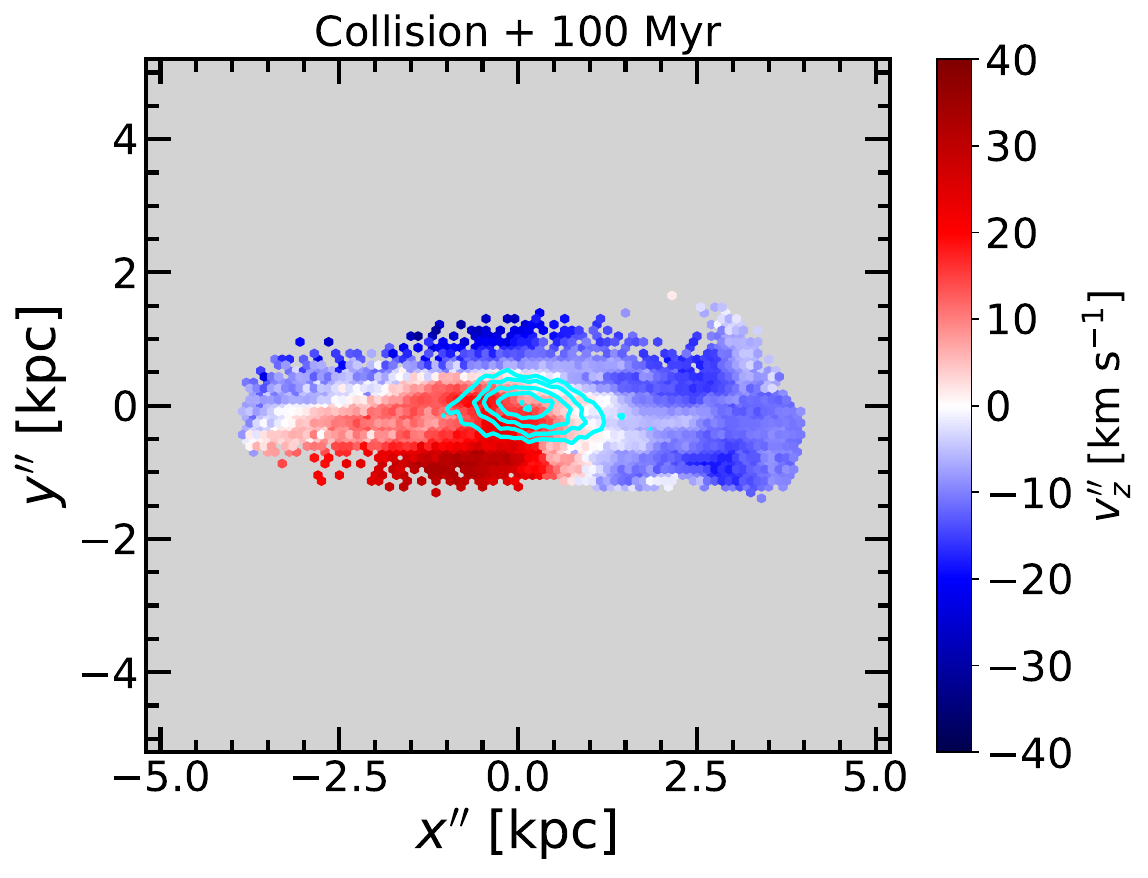}
    \includegraphics[width=0.32\textwidth]{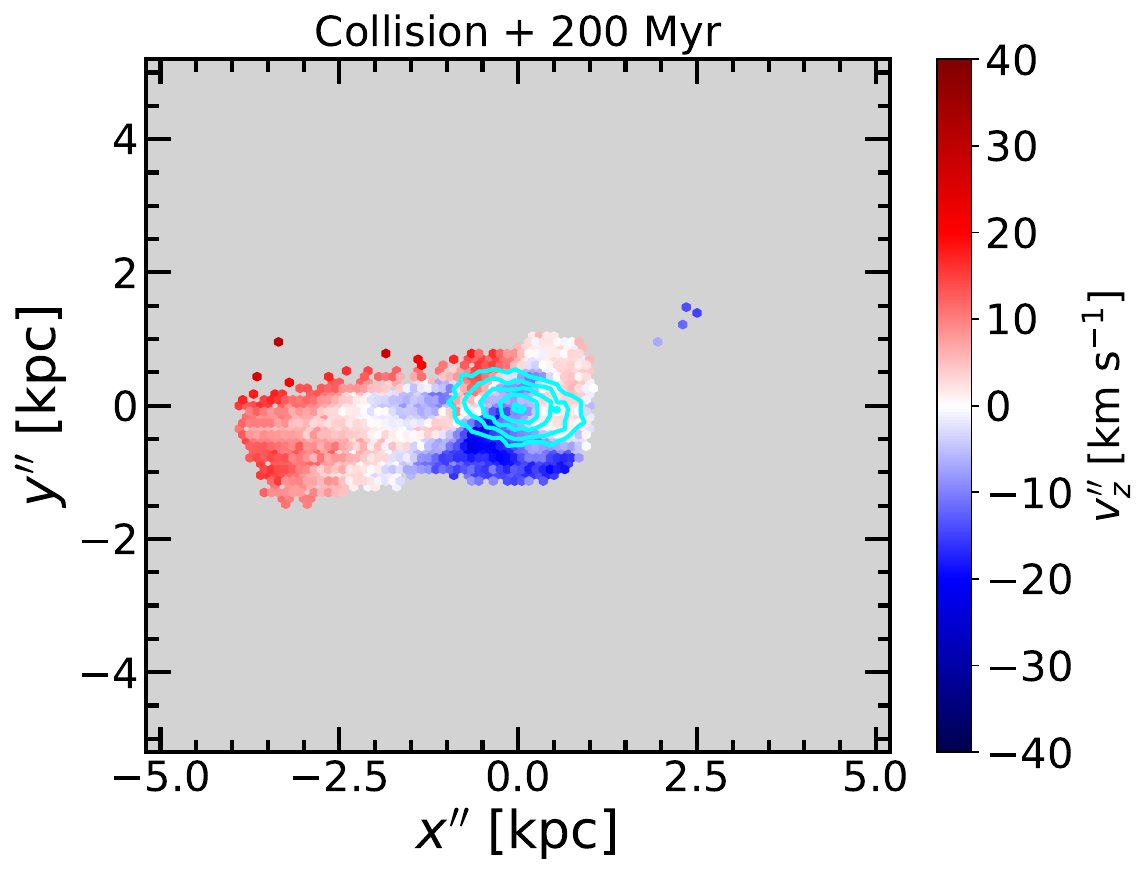}
    \includegraphics[width=0.32\textwidth]{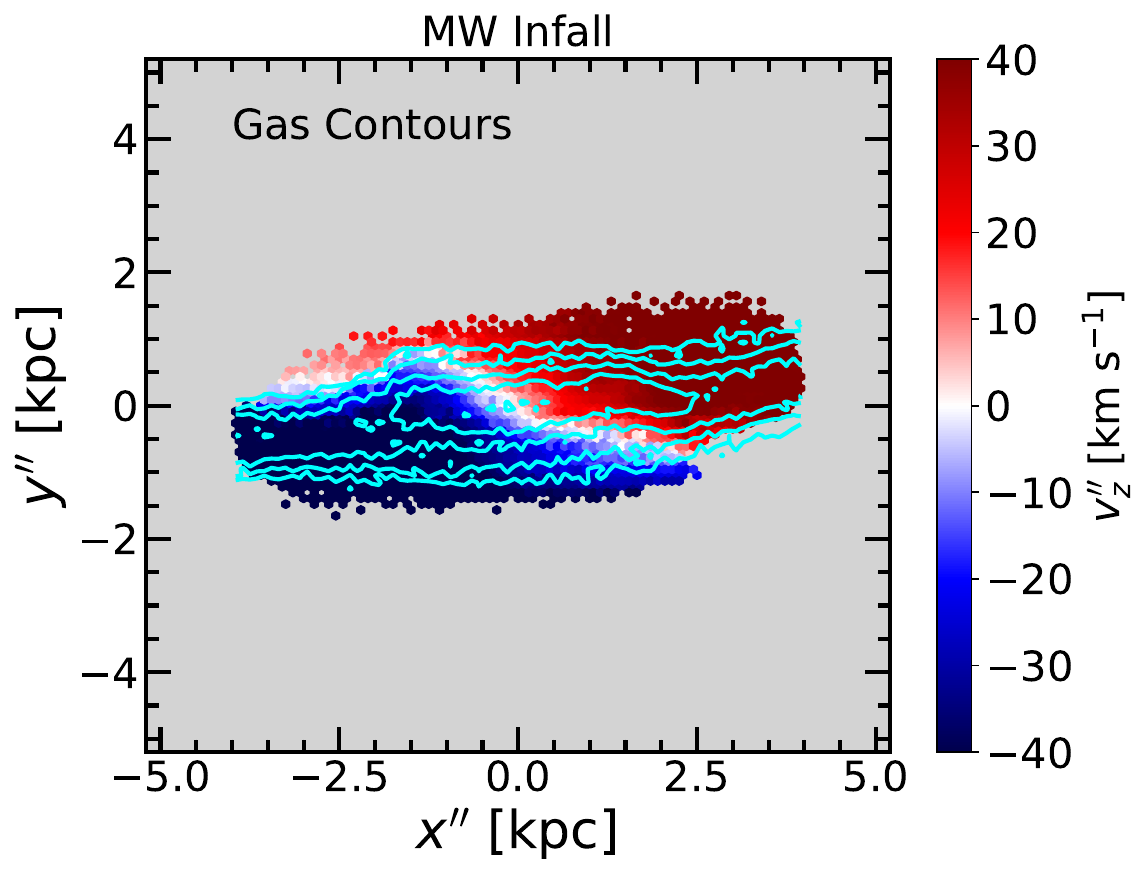}
    \includegraphics[width=0.32\textwidth]{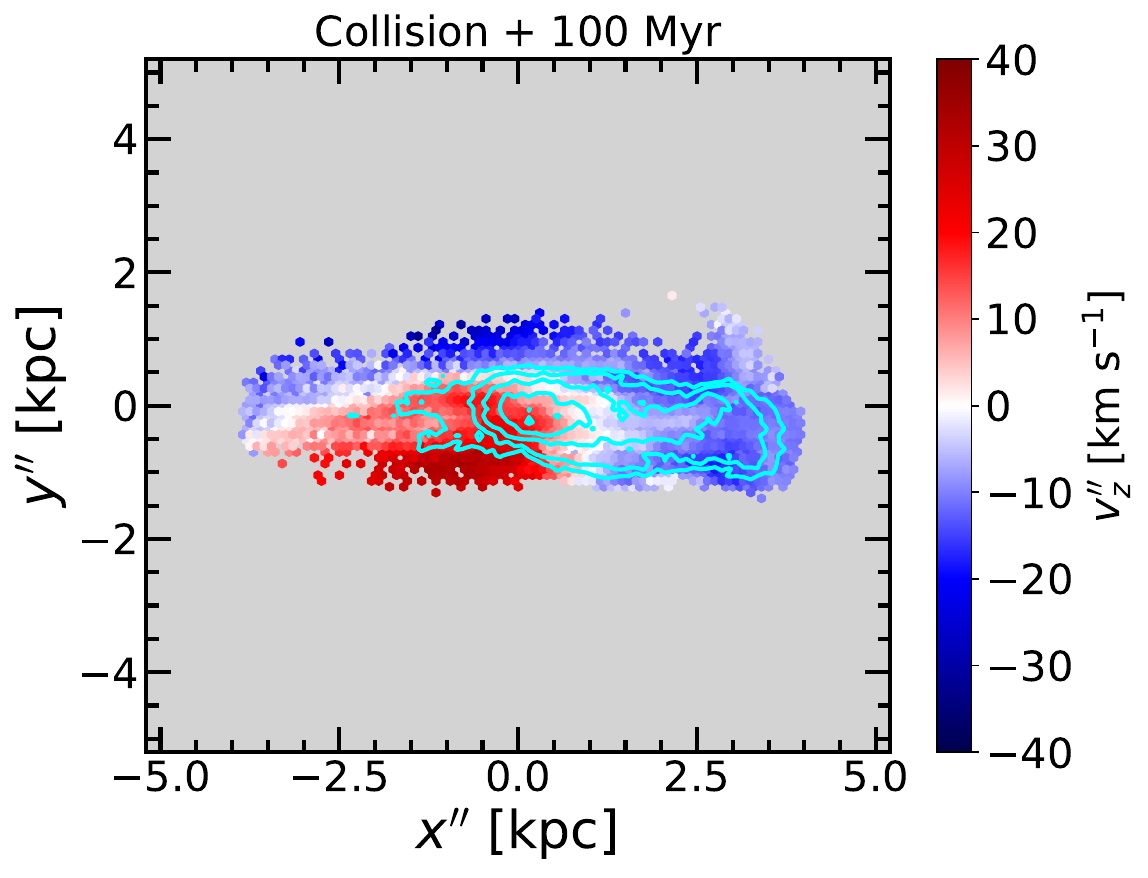}
    \includegraphics[width=0.32\textwidth]{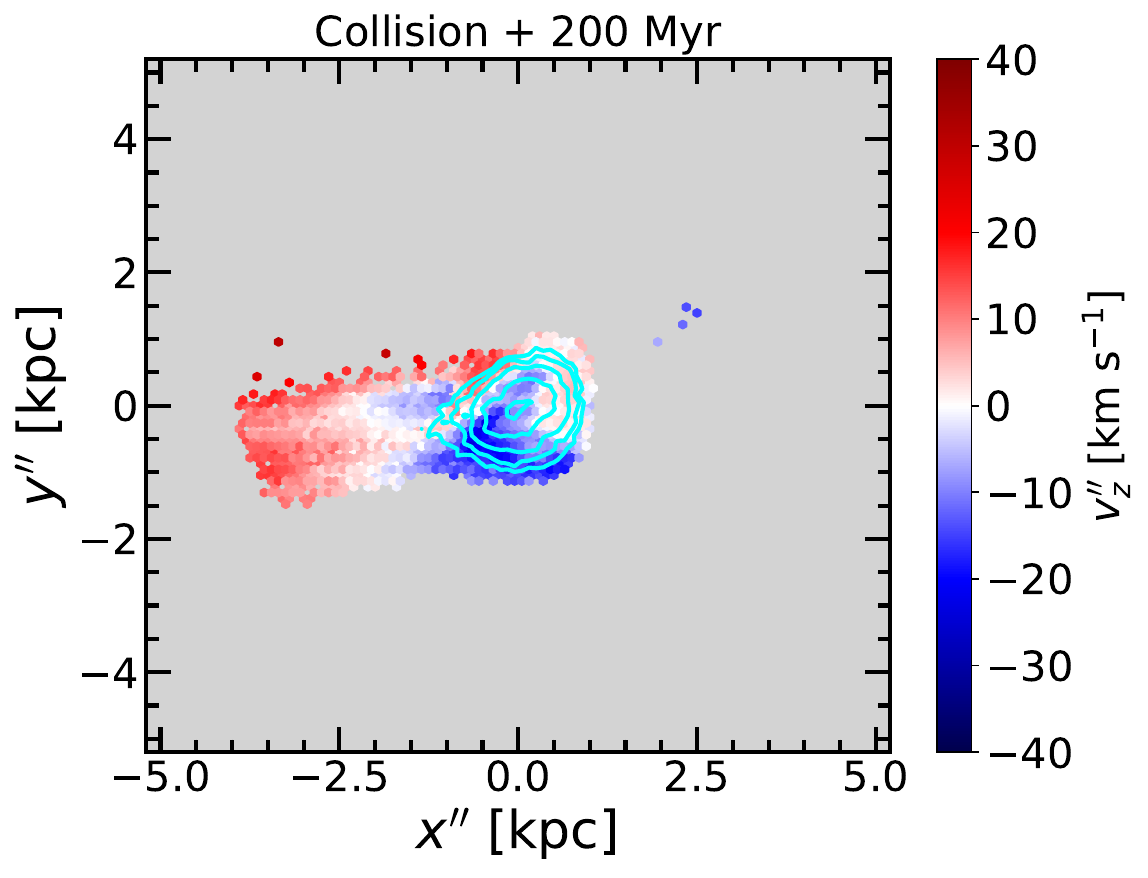}
    \caption{Gas kinematic maps for the Model 2 simulated SMC at three epochs: MW infall; 100 Myr post LMC-SMC collision; and 200 Myr post collision. The coordinate axes are aligned with the principal axes of the stellar distribution (x'', y''), with the z''-axis being the longest axis. The color map depicts the velocity along the z''-axis, mimicking the LoS velocity field. Contours of the stellar (gas) density field are shown in the {\em top} ({\em bottom}) row. The innermost contour has a density of 95\% of the peak density, and each level corresponds to a 1.5 times decrease in density. Apart from the contours, both rows are the same. Pre-collision, the SMC's $v_z$'' gas velocity field exhibits a large gradient (80-100 km s$^{-1}$), and the gradient is due to a coherent in-plane gas rotation field (Figure \ref{fig:gas_vf_rot_curve} {\em top left}). Post-collision, the $v_z$'' gas velocity field is significantly disturbed, but a gradient of 40 - 80 km s$^{-1}$ is still present. However, the post-collision gradient is due to radially outward gas motions as opposed to rotation (Figure \ref{fig:gas_vf_rot_curve} {\em top right and bottom left}). Pre-collision (MW Infall), the SMC's gas zero velocity curve (white color) is centered with both the stellar and gas density contours. However, post-collision, the gas zero velocity curve is displaced from the innermost stellar and gas density contours by $\sim 1$ kpc, consistent with observations. 
    }
    \label{fig:gas_kine_map}
\end{figure*}

Observationally, gas 3D kinematics are unknown, leaving only the LoS velocities to study the internal gas kinematics of the SMC. The observed SMC shows a HI LoS velocity gradient of $60 - 100$ km s$^{-1}$ across an on-sky extent of $\approx 8^\circ$ \citep{DiTeodoro2019}. Rotating disk models \citep[e.g.][]{Marel2001, Marel2002} have been fit to the LoS velocity field, and a rotation curve is inferred from the data, with a peak amplitude of $\approx 50$ km s$^{-1}$ \citep{DiTeodoro2019}.

However, several recent observational studies \citep[e.g.][]{Murray2019, Nakano2025a, Nakano2025b} have used young stars (age $\lesssim$ 50 Myr) as tracers of the SMC's gas kinematics. \cite{Murray2019} used a sample of O and B type stars (age $<$ 10 Myr) with PMs from Gaia and LoS velocities from a variety of spectroscopic surveys. They selected O and B stars with LoS velocities similar to the HI, and found that the young stellar PM field is discrepant from a rotating disk model. Instead, the velocity vectors of the young stars have significant radially outward motions. \cite{Nakano2025a, Nakano2025b} inferred the velocity field of a sample of massive stars (mass $>$ 8$M_\odot$) with PMs from Gaia DR3, which also included Cepheid variables with reliable distance estimates. They found that the massive star kinematics are also dominated by radially outward motions, in a direction broadly consistent with the LMC-SMC position vector. 

In this section, we aim to interpret the observations by studying the effect of the LMC-SMC collision on the simulated SMC's gas kinematics, and attempt to reconcile the observed discrepancy between the stellar and gas kinematics (e.g., Figure \ref{fig:btfr} {\em left panel}).

Pre-collision, we found that the stellar kinematic center, stellar systemic velocity, and the stellar disk angular momentum vector are adequate representations of the gas kinematics in both Model 2 and Model 1. However, post-collision in Model 2, we were unable to reliably determine the SMC gas kinematic center and the gas angular momentum vector. Despite our best efforts at fine tuning the iterative algorithms described in section \ref{sec:centering}, we failed to converge. Moreover, the inferred gas kinematic center and the gas angular momentum vector are very sensitive to the choice of hyper-parameters in the iterative algorithms. Being unsuccessful at centering and aligning the gas distribution already indicates that the gas velocity field is significantly more disturbed than the stellar velocity field. To analyze the SMC's gas kinematics, we choose the stellar kinematic center to center the gas spatial distribution post-collision. We have checked that the choice of centering does not affect our final conclusions for this section.

In Figure \ref{fig:gas_vf_rot_curve}, we plot the SMC's gas velocity field in the stellar plane of rotation. We consider only those SPH particles that are within 2 kpc of the plane of rotation, to avoid contribution from the tidal bridge and tail.

Pre-collision (MW Infall), the SMC has a well defined gas rotation field, by design. Post-collision, there is negligible rotation in gas. Most of the gas motions are radially outwards. Further, the direction of maximal radial motions is consistent with $\vect{R_{\rm LMC-SMC}}$, indicating that these motions are caused by the LMC. Moreover, unlike stars, there is no signature of remnant gas rotation in the inner 1 kpc. Hence, defining a gas kinematic center is not physically meaningful. These findings indicate that the stellar photometric center should be used for observational studies of {\it all} baryonic components of the SMC that rely on a centering choice. 

The azimuthally averaged gas rotation curve for different epochs is shown in the {\em bottom right} panel of Figure \ref{fig:gas_vf_rot_curve}. The rotation curve is well defined pre-collision (MW Infall), and peaks at $\approx 60$ km s$^{-1}$. Model 1 shows a similarly well defined and coherent rotation field.  The rotation curve starts to disrupt at the collision epoch, and the rotation amplitude is negligible post-collision. In particular, post-collision, there is no rise in the rotation curve in the inner 2 kpc.   

Since the post-collision gas kinematics are dominated by radially outward motions, viewing such radial motions from an inclined perspective can result in a LoS velocity gradient that could be confused as a rotation signature \citep[see][]{SylosLabini2019}. Next, we investigate whether the post-collision simulated SMC's gas shows LoS velocity gradients at a particular viewing perspective.

We assume a viewing perspective where the longest stellar axis (section \ref{sec:stellar_struc}) is a proxy for the LoS. Such a viewing perspective is consistent with the observed SMC's large LoS depth. We use the transformation matrix $\mathbf{T}$ (eq. \ref{eq:transformation}) to accordingly rotate the gas distribution and kinematics. The new coordinate axes are called $(x'', y'', z'')$, where $z''$ is along the longest stellar axis and serves as a proxy for the LoS. 

The simulated LoS gas velocity field is shown in Figure \ref{fig:gas_kine_map}. To minimize contribution from the tidal bridge and tail, we only select those SPH particles that reside within an enclosed radius of 4 kpc from the SMC's stellar kinematic center. In the {\em top row} ({\em bottom row}) of the figure, the contours represent stellar (gas) density. 

Prior to the collision (e.g. MW Infall/Model 1 control), the simulated SMC has a well defined velocity gradient of $\approx 100$ km s$^{-1}$. This is expected, as the SMC is initialized as a gas disk with coherent rotation (amplitude $\approx 50$ km s$^{-1}$) and the longest axis resides in the disk plane. We are seeing the disk, edge-on, resulting in a maximal velocity gradient.   

Post-collision, the LoS velocity field is significantly disturbed. However, a gradient of $40 - 80$ km s$^{-1}$ can still be discerned, with the exact value depending on time elapsed since the collision and the axis chosen to measure the gradient. However, in Figure \ref{fig:gas_vf_rot_curve}, we showed that the post-collision SMC's gas kinematics are dominated by radially outward motions and not rotation. Based on these findings, we conclude that the observed SMC's gas LoS velocity gradient is likely a consequence of radially outward gas motions.  

Contours of both stellar and gas densities are shown in Figure \ref{fig:gas_kine_map}. At MW Infall, the gas density contours are significantly more extended as compared to the stellar density contours, largely due to the more extended gas disk adopted in the initial conditions (see section \ref{sec:sims}). Pre-collision, the gas zero velocity curve (defined by the locus of points in the $x'' - y''$ plane that have an LoS velocity equal to the systemic velocity) is well centered with both the stellar and gas density contours. However, post-collision, the zero velocity curve is displaced from the innermost stellar and gas density contour by $\sim 1$ kpc (see the Collision + 200 Myr epoch in particular). Hence, the SMC-LMC collision can explain the separation between the observed SMC’s gas LoS zero velocity curve and its stellar photometric center (see Figure \ref{fig:btfr}, {\em right panel}). 

Note that the behavior of gas is different from stars; we have verified that the stellar kinematic center and the stellar density center remain coincident post-collision. However, the gas $v_{z''}$ zero velocity curve and the gas density contours are not coincident post-collision.

If the observed SMC's gas kinematics does not have a significant rotation, then it is not reasonable to use the peak of the gas velocity gradient as a rotation signature in the BTFR. This would resolve the discrepancy between the SMC's gas and stellar kinematics. We conclude that the SMC does not reside on the BTFR, irrespective of whether stellar or gas kinematics is considered. The SMC is a galaxy with a very high degree of kinematic disequilibrium (see also \citealt{DeLeo2020}).

In this section, we have seen that post SMC-LMC collision, the simulated SMC's gas is significantly more disturbed as compared to the SMC's stars. Clearly, there must be forces apart from the LMC's tidal forces that preferentially act on the SMC's gas and not stars. These forces are hydrodynamic forces, which are a consequence of the collisional nature of the galaxies' gas components. The role of hydrodynamics during the collision is discussed in section \ref{sec:ram_pressure}.

\section{Discussion} \label{ref:discussion}

We have shown using the B12 hydrodynamic simulations that the recent SMC-LMC collision can explain observations of the SMC's large LoS depth, lack of stellar rotation, and discrepant stellar and gas centers. Post-collision, the simulated SMC's gas kinematics has significant radially outward motions, which can manifest as a LoS velocity gradient if viewed at an inclined perspective. Such a gradient can be misinterpreted as a rotating disk. Consequently, the SMC does not reside on the BTFR, and is in a very high state of disequilibrium. Next, we discuss the following additional aspects of the SMC's disequilibrium: (i) effect of LMC's ram pressure on the SMC's gas (section \ref{sec:ram_pressure}); (ii) The distribution of SMC's gas along the LoS (section \ref{sec:gas_los}); and (iii) implications for measuring the SMC's DM content (section \ref{sec:virial}). We also compare our analyses and results with previous works in section \ref{sec:prev}. Finally, we present the limitations of our work and discuss future prospects (section \ref{sec:future}). 

\subsection{Impact of LMC Ram Pressure on the SMC's Gas} \label{sec:ram_pressure}

\begin{figure}
    \centering
    \includegraphics[width=\columnwidth]{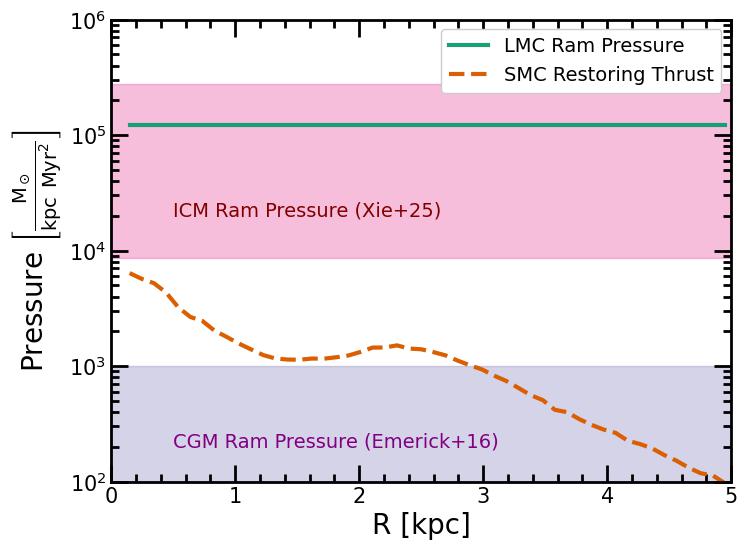}
    \caption{Ram pressure (green solid line) exerted on the SMC's gas disk by the LMC's gas disk during an impulsive SMC-LMC collision. The orange dashed line denotes the restoring thrust applied by the SMC's mass distribution on its gas. The ram pressure from the LMC is more than an order of magnitude larger than the SMC's restoring thrust. Due to ram pressure, the SMC's gas gets an impulsive velocity kick of $\approx 30$ km s$^{-1}$. Hence, ram pressure can potentially explain the lack of remnant rotation in gas (Figure \ref{fig:gas_vf_rot_curve}), and the spatial offset between the SMC's stellar density center and the gas LoS zero velocity curve (see Figure \ref{fig:btfr} {\em right panel} and Figure \ref{fig:gas_kine_map}). The ram pressure exerted by the LMC's ISM on the SMC during the SMC-LMC collision is much larger than the typical ram pressure values in galaxy CGMs \citep{Emerick2016} and on the upper end of the ram pressure exerted by ICMs \citep{Xie2025}.
    }
    \label{fig:ram_pressure}
\end{figure}

In this section, we explore the role of ram pressure exerted by the LMC's gas disk in shaping the post-collision kinematics of the SMC's gas.

Note that exploring LMC's ram pressure in detail is not within the scope of this paper. Further, the SPH nature of the B12 simulation limits the ability to fully resolve the hydrodynamic forces like ram pressure \citep{Agertz2007, Sijacki2012, Hayward2014}, and more advanced mesh-based simulations with high spatial resolution and a truly multi-phase ISM are needed, in order to properly resolve the physical processes at fluid interfaces (like the Kelvin-Helmholtz instability). However, here we present some order of magnitude analytical calculations to help understand the role of ram pressure.

We use the analytical formulation of \citet[][hereafter GG72]{Gunn1972} to estimate the relative strength of the LMC's ram pressure and the SMC's own gravitational restoring thrust on the SMC's gas. The GG72 formulation is routinely used \citep[e.g.][]{Roediger2005, Roediger2006, Tonnesen2009, Salem2015} to obtain order of magnitude estimates of the ram pressure experienced by satellite galaxies as they infall into a circumgalactic medium (CGM) or an intra-cluster medium (ICM). We apply the GG72 formulation to the SMC's gas as it passes through the LMC's ISM during a recent collision, as described below.

The LMC's ram pressure on the SMC is:
\begin{equation} \label{eq:lmc_ram_pressure}
    P_{\rm ram} = \rho \: v^2
\end{equation}
\noindent where $\rho$ is the volume density of the LMC's gas computed at the LMC-SMC impact parameter ($\approx 2$ kpc), and $v$ is the SMC-LMC relative speed during the collision ($\approx200$ km s$^{-1}$). We use the LMC's gas distribution at MW Infall to compute $\rho$, where we divide the surface density of the gas disk by the scale height ($h = 0.34$ kpc). 

Following \cite{Salem2015}, the SMC's gravitational restoring thrust (restoring force per unit area; $P_{SMC}$), at a radial location $R$ inside its disk, is computed as:
\begin{equation}\label{eq:smc_rest}
    P_{\rm SMC} (R) = 2\pi G[\Sigma_{\ast}(R) + \Sigma_{\rm gas}(R)] \Sigma_{\rm gas}(R)
\end{equation}
\noindent where $\Sigma_{\ast}(R)$ and $\Sigma_{\rm gas}(R)$ denote the surface densities of SMC's stellar and gas distribution, respectively.

Figure \ref{fig:ram_pressure} shows $P_{\rm ram}$ and $P_{\rm SMC}(R)$. {\em The ram pressure is more than an order of magnitude larger than the restoring thrust at all radii.} In the above arguments, we have not accounted for the ram pressure on the SMC's gas from the MW's CGM. Further, Ultra-Violet absorption spectroscopic surveys suggest the presence of a hot (at least $\sim 10^4$ K) gas component associated with the LMC as well \citep{Krishnarao2022}. However, ram pressure from a host galaxy's CGM on a satellite ranges between $\mathcal{O}(10^2) - \mathcal{O}(10^3)$ M$_\odot$ kpc Myr$^{-2}$ \citep{Salem2015, Emerick2016}, whereas the ram pressure from the LMC's gas disk on the SMC is much larger ($\mathcal{O}(10^5)$ M$_\odot$ kpc Myr$^{-2}$). Only extreme extra-galactic environments can lead to such large ram pressures, like the central regions of massive galaxy clusters ($10^{15}$ M$_\odot$) where galaxies move at speeds exceeding 1000 km s$^{-1}$ \citep{Xie2025}.

Consequently, the ram pressure from the LMC's gas disk is the most significant external hydrodynamic process affecting the SMC's gas kinematics. Next, we obtain a rough estimate of how much the LMC's ram pressure can affect the SMC's gas during the timescale of the collision.

We assume that the SMC's gas is subject to the LMC's ram pressure only during the time it takes for the SMC to cross the LMC's disk scale height. This timescale is $\Delta t \sim h/v \sim 2 \: \rm{Myr}$. Further, we assume that the ram pressure is impulsive, since the LMC disk crossing timescale is much smaller than any dynamical timescale associated with the Clouds. We ignore the SMC's restoring thrust, since it is subdominant compared to the ram pressure. The impulsive velocity kick ($\Delta v$) to SMC gas within $1$ kpc of the SMC's stellar density center is then given by:
\begin{equation}
    \Delta v \sim \frac{P_{\rm ram}}{\langle\Sigma_{\rm gas}\rangle} \Delta t  \approx 30 \,{\rm km \,s}^{-1}
\end{equation}
\noindent where $\langle\Sigma_{\rm gas}\rangle$ denotes the average SMC gas surface density within a radius of 1 kpc.

From Figure \ref{fig:gas_vf_rot_curve}, the pre-collision (MW Infall) SMC's gas rotation amplitude is $\approx 30$ km s$^{-1}$ at 1 kpc. Hence, an impulsive velocity kick of $\approx 30$ km s$^{-1}$ due to ram pressure is sufficient to wipe out any gas rotation in the SMC's inner 1 kpc. The hydrodynamic interaction between the LMC and SMC during a direct collision can therefore explain the lack of remnant rotation in the SMC's gas at present day (Figure \ref{fig:gas_vf_rot_curve}). Moreover, because of the ram pressure kick, the SMC's gas zero velocity curve will be different from the stellar kinematic center (or equivalently the stellar density center). This scenario provides a natural explanation for why the observed SMC's HI LoS velocity field center is not coincident with the stellar photometric center (Figure \ref{fig:btfr}). An additional ram pressure kick would make the systemic velocity of gas different from the systemic velocity of the stars.

This calculation highlights a physical effect that is not generally considered in studies of satellite galaxy evolution. Theoretical studies find it challenging to remove enough gas from the inner regions of satellites in order to explain their quenching timescales of $\sim 2$ Gyr \citep[e.g.][]{Emerick2016}. Here, we illustrate that collisions between low mass galaxies (e.g. pre-processing in low mass groups) can help provide an extra kick to the gas, which aids in subsequent gas removal through other environmental processes (e.g. ram pressure from the host CGM, host tidal field). Hence, such ram pressure kicks during galaxy collisions are expected to accelerate dIrr to dE/dSph transition. The strength of ram pressure is equivalent to an ICM environment, where dEs are known to exist. Further, these kicks are impulsive, and hence they are particularly significant for low mass galaxies with a shallow potential depth. Additionally, such ram pressure kicks from the primary galaxy's ISM during close satellite interactions (see also \citealt{Matijevic2025}) need to be taken into account for interpreting the existence of dEs around M31 (e.g. NGC 147, NGC 185, NGC 205) and the evolution of Sagittarius dSph as it orbits around the MW.

\subsection{The SMC's Gas LoS Distribution} \label{sec:gas_los}

\begin{figure*}
    \centering
    \includegraphics[width=\textwidth]{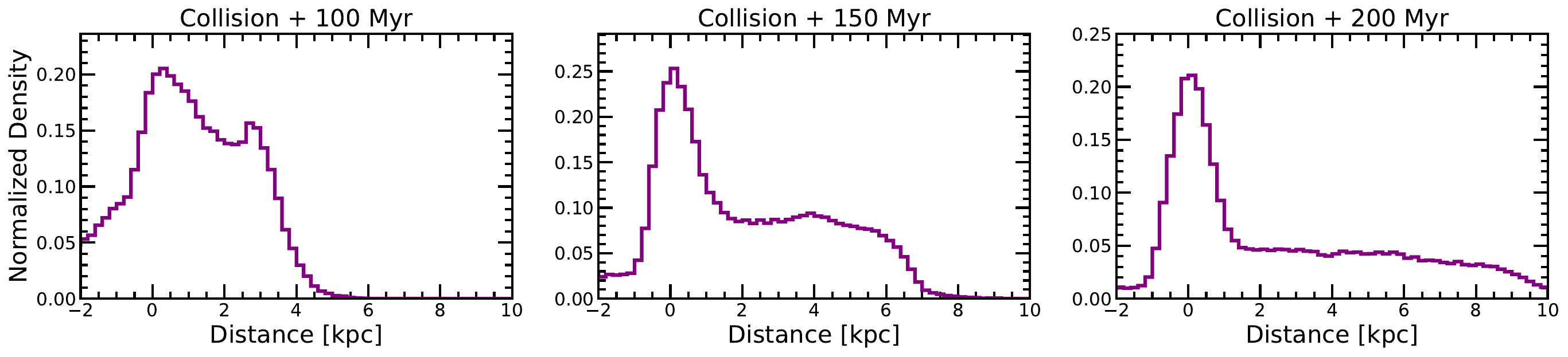}
    \caption{The distribution of gas as a function of distance along the longest axis of the stellar distribution in the Model 2 simulated SMC. A distance of 0 kpc on the x-axis corresponds to the SMC's stellar density center, and positive values are in the direction of the tidal tail. Gas particles are selected to reside within 4 kpc cylinder about the longest axis. Three epochs are shown - 100 Myr ({\em left panel}), 150 Myr ({\em middle panel}) and 200 Myr ({\em right panel}) post SMC-LMC collision. At all times, the gas distribution peaks at the stellar density center. 100 Myr post-collision, the gas distribution has a prominent secondary peak at 3-4 kpc. This secondary peak is due to a gas tidal tail that develops from the SMC-LMC collision. With time, the amplitude of the secondary peak decreases and the gas distribution in the tidal tail becomes more elongated. The gas present along the SMC's tidal tail needs to be taken into account when interpreting the observed SMC's HI LoS distance bi-modality \citep{Murray2024}.}
    \label{fig:gas_los}
\end{figure*}

As mentioned in section \ref{sec:intro}, several observational studies indicate that the SMC's HI shows two prominent velocity components along the LoS. These components have similar brightness temperatures, indicating they have similar HI mass. \cite{Murray2024} selected a sample of young stars (age of 10-100 Myr) from the Gaia and APOGEE surveys that have similar LoS velocities as the HI components. After accounting for dust extinction, they find that the stars associated with the HI components are separated by 5 - 10 kpc along the LoS. Hence, the SMC's HI LoS velocity bimodality is likely due to a bimodality in the gas distribution along the LoS. In this section, using the B12 Model 2 simulation, we attempt to provide an interpretation of the complex LoS structure of the SMC's ISM.

We showed in section \ref{sec:stellar_struc} that the SMC's stellar tidal tail born out of the SMC-LMC collision can explain the observed SMC's stellar LoS depth. The gas will also form a tidal tail similar to the stars. We investigate whether the elongated structure of the SMC's gas tidal tail can help us interpret the observed HI distance bimodality.

The Model 2 simulated SMC's gas distribution is centered at the stellar density center. The coordinate axes are aligned with the principal axes of the stellar distribution (see section \ref{sec:stellar_struc}). As such, the longest stellar axis is used as a proxy for the LoS. 

We have verified that the gas tidal tail and the stellar tidal tail in the simulation are approximately co-located. We consider gas particles within a cylinder of radius 4 kpc along the longest stellar axis. In Figure \ref{fig:gas_los}, the normalized density distribution of these gas particles is plotted as a function of distance along the longest stellar axis. Three simulation epochs are shown - 100 Myr, 150 Myr and 200 Myr, post SMC-LMC collision.

100 Myr post-collision, the SMC's gas distribution has a prominent bimodality. The secondary peak is 3-4 kpc from the stellar density center, coincident with the formation of the tidal tail. With time, the amplitude of the secondary peak decreases as the tidal tail becomes more elongated. 

Hence, the gas distribution along the SMC's tidal tail needs to be taken into account when interpreting the LoS structure of the observed SMC's HI. Such a tidal gas tail can result in a bimodal gas distribution along the LoS, where the primary gas component corresponds to the SMC's main body, and the secondary component corresponds to the extended gas distribution along the tail. These findings are also consistent with the hypothesis of \cite{Murray2024}. In particular, we have shown that the SMC's complex LoS gas distribution can be explained {\em without} requiring a superposition of an independent gas component (like another dwarf galaxy) along the LoS.

As mentioned at the beginning of this section, the two HI components of the observed SMC have a similar mass. In the simulation, the amplitude of the secondary gas peak at 100 Myr post-collision is similar to that of the primary peak. However, the ratio of the amplitudes of the secondary peak to the primary peak has decreased by almost a factor of 2 in a time-span of 50 Myr. So, either the observed gas distance bimodality is very sensitive to the time elapsed since the collision, or there are other physical effects not captured in the B12 simulation that impact the evolution of the gas. For example, the LMC and MW CGM is not included in these simulations. As such, hydrodynamic instabilities that can impact the gas density (e.g. Kelvin-Helmholtz instabilities) are not included. Also, supernova feedback is not included in the B12 simulation. The observed SMC's HI components harbor young and massive O and B type stars with ages $\sim 10$ Myr. These stars can cause a significant amount of feedback over timescales less than 50 Myr, causing the LoS gas distribution to be more clumpy.

The formation of a gaseous tidal tail after the LMC-SMC collision is not impacted by the above caveats. However,
more advanced hydrodynamic simulations with supernova feedback and a MW/LMC CGM are needed to better capture the evolution of this gaseous tidal tail in order to compare with observations in detail. 

\subsection{Implications for Dynamical Mass Measurements of the SMC} \label{sec:virial}

\begin{figure}
    \centering
    \includegraphics[width=\columnwidth]{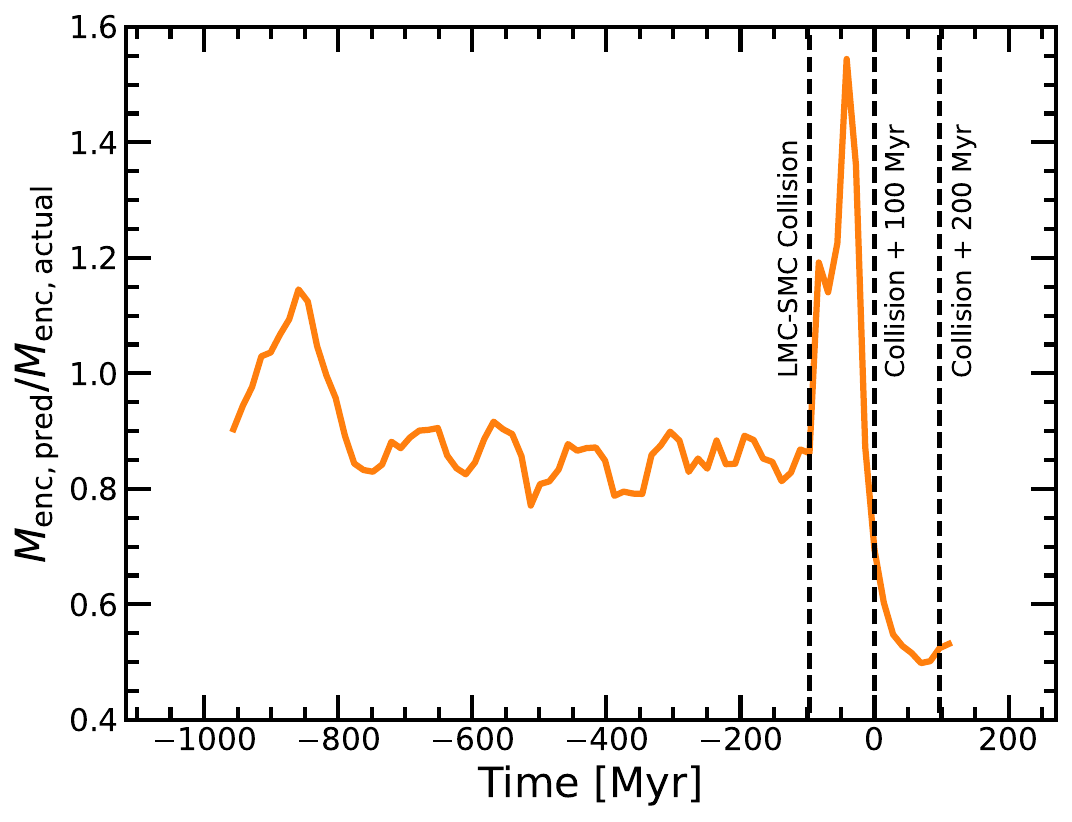}
    \caption{Testing the accuracy of the Virial estimator for measuring the Model 2 simulated SMC’s enclosed mass within 2 kpc of its stellar density center. The Virial estimator used has been calibrated with the Model 1 simulation. The ratio of the enclosed mass predicted by the Virial estimator and the actual enclosed mass in the simulation (y-axis) is plotted as a function of time. Pre-collision, the Virial estimator reasonably recovers the SMC’s enclosed mass. Post-collision, the Virial mass estimate can over or under estimate the actual enclosed mass by a factor of $\sim$2, depending on the time elapsed since the collision. The discrepancy is likely to be higher in observations, where the Virial estimator cannot be precisely calibrated. Approaches that do not rely on equilibrium assumptions are needed to measure the SMC's dynamical mass.}
    \label{fig:virial}
\end{figure}

Precise measurement of the SMC's DM content is important for several purposes. These include constraining the SMC-LMC-MW orbit \citep{Kallivayalil2013, Zivick2018, Patel2020} and understanding the effect of the SMC on the LMC's disk structure \citep[e.g.][]{Pardy2016, Arranz2025, Rathore2025b}, internal kinematics \citep{Choi2022}, star formation history \citep{Massana2022,Harris2009}, and dark matter halo ({\em H. Foote et al. in prep}). Further, precise measurement of the SMC's DM mass is needed to accurately constrain the low mass end of the cosmological stellar mass - halo mass relation \citep[e.g.][]{Wechsler2002, Behroozi2019, Bowden2023} and to ultimately use the SMC-LMC-MW system as a testbed for DM physics \citep{Foote2023}.

However, the SMC has a very high degree of disequilibrium due to its recent collision with the LMC, which makes it very challenging to precisely measure the SMC's mass using methods that rely on the assumption of equilibrium.

Equilibrium is typically assumed in the majority of existing measurements of the SMC's dynamical mass. For example, studies have assumed that the SMC has a stable rotation curve \citep{Stanimirovic1999, Stanimirovic2004, Bekki2009, DiTeodoro2019} or that the Virial mass estimator is valid \citep{Dopita1985, Harris2006}. 

If the SMC is assumed to be a rotating disk, rotation curves can be fit to the velocity data to derive a mass profile. For example, \cite{DiTeodoro2019} applied this method to derive a total mass of total mass of $\sim 2.4 \times 10^9$ M$_\odot$ within 4 kpc of the SMC's HI kinematic center. If the SMC is assumed to be in virial equilibrium, virial mass estimators can be applied to the stellar velocity data to derive a dynamical mass. For example, HZ06 used LoS velocities of red giant stars and determined the SMC's total enclosed mass within $\approx 2$ kpc to be $1.4 - 1.9 \times 10^9$ M$_\odot$, and within $\approx 3$~kpc to be $2.7 - 5.1 \times 10^9$ M$_\odot$. 

We have shown that the SMC's gas velocity field likely has significant radially outward motions that result in a LoS velocity gradient that can mimic a rotation curve. The fact that the inferred rotation curve peaks at a value consistent with expectations from the BTFR is likely a coincidence. While there may be information about the SMC's mass profile within the HI velocity data, a rotation curve analysis is not the appropriate approach to constrain the mass profile. 

We have further demonstrated that the SMC's stellar kinematics can be described as a bound core of radius $\approx$2-4 kpc, surrounded by a radially expanding velocity field. Here, we test the accuracy of the Viral mass estimator to measure the post-collision SMC's mass.

Virial mass estimators have been shown to reliably recover the mass profile for galaxies facing significant external perturbations and interactions \citep[e.g.][]{Lynden-Bell1967, Silva2017, Hamilton2024}. This is because these galaxies achieve Virial equilibrium within a few dynamical timescales. However, the SMC-LMC collision occurred $\sim$100 Myr ago, which is less than a dynamical time, even in the inner regions of the SMC \citep{Rathore2025b}. This fact calls to question the validity of using Virial mass estimators in the SMC. 

The Virial mass estimator is defined as follows:
\begin{equation} \label{eq:virial}
    M_{\rm enc, pred} = \frac{R_{\rm enc} (v_{\rm rot}^2 + \sigma^2)}{f G}
\end{equation}
\noindent where, $M_{\rm enc, pred}$ is the predicted enclosed mass within a radius of $R_{\rm enc}$, $v_{\rm rot}$ is the maximum azimuthal velocity within $R_{\rm enc}$ (e.g. Figure \ref{fig:vel_field}), $\sigma$ is the total velocity dispersion within $R_{\rm enc}$, and $G$ is the gravitational constant. We choose $R_{\rm enc} = 2$ kpc to be consistent with the on-sky field of view used in HZ06, where the simulated SMC is centered at the stellar density center.

The parameter $f$ is the \enquote{Virial factor}, which corresponds to the ratio of the gravitational self energy of the galaxy within $R_{\rm enc}$ and the gravitational self energy of a homogeneous solid sphere with mass $M_{\rm enc}$ and radius $R_{\rm enc}$. $f$ is usually of order unity, and is typically the main source of uncertainty in Virial mass estimates. Hence, even for relatively undisturbed galaxies, Virial mass estimates are uncertain up to a factor of a few, since $f$ is difficult to constrain observationally.

Here, we calibrate $f$ using the Model 1 simulation (where the SMC and LMC do not collide and the SMC is relatively undisturbed), by comparing $M_{\rm enc, pred}$ and the actual simulated total enclosed mass over time. We find $f = 0.9$. Thus, $f$ corresponds to the gravitational self energy of a mass configuration having the same geometry as the pre-collision SMC. 

Figure \ref{fig:virial} shows the ratio of the predicted enclosed mass computed using the Virial estimator ($M_{\rm enc, pred}$) and the actual enclosed mass ($M_{\rm enc, actual}$) for the Model 2 simulated SMC as a function of time. Pre-collision, this ratio is close to unity, indicating that the Virial estimator works very well. 

However, post-collision, the ratio varies by a factor of 3. The uncertainty will be higher when applied to observational data, as $f$ cannot be calibrated. For instance, if $f$ itself is uncertain to a factor of 2, the uncertainty in the observed SMC's Virial mass estimates can be up to a factor of 6. This level of uncertainty is sufficient to change the response of the LMC disk to the SMC's collision \citep[e.g., the LMC's bar tilt][]{Rathore2025b} and the SMC's orbital history \citep{Arranz2024}. Fundamentally, this level of uncertainty is not sufficient to constrain the inner DM density profile of the SMC, which is a critical test of DM physics \citep{Rathore2025b}.  

The inadequacy of the Virial mass estimator for precisely measuring the SMC's dynamical mass directly reflects the high degree of disequilibrium in the SMC. The analyses of the simulated SMC's morphology and kinematics conducted so far in this manuscript suggests that the SMC's disequilibrium persists for at least 200 Myr post SMC-LMC collision.

To precisely measure the SMC's DM content, we need approaches that do not rely on equilibrium assumptions. A way forward is to use the SMC's observable perturbations on the LMC's disk during the SMC-LMC collision to measure the SMC's DM content. R25 showed that morphological perturbations in the LMC, such as the LMC's bar tilt ($5^\circ - 15^\circ$) \citep{Choi2018, Arranz2025b}, are caused by the SMC's torques during the collision. R25 estimated the SMC's pre-collision total enclosed mass within $R_{\rm enc} = 2$ kpc to be $0.8 - 2.4 \times 10^9$ M$_{\odot}$ by modeling the SMC's gravitational torques on the LMC's bar. Further, they suggest that the precision in the SMC's mass measurement in their framework is largely limited by the observational uncertainty in the LMC bar's tilt. Hence, it is important to get more precise observational constraints on the disequilibrium morphology and kinematics of the LMC and SMC to enable more precise determinations of the SMC's mass, without relying on equilibrium assumptions. 

Recently, \cite{DeLeo2024}, measured the SMC's dynamical mass by applying Jeans Modeling \citep{BT2008} on a bound sample of SMC stars obtained after removing stars that were unbound due to the LMC's tides. They inferred a total enclosed mass within 3 kpc to be $2.29 \pm 0.46 \times 10^9$ M$_\odot$, which is consistent with the R25 estimate.

\subsection{Comparison with Previous Studies} \label{sec:prev}

Several works have presented simulations of the SMC-LMC-MW interactions \citep[e.g.][]{Gardiner-Noguchi1996, Connors2006, Besla2010, DB2012, Pardy2018, Cullinane2022, Lucchini2024, Arranz2024}. However, the majority of the aforementioned studies have focused on either the LMC's internal structure and kinematics, or the MS. Further, most previous studies have modeled the SMC with either static potentials or spherical halos, which limits detailed investigations of the SMC's internal structure and kinematics as the SMC's halo will be significantly deformed throughout its interaction with the LMC (Foote et al. {\em in prep}). To the best of our knowledge, the most recent work that has investigated the internal structure and kinematics of the SMC with simulations is \citet[hereafter DB12,][]{DB2012}. Hence, we primarily compare our analysis and results with DB12.

The primary N-body model of DB12 represents the SMC with a live DM halo, a live extended disk (which they treat as the gas disk), and a live stellar spheroid. The DB12 simulations do not include hydrodynamics and their LMC and MW are represented with static potentials. DB12's  adoption of a \enquote{disk+spheroid} model for the SMC initial conditions was motivated by the observations of the distinct kinematics exhibited by the old stars vs. the HI gas.   

Note that the DB12 simulation setup is significantly different from the setup of B12. In B12 both the SMC and LMC are represented by live DM halos, live stellar disks, live gas disks using SPH gas particles, but the MW is still a static potential. In particular, the B12 simulations do not initialize the SMC with a separate stellar spheroidal component. 

Importantly, the SMC-LMC orbital geometry is also distinct in the two studies. In the SMC-LMC orbit of DB12, the SMC-LMC separations during the last two pericentric passages are $\approx 6$ kpc. In B12, the last two pericentric passages were at a separation of $\approx 2$ kpc and $\approx 10$ kpc, respectively. The most recent pericenter in B12 ($\approx 2$ kpc) was a direct collision between the Clouds. Hence, the SMC and LMC interact more strongly in the B12 simulation than in DB12. 

During the most recent SMC-LMC pericenter in DB12, two tidal features form due to the LMC's tidal forces on the SMC. The first feature is a bridge of stars between the SMC and LMC. The second feature is a tidal tail of stars, which DB12 refer to as the \enquote{counter-bridge} \citep[see also][]{Connors2006}. DB12 argue that the large observed LoS depth results from a viewing perspective that aligns along the \enquote{counter-bridge}. In our study, we have similarly argued for this scenario, and shown that a tidal tail can also be generated from a direct collision where the impact parameter is much smaller. 

DB12 analyze the LoS velocity field of the simulated SMC at present day. They find that the initialized stellar spheroid, remains a spheroid, consistent with observations of the old stars. In addition, the SMC's disk is still rotating, producing a velocity gradient of $\approx 80$ km s$^{-1}$, which they compare to observations of the HI. The DB12 findings of agreement with the observations of distinct kinematics between the stars and the gas is a product of their initial conditions.

With the B12 simulations, we have shown that the SMC can have a rotating old stellar disk and gas disk prior to the collision. This is consistent with the known morphology of isolated dIrr galaxies \citep{McGaugh2021}. Through a collision, strong LMC tides can transform the SMC's rotating stellar disk to a dispersion dominated system, consistent with observations.

To summarize, both the B12 and DB12 simulations agree regarding the origin of the observed SMC's LoS depth. However, given the strong observational evidence for the SMC-LMC collision and the SMC's high gas fraction, we conclude that the SMC's old stellar component was likely a rotating disk pre-collision, which has been transformed into a dispersion dominated system due to the LMC's tides. Further, we conclude that the observed SMC's gas kinematics is likely dominated by radially outward motions rather than disk rotation. 

\subsection{Limitations and Future Work} \label{sec:future}

In this subsection, we present pertinent limitations of the B12 simulation and how it affects our analysis, as well as prospects for improving on those limitations in future simulations.

A limitation with the B12 simulation is the limited DM mass resolution ($10^6$ M$_{\odot}$ per particle). However, as shown in \cite{Rathore2025b}, this is not a concern for understanding the morphological and kinematic transformation of the Clouds driven by their recent collision. 

The LMC's galactocentric velocity at \enquote{present day} in the B12 simulation has been matched to the Hubble Space Telescope (HST) PMs from \citep{Kallivayalil2006, Kallivayalil2006b}. However, the observed LMC PMs have since been updated with third epoch HST observations \citep{Kallivayalil2013}, and further refined by Gaia \citep{Luri2021}. The galactocentric position and velocity of the SMC at present day are also not quite correct in the B12 simulation. Further, the maximum post-collision SMC-LMC separation in the simulation is $\approx 10$ kpc, whereas the observed 3D separation between the Clouds is $\approx 20$ kpc. The inability of the simulation to fully match the observed position and velocity of the Clouds limits direct comparisons with observations. However, direct comparison with observations is not the goal of our study. Rather, as mentioned in section \ref{sec:sims}, our goal is to understand the physical principles that operate to transform the SMC's structure and kinematics due to the recent SMC-LMC collision, and guide the interpretation of the observational data. We also refrain from using the word \enquote{present day} for labeling different epochs in the simulation, and rather perform a generic study of the SMC's internal state as a function of time elapsed since the collision. 

The B12 simulation includes neither supernova feedback nor a MW/LMC CGM nor reionization heating of the Clouds' gas. All of these factors can significantly affect the gas morphology and kinematics of the SMC \citep{Lucchini2021}. Given the SMC's high star-formation rate \citep[][]{Harris2006, Rubele2018, Massana2022}, feedback is expected to be significant over the past $\sim 100$ Myr. However, neither feedback nor ram pressure from the LMC/MW CGM nor reionization heating will enable the SMC to retain its gas or support the persistence of internal rotation (see also \citealt{Gatto2013, Mayer2006}). As such, the inclusion of these physical effects would only strengthen the arguments made in this paper. Moreover, we have demonstrated that the ram pressure exerted by the LMC's gas disk on the SMC is significantly stronger than that expected by the LMC or MW CGM.  

There is a large amount of observational evidence (mostly) related to the LMC's disturbed morphology and kinematics that suggests a recent ($\sim 100$ Myr ago) close encounter ($\sim 2$ kpc) between the Clouds (see section \ref{sec:intro}). However, the SMC-LMC orbit is not tightly constrained and a collision, while allowed within the proper motion error space, is not the mean result. Different choices of the SMC center (see right panel in Figure \ref{fig:btfr}) result in differences in the SMC 3D velocity measurements reported using HST \citep{Zivick2018} and Gaia \citep{Luri2021}. In this work, we argue that the stellar photometric center, rather than the HI kinematic center should be used to determine the 3D velocity of the SMC. However, it remains to be tested how a collision impacts the measurements of the SMC's stellar proper motions compared to the center of mass motion of the SMC's DM.  

Further, it is likely that the DM halos of the LMC and SMC are significantly deformed due to their interactions with each other (H. Foote et al. 2025 {\em in prep}) and interactions with the MW \citep{GC2019}. Thus, it is likely that analytical orbits are not able to capture the full complexity of the SMC-LMC interaction and need to be significantly revised using time-evolving potentials \citep{Foote2026}.

Finally, the present work primarily focusses on the SMC's main body ($R < 4$ kpc), and not the outskirts ($R > 4$ kpc). The SMC's outskirts harbor several interesting tidal features like the \enquote{SMC wing} (see \citealt{Almeida2024} and references therein, Escala et al. 2026 {\em in prep}), which could also provide important constraints on the LMC-SMC interactions. This will be a subject of future analyses.

To conclude, observationally fine-tuned high resolution hydrodynamical simulations of the SMC-LMC-MW interaction history are needed to enable direct comparisons with the SMC's observational data, and use this galaxy as a new laboratory for understanding ISM physics, star-formation, galactic interactions, and the properties of dark matter. These modeling efforts may need sophisticated hydrodynamic frameworks \citep{Tepper-Garcia2024, Bland-Hawthorn2024} to rigorously treat the high gas fractions of the Clouds. 

\section{Conclusions} \label{sec:conc}

With a wealth of astrometric, photometric, and spectroscopic data, the SMC offers a unique opportunity to study ISM physics and star formation in a low metallicity environment. Excitingly, the precision of astrometric data sets the stage for the SMC to be a laboratory for DM physics. 

However, the SMC and LMC have likely recently collided ($< 200$ Myr ago), complicating the interpretation of the SMC's observed structure and kinematics. For example, the SMC's HI LoS velocity maps show a gradient of $60 - 100$ km s$^{-1}$ which has been interpreted as a rotating HI disk. However, the SMC's old (age $>$ 1 Gyr) stellar kinematics is dispersion dominated with barely any rotation ($< 10$ km s$^{-1}$). Despite a small on-sky extent ($\sim $4 kpc), the SMC has a large LoS depth ($10 - 20$ kpc). The SMC's gas distribution is also complex, with multiple velocity components separated by 5 - 10 kpc along the LoS. Also, the SMC's stellar photometric center is separated from the HI kinematic center by $\sim 1$ kpc. The SMC's intriguing structure and kinematics indicate that this galaxy is in a high state of disequilibrium.

In this study, we investigated whether a recent SMC-LMC collision can explain the observed SMC's disturbed morphology and kinematics. We used hydrodynamic simulations of the SMC-LMC-MW interaction history from B12. In these simulations, the initial SMC and LMC are modeled with live exponential stellar disks, SPH gas disks (with radiative cooling and star-formation) and live DM halos. We studied two scenarios - Model 1 and Model 2 (see Figure \ref{fig:orbit}). In Model 1, the SMC and LMC remain well-separated, with their closest pericenter being $> 25$ kpc. In Model 2, the most recent pericenter ($\sim$100 Myr ago) between the Clouds was a direct collision (impact parameter of $\sim$2 kpc). Our focus is to use the B12 simulations to understand the impact of a direct collision on the morphological and kinematic evolution of old stars (age $>$ 1 Gyr) and gas in the SMC. Our main results are summarized below:

\begin{enumerate}
    \item {\em The SMC's tidal tail that formed from a direct SMC-LMC collision can explain the SMC's large LoS depth and stellar axis ratio:} After the SMC-LMC collision, the LMC's tides cause a break in the SMC's stellar density profile at $R < 4$ kpc (Figure \ref{fig:smc_m2_dens_profile}) and a tidal tail and bridge are formed. The stellar tidal tail explains the observed SMC's large axis ratio ($> 4:1$, Figure \ref{fig:ax_ratio_time}), which is not achieved without a direct collision. The stellar LoS depth of the SMC is naturally explained if the tidal tail is oriented along the LoS (Figure \ref{fig:los_vis} {\em right panel}) \citep[see also,][]{DB2012} . Further, the tidal elongation has a significant component perpendicular to the SMC's plane of rotation (Figure \ref{fig:ax_angle}), and the orientation of the tail cannot be used to constrain the SMC's inclination. The formation of an SMC tidal tail is a generic prediction for all tidal SMC-LMC simulations \citep[e.g.][]{Connors2006}. We have demonstrated that a direct collision can also create this morphology. Further, the gas tidal tail born out of the SMC-LMC collision can in-part explain the existence of gas components at different LoS distances in the observed SMC (Figure \ref{fig:gas_los}).
    \item {\em The SMC-LMC collision converts the SMC from a rotationally supported disk galaxy to a dispersion dominated galaxy}: The pre-collision simulated SMC has a coherent stellar rotation field by design. Post-collision, the simulated SMC's stars at $R > 2$ kpc exhibit radial outward motions (Figure \ref{fig:vel_field}), with a small ($< 10$ km s$^{-1}$) remnant rotation in the inner $R < 2$ kpc. The post-collision simulated SMC's internal stellar velocity field is consistent with observations \citep{Niederhofer2021, Zivick2021}. Further, the SMC's stellar velocity field in the scenario where the Clouds do not collide, remains a rotating disk (peak velocity $\sim40$ km s$^{-1}$). This is inconsistent with observations, indicating that strong LMC tides (like in a direct collision) are needed to sufficiently disrupt the SMC's internal velocity field. The SMC's rotation to dispersion ratio significantly reduces post-collision ($< 0.4$, Figure \ref{fig:v_sigma}), indicating the SMC is in the midst of a transformation from a rotationally supported disk to a dispersion dominated spheroid.
    \item {\em The SMC's gas kinematics is dominated by radially outward motions as opposed to rotation:} Pre-collision, the simulated SMC has a coherent gas rotation field, by design. Post-collision, the simulated SMC's gas also exhibits radially outward motions and the rotation field is destroyed (Figure \ref{fig:gas_vf_rot_curve}). The resulting diverging velocity field manifests as a velocity gradient ($40 - 80$ km s$^{-1}$) when viewed at an inclination. Hence, the SMC's observed HI velocity gradient is likely due to radially outward motions of gas viewed at an inclination, and is not a signature of rotation. This finding is consistent with the kinematics of the SMC's young stellar populations \citep[$<100$ Myr,][]{Murray2019,Nakano2025a, Nakano2025b}. The SMC's gas kinematics cannot be used to place the SMC on the BTFR (Figure \ref{fig:btfr} {\rm left panel}).
    \item {\em Discrepancies in the SMC's stellar and gas kinematics can be explained by the ram pressure from the LMC's gas disk during a direct collision:} The ram pressure from the LMC's gas disk on the SMC's gas during the SMC-LMC collision is almost an order of magnitude larger than the SMC's gravitational restoring force (Figure \ref{fig:ram_pressure}). During the collision, the LMC's ram pressure gives an additional $\sim 30$ km s$^{-1}$ kick to the SMC's gas, which is not experienced by the stars. Consequently, the gas and stars in the SMC have different systemic velocities, naturally explaining the offset between the observed SMC's stellar photometric center and the LoS gas zero velocity curve (Figure \ref{fig:gas_kine_map} and Figure \ref{fig:btfr} {\rm right panel}). As such, the SMC gas kinematic center should not be used in observational studies, as it neither tracks gas rotation, nor the SMC's center of mass motion. The ram pressure from the LMC's ISM during the SMC-LMC collision is more than an order of magnitude larger than ram pressure from typical galaxy CGMs, and is comparable to extreme ram pressures in massive ICMs.   
    \item{\em Frameworks that account for the SMC as a galaxy in the midst of a transformation are needed to precisely measure the SMC's DM profile:} Given the SMC's high state of disequilibrum, models relying on equilibrium assumptions (like the Virial theorem or gas rotation curve modeling) are not appropriate for obtaining precise estimates (within a factor of 2) of the SMC's total mass profile or DM mass profile (Figure \ref{fig:virial}). Instead, using the SMC's observable perturbations on the LMC's disk (e.g. R25) is a promising way to precisely constrain the SMC's DM content. 
\end{enumerate}

We conclude that a recent ($< 200$ Myr ago) SMC-LMC, direct collision (impact parameter $\sim$ 2 kpc) is necessary to explain the observed SMC's disequilibrium. Further, the SMC's disequilibrium is strongly dependent on the time elapsed since the collision, and the disequilibrium persists for at least 200 Myr post-collision.

Our work highlights the crucial role of galactic collisions in transforming dIrr type galaxies to dE/dSph type over short timescales ($\sim 100$ Myr). Further, ram pressure exerted by the primary galaxy's ISM during such close encounters can provide a significant velocity kick to the secondary's gas, assisting the removal of gas and contributing to quenching of low mass galaxies. The SMC gives a front row view of group processes driving dramatic morphological and kinematic transformations, without requiring the presence of sustained environmental effects (like a cluster ICM). Such group processing may explain the how dEs can exist around massive galaxies like M31. Our work also suggests that caution must be exercised in interpreting the measured HI density profiles and velocity profiles of SMC like galaxies around MW like hosts in the local universe \citep{Geha2024, Zhu2025}, since those profiles can be significantly affected by tidal interactions in galaxy groups occurring prior to the group's infall.

State of the art hydrodynamic simulations capturing the full complexity of the SMC-LMC-MW interactions, including a direct SMC-LMC collision, are required to realize the potential of the SMC as a laboratory for ISM physics, DM physics, and as a local reference for studies of high redshift galaxy evolution. 

\begin{acknowledgements}
Himansh Rathore would like to thank Annapurni Subramaniam, Smitha Subramanian, Claire Murray, Paul Zivick, Andrew Fox, Ivanna Escala, Nico Garavito Camargo, Kathryne J. Daniel, S.R. Dhanush, Mathieu Renzo, Yumi Choi, Dennis Zaritsky, Mariarosa Cioni, Lara Cullinane and Hayden Foote for interesting scientific discussions. We would like to thank the anonymous referee for providing insightful comments that improved the quality as well as clarity of the paper. Himansh Rathore and Gurtina Besla are supported by NASA FINESST 80NSSC24K1469, NASA ATP 80NSSC24K1225 and NSF CAREER AST 1941096. This work utilized the Puma and ElGato High Performance Computing clusters at the University of Arizona. The Theoretical Astrophysics Program (TAP) at the University of Arizona provided resources to support this work. We respectfully acknowledge the University of Arizona is on the land and territories of Indigenous peoples. Today, Arizona is home to 22 federally recognized tribes, with Tucson being home to the O’odham and the Yaqui. The university strives to build sustainable relationships with sovereign Native Nations and Indigenous communities through education offerings, partnerships, and community service.

\software{This work made use of python, and its packages like numpy \citep{vanderWalt2011, Harris2020}, scipy \citep{Virtanen2020} and matplotlib \citep{Hunter2007}.}
\end{acknowledgements}

\bibliography{references}{}

@ARTICLE{Power2003,
       author = {{Power}, C. and {Navarro}, J.~F. and {Jenkins}, A. and {Frenk}, C.~S. and {White}, S.~D.~M. and {Springel}, V. and {Stadel}, J. and {Quinn}, T.},
        title = "{The inner structure of {\ensuremath{\Lambda}}CDM haloes - I. A numerical convergence study}",
      journal = {\mnras},
         year = 2003,
        month = jan,
       volume = {338},
       number = {1},
        pages = {14-34},
          doi = {10.1046/j.1365-8711.2003.05925.x},
archivePrefix = {arXiv},
       eprint = {astro-ph/0201544},
 primaryClass = {astro-ph},
       adsurl = {https://ui.adsabs.harvard.edu/abs/2003MNRAS.338...14P}
}

@ARTICLE{GC2019,
       author = {{Garavito-Camargo}, Nicolas and {Besla}, Gurtina and {Laporte}, Chervin F.~P. and {Johnston}, Kathryn V. and {G{\'o}mez}, Facundo A. and {Watkins}, Laura L.},
        title = "{Hunting for the Dark Matter Wake Induced by the Large Magellanic Cloud}",
      journal = {\apj},
         year = 2019,
        month = oct,
       volume = {884},
       number = {1},
          eid = {51},
        pages = {51},
          doi = {10.3847/1538-4357/ab32eb},
archivePrefix = {arXiv},
       eprint = {1902.05089},
 primaryClass = {astro-ph.GA},
       adsurl = {https://ui.adsabs.harvard.edu/abs/2019ApJ...884...51G}
}

@ARTICLE{Rathore2025a,
       author = {{Rathore}, Himansh and {Choi}, Yumi and {Olsen}, Knut A.~G. and {Besla}, Gurtina},
        title = "{Precise Measurements of the LMC Bar's Geometry with Gaia DR3 and a Novel Solution to Crowding-induced Incompleteness in Star Counting}",
      journal = {\apj},
         year = 2025,
        month = jan,
       volume = {978},
       number = {1},
          eid = {55},
        pages = {55},
          doi = {10.3847/1538-4357/ad93ae},
archivePrefix = {arXiv},
       eprint = {2410.18182},
 primaryClass = {astro-ph.GA},
       adsurl = {https://ui.adsabs.harvard.edu/abs/2025ApJ...978...55R}
}

@ARTICLE{Rathore2025b,
       author = {{Rathore}, Himansh and {Besla}, Gurtina and {Daniel}, Kathryne J. and {Beraldo e Silva}, Leandro},
        title = "{Response of the LMC's Bar to a Recent SMC Collision and Implications for the SMC's Dark Matter Profile}",
      journal = {\apj},
         year = 2025,
        month = jul,
       volume = {988},
       number = {1},
          eid = {79},
        pages = {79},
          doi = {10.3847/1538-4357/ade0ae},
archivePrefix = {arXiv},
       eprint = {2504.16163},
 primaryClass = {astro-ph.GA},
       adsurl = {https://ui.adsabs.harvard.edu/abs/2025ApJ...988...79R}
}

@ARTICLE{Zivick2021,
       author = {{Zivick}, Paul and {Kallivayalil}, Nitya and {van der Marel}, Roeland P.},
        title = "{Deciphering the Kinematic Structure of the Small Magellanic Cloud through Its Red Giant Population}",
      journal = {\apj},
         year = 2021,
        month = mar,
       volume = {910},
       number = {1},
          eid = {36},
        pages = {36},
          doi = {10.3847/1538-4357/abe1bb},
archivePrefix = {arXiv},
       eprint = {2011.02525},
 primaryClass = {astro-ph.GA},
       adsurl = {https://ui.adsabs.harvard.edu/abs/2021ApJ...910...36Z}
}

@ARTICLE{Murray2024,
       author = {{Murray}, Claire E. and {Hasselquist}, Sten and {Peek}, Joshua E.~G. and {Lindberg}, Christina Willecke and {Almeida}, Andres and {Choi}, Yumi and {Craig}, Jessica E.~M. and {D{\'e}nes}, Helga and {Dickey}, John M. and {Di Teodoro}, Enrico M. and {Federrath}, Christoph and {Gerrard}, Isabella. A. and {Gibson}, Steven J. and {Leahy}, Denis and {Lee}, Min-Young and {Lynn}, Callum and {Ma}, Yik Ki and {Marchal}, Antoine and {McClure-Griffiths}, N.~M. and {Nidever}, David and {Nguyen}, Hiep and {Pingel}, Nickolas M. and {Tarantino}, Elizabeth and {Uscanga}, Lucero and {van Loon}, Jacco Th.},
        title = "{A Galactic Eclipse: The Small Magellanic Cloud Is Forming Stars in Two Superimposed Systems}",
      journal = {\apj},
         year = 2024,
        month = feb,
       volume = {962},
       number = {2},
          eid = {120},
        pages = {120},
          doi = {10.3847/1538-4357/ad1591},
archivePrefix = {arXiv},
       eprint = {2312.07750},
 primaryClass = {astro-ph.GA},
       adsurl = {https://ui.adsabs.harvard.edu/abs/2024ApJ...962..120M}
}

@ARTICLE{Crowl2001,
       author = {{Crowl}, Hugh H. and {Sarajedini}, Ata and {Piatti}, Andr{\'e}s E. and {Geisler}, Doug and {Bica}, Eduardo and {Clari{\'a}}, Juan J. and {Santos}, Jr., Jo{\~a}o F.~C.},
        title = "{The Line-of-Sight Depth of Populous Clusters in the Small Magellanic Cloud}",
      journal = {\aj},
         year = 2001,
        month = jul,
       volume = {122},
       number = {1},
        pages = {220-231},
          doi = {10.1086/321128},
archivePrefix = {arXiv},
       eprint = {astro-ph/0104227},
 primaryClass = {astro-ph},
       adsurl = {https://ui.adsabs.harvard.edu/abs/2001AJ....122..220C}
}

@ARTICLE{McGaugh2000,
       author = {{McGaugh}, S.~S. and {Schombert}, J.~M. and {Bothun}, G.~D. and {de Blok}, W.~J.~G.},
        title = "{The Baryonic Tully-Fisher Relation}",
      journal = {\apjl},
         year = 2000,
        month = apr,
       volume = {533},
       number = {2},
        pages = {L99-L102},
          doi = {10.1086/312628},
archivePrefix = {arXiv},
       eprint = {astro-ph/0003001},
 primaryClass = {astro-ph},
       adsurl = {https://ui.adsabs.harvard.edu/abs/2000ApJ...533L..99M}
}

@ARTICLE{Harris2004,
       author = {{Harris}, Jason and {Zaritsky}, Dennis},
        title = "{The Star Formation History of the Small Magellanic Cloud}",
      journal = {\aj},
         year = 2004,
        month = mar,
       volume = {127},
       number = {3},
        pages = {1531-1544},
          doi = {10.1086/381953},
archivePrefix = {arXiv},
       eprint = {astro-ph/0312100},
 primaryClass = {astro-ph},
       adsurl = {https://ui.adsabs.harvard.edu/abs/2004AJ....127.1531H}
}

@ARTICLE{Stanimirovic1999,
       author = {{Stanimirovic}, S. and {Staveley-Smith}, L. and {Dickey}, J.~M. and {Sault}, R.~J. and {Snowden}, S.~L.},
        title = "{The large-scale HI structure of the Small Magellanic Cloud}",
      journal = {\mnras},
         year = 1999,
        month = jan,
       volume = {302},
       number = {3},
        pages = {417-436},
          doi = {10.1046/j.1365-8711.1999.02013.x},
       adsurl = {https://ui.adsabs.harvard.edu/abs/1999MNRAS.302..417S}
}

@ARTICLE{Cioni2000,
       author = {{Cioni}, M. -R.~L. and {van der Marel}, R.~P. and {Loup}, C. and {Habing}, H.~J.},
        title = "{The tip of the red giant branch and distance of the Magellanic Clouds: results from the DENIS survey}",
      journal = {\aap},
         year = 2000,
        month = jul,
       volume = {359},
        pages = {601-614},
          doi = {10.48550/arXiv.astro-ph/0003223},
archivePrefix = {arXiv},
       eprint = {astro-ph/0003223},
 primaryClass = {astro-ph},
       adsurl = {https://ui.adsabs.harvard.edu/abs/2000A&A...359..601C}
}

@ARTICLE{Hunter2012,
       author = {{Hunter}, Deidre A. and {Ficut-Vicas}, Dana and {Ashley}, Trisha and {Brinks}, Elias and {Cigan}, Phil and {Elmegreen}, Bruce G. and {Heesen}, Volker and {Herrmann}, Kimberly A. and {Johnson}, Megan and {Oh}, Se-Heon and {Rupen}, Michael P. and {Schruba}, Andreas and {Simpson}, Caroline E. and {Walter}, Fabian and {Westpfahl}, David J. and {Young}, Lisa M. and {Zhang}, Hong-Xin},
        title = "{Little Things}",
      journal = {\aj},
         year = 2012,
        month = nov,
       volume = {144},
       number = {5},
          eid = {134},
        pages = {134},
          doi = {10.1088/0004-6256/144/5/134},
archivePrefix = {arXiv},
       eprint = {1208.5834},
 primaryClass = {astro-ph.GA},
       adsurl = {https://ui.adsabs.harvard.edu/abs/2012AJ....144..134H}
}

@ARTICLE{Johnson2012,
       author = {{Johnson}, Megan and {Hunter}, Deidre A. and {Oh}, Se-Heon and {Zhang}, Hong-Xin and {Elmegreen}, Bruce and {Brinks}, Elias and {Tollerud}, Erik and {Herrmann}, Kimberly},
        title = "{The Stellar and Gas Kinematics of the LITTLE THINGS Dwarf Irregular Galaxy NGC 1569}",
      journal = {\aj},
         year = 2012,
        month = nov,
       volume = {144},
       number = {5},
          eid = {152},
        pages = {152},
          doi = {10.1088/0004-6256/144/5/152},
archivePrefix = {arXiv},
       eprint = {1209.2453},
 primaryClass = {astro-ph.CO},
       adsurl = {https://ui.adsabs.harvard.edu/abs/2012AJ....144..152J}
}

@ARTICLE{Johnson2015,
       author = {{Johnson}, Megan C. and {Hunter}, Deidre and {Wood}, Sarah and {Oh}, Se-Heon and {Zhang}, Hong-Xin and {Herrmann}, Kimberly A. and {Levine}, Stephen E.},
        title = "{The Shape of LITTLE THINGS Dwarf Galaxies DDO 46 and DDO 168: Understanding the Stellar and Gas Kinematics}",
      journal = {\aj},
         year = 2015,
        month = jun,
       volume = {149},
       number = {6},
          eid = {196},
        pages = {196},
          doi = {10.1088/0004-6256/149/6/196},
archivePrefix = {arXiv},
       eprint = {1504.06673},
 primaryClass = {astro-ph.GA},
       adsurl = {https://ui.adsabs.harvard.edu/abs/2015AJ....149..196J}
}

@ARTICLE{Geha2006,
       author = {{Geha}, M. and {Blanton}, M.~R. and {Masjedi}, M. and {West}, A.~A.},
        title = "{The Baryon Content of Extremely Low Mass Dwarf Galaxies}",
      journal = {\apj},
         year = 2006,
        month = dec,
       volume = {653},
       number = {1},
        pages = {240-254},
          doi = {10.1086/508604},
archivePrefix = {arXiv},
       eprint = {astro-ph/0608295},
 primaryClass = {astro-ph},
       adsurl = {https://ui.adsabs.harvard.edu/abs/2006ApJ...653..240G}
}

@ARTICLE{TF1977,
       author = {{Tully}, R.~B. and {Fisher}, J.~R.},
        title = "{A new method of determining distances to galaxies.}",
      journal = {\aap},
         year = 1977,
        month = feb,
       volume = {54},
        pages = {661-673},
       adsurl = {https://ui.adsabs.harvard.edu/abs/1977A&A....54..661T}
}

@ARTICLE{Begum2008,
       author = {{Begum}, Ayesha and {Chengalur}, Jayaram N. and {Karachentsev}, I.~D. and {Sharina}, M.~E.},
        title = "{Baryonic Tully-Fisher relation for extremely low mass Galaxies}",
      journal = {\mnras},
         year = 2008,
        month = may,
       volume = {386},
       number = {1},
        pages = {138-144},
          doi = {10.1111/j.1365-2966.2008.13010.x},
archivePrefix = {arXiv},
       eprint = {0801.3606},
 primaryClass = {astro-ph},
       adsurl = {https://ui.adsabs.harvard.edu/abs/2008MNRAS.386..138B}
}

@ARTICLE{McGaugh2021,
       author = {{McGaugh}, Stacy S. and {Lelli}, Federico and {Schombert}, James M. and {Li}, Pengfei and {Visgaitis}, Tiffany and {Parker}, Kaelee S. and {Pawlowski}, Marcel S.},
        title = "{The Baryonic Tully-Fisher Relation in the Local Group and the Equivalent Circular Velocity of Pressure-supported Dwarfs}",
      journal = {\aj},
         year = 2021,
        month = nov,
       volume = {162},
       number = {5},
          eid = {202},
        pages = {202},
          doi = {10.3847/1538-3881/ac2502},
archivePrefix = {arXiv},
       eprint = {2109.03251},
 primaryClass = {astro-ph.GA},
       adsurl = {https://ui.adsabs.harvard.edu/abs/2021AJ....162..202M}
}

@ARTICLE{Harris2006,
       author = {{Harris}, Jason and {Zaritsky}, Dennis},
        title = "{Spectroscopic Survey of Red Giants in the Small Magellanic Cloud. I. Kinematics}",
      journal = {\aj},
         year = 2006,
        month = may,
       volume = {131},
       number = {5},
        pages = {2514-2524},
          doi = {10.1086/500974},
archivePrefix = {arXiv},
       eprint = {astro-ph/0601025},
 primaryClass = {astro-ph},
       adsurl = {https://ui.adsabs.harvard.edu/abs/2006AJ....131.2514H}
}

@ARTICLE{Dobbie2014,
       author = {{Dobbie}, P.~D. and {Cole}, A.~A. and {Subramaniam}, A. and {Keller}, S.},
        title = "{Red giants in the Small Magellanic Cloud - I. Disc and tidal stream kinematics}",
      journal = {\mnras},
         year = 2014,
        month = aug,
       volume = {442},
       number = {2},
        pages = {1663-1679},
          doi = {10.1093/mnras/stu910},
archivePrefix = {arXiv},
       eprint = {1405.3705},
 primaryClass = {astro-ph.GA},
       adsurl = {https://ui.adsabs.harvard.edu/abs/2014MNRAS.442.1663D}
}

@ARTICLE{Dhanush2025,
       author = {{Dhanush}, S.~R. and {Subramaniam}, A. and {Subramanian}, S.},
        title = "{Unraveling the Kinematic and Morphological Evolution of the Small Magellanic Cloud}",
      journal = {\apj},
         year = 2025,
        month = feb,
       volume = {980},
       number = {1},
          eid = {73},
        pages = {73},
          doi = {10.3847/1538-4357/ada55f},
archivePrefix = {arXiv},
       eprint = {2501.00788},
 primaryClass = {astro-ph.GA},
       adsurl = {https://ui.adsabs.harvard.edu/abs/2025ApJ...980...73D}
}

@ARTICLE{Stanimirovic2004,
       author = {{Stanimirovi{\'c}}, S. and {Staveley-Smith}, L. and {Jones}, P.~A.},
        title = "{A New Look at the Kinematics of Neutral Hydrogen in the Small Magellanic Cloud}",
      journal = {\apj},
         year = 2004,
        month = mar,
       volume = {604},
       number = {1},
        pages = {176-186},
          doi = {10.1086/381869},
archivePrefix = {arXiv},
       eprint = {astro-ph/0312223},
 primaryClass = {astro-ph},
       adsurl = {https://ui.adsabs.harvard.edu/abs/2004ApJ...604..176S}
}

@ARTICLE{DiTeodoro2019,
       author = {{Di Teodoro}, E.~M. and {McClure-Griffiths}, N.~M. and {Jameson}, K.~E. and {D{\'e}nes}, H. and {Dickey}, John M. and {Stanimirovi{\'c}}, S. and {Staveley-Smith}, L. and {Anderson}, C. and {Bunton}, J.~D. and {Chippendale}, A. and {Lee-Waddell}, K. and {MacLeod}, A. and {Voronkov}, M.~A.},
        title = "{On the dynamics of the Small Magellanic Cloud through high-resolution ASKAP H I observations}",
      journal = {\mnras},
         year = 2019,
        month = feb,
       volume = {483},
       number = {1},
        pages = {392-406},
          doi = {10.1093/mnras/sty3095},
archivePrefix = {arXiv},
       eprint = {1811.09627},
 primaryClass = {astro-ph.GA},
       adsurl = {https://ui.adsabs.harvard.edu/abs/2019MNRAS.483..392D}
}

@ARTICLE{Pingel2022,
       author = {{Pingel}, N.~M. and {Dempsey}, J. and {McClure-Griffiths}, N.~M. and {Dickey}, J.~M. and {Jameson}, K.~E. and {Arce}, H. and {Anglada}, G. and {Bland-Hawthorn}, J. and {Breen}, S.~L. and {Buckland-Willis}, F. and {Clark}, S.~E. and {Dawson}, J.~R. and {D{\'e}nes}, H. and {Di Teodoro}, E.~M. and {For}, B. -Q. and {Foster}, Tyler J. and {G{\'o}mez}, J.~F. and {Imai}, H. and {Joncas}, G. and {Kim}, C. -G. and {Lee}, M. -Y. and {Lynn}, C. and {Leahy}, D. and {Ma}, Y.~K. and {Marchal}, A. and {McConnell}, D. and {Miville-Desch{\`e}nes}, M. -A. and {Moss}, V.~A. and {Murray}, C.~E. and {Nidever}, D. and {Peek}, J. and {Stanimirovi{\'c}}, S. and {Staveley-Smith}, L. and {Tepper-Garcia}, T. and {Tremblay}, C.~D. and {Uscanga}, L. and {van Loon}, J. Th. and {V{\'a}zquez-Semadeni}, E. and {Allison}, J.~R. and {Anderson}, C.~S. and {Ball}, Lewis and {Bell}, M. and {Bock}, D.~C. -J. and {Bunton}, J. and {Cooray}, F.~R. and {Cornwell}, T. and {Koribalski}, B.~S. and {Gupta}, N. and {Hayman}, D.~B. and {Harvey-Smith}, L. and {Lee-Waddell}, K. and {Ng}, A. and {Phillips}, C.~J. and {Voronkov}, M. and {Westmeier}, T. and {Whiting}, M.~T.},
        title = "{GASKAP-HI pilot survey science I: ASKAP zoom observations of HI emission in the Small Magellanic Cloud}",
      journal = {\pasa},
         year = 2022,
        month = feb,
       volume = {39},
          eid = {e005},
        pages = {e005},
          doi = {10.1017/pasa.2021.59},
archivePrefix = {arXiv},
       eprint = {2111.05339},
 primaryClass = {astro-ph.GA},
       adsurl = {https://ui.adsabs.harvard.edu/abs/2022PASA...39....5P}
}

@ARTICLE{Hindman1967,
       author = {{Hindman}, J.~V.},
        title = "{A high resolution study of the distribution and motions of neutral hydrogen in the Small Cloud of Magellan}",
      journal = {Australian Journal of Physics},
         year = 1967,
        month = jan,
       volume = {20},
        pages = {147},
          doi = {10.1071/PH670147},
       adsurl = {https://ui.adsabs.harvard.edu/abs/1967AuJPh..20..147H}
}

@ARTICLE{Arranz2023,
       author = {{Jim{\'e}nez-Arranz}, {\'O}. and {Romero-G{\'o}mez}, M. and {Luri}, X. and {Masana}, E.},
        title = "{Application of a neural network classifier for the generation of clean Small Magellanic Cloud stellar samples}",
      journal = {\aap},
         year = 2023,
        month = apr,
       volume = {672},
          eid = {A65},
        pages = {A65},
          doi = {10.1051/0004-6361/202245720},
archivePrefix = {arXiv},
       eprint = {2301.08494},
 primaryClass = {astro-ph.GA},
       adsurl = {https://ui.adsabs.harvard.edu/abs/2023A&A...672A..65J}
}

@ARTICLE{Besla2012,
       author = {{Besla}, Gurtina and {Kallivayalil}, Nitya and {Hernquist}, Lars and {van der Marel}, Roeland P. and {Cox}, T.~J. and {Kere{\v{s}}}, Du{\v{s}}an},
        title = "{The role of dwarf galaxy interactions in shaping the Magellanic System and implications for Magellanic Irregulars}",
      journal = {\mnras},
         year = 2012,
        month = apr,
       volume = {421},
       number = {3},
        pages = {2109-2138},
          doi = {10.1111/j.1365-2966.2012.20466.x},
archivePrefix = {arXiv},
       eprint = {1201.1299},
 primaryClass = {astro-ph.GA},
       adsurl = {https://ui.adsabs.harvard.edu/abs/2012MNRAS.421.2109B}
}

@ARTICLE{Niederhofer2021,
       author = {{Niederhofer}, Florian and {Cioni}, Maria-Rosa L. and {Rubele}, Stefano and {Schmidt}, Thomas and {Diaz}, Jonathan D. and {Matijev{\u{i}}c}, Gal and {Bekki}, Kenji and {Bell}, Cameron P.~M. and {de Grijs}, Richard and {El Youssoufi}, Dalal and {Ivanov}, Valentin D. and {Oliveira}, Joana M. and {Ripepi}, Vincenzo and {Subramanian}, Smitha and {Sun}, Ning-Chen and {van Loon}, Jacco Th},
        title = "{The VMC survey - XLI. Stellar proper motions within the Small Magellanic Cloud}",
      journal = {\mnras},
         year = 2021,
        month = apr,
       volume = {502},
       number = {2},
        pages = {2859-2878},
          doi = {10.1093/mnras/stab206},
archivePrefix = {arXiv},
       eprint = {2101.09099},
 primaryClass = {astro-ph.GA},
       adsurl = {https://ui.adsabs.harvard.edu/abs/2021MNRAS.502.2859N}
}

@ARTICLE{Nidever2013,
       author = {{Nidever}, David L. and {Monachesi}, Antonela and {Bell}, Eric F. and {Majewski}, Steven R. and {Mu{\~n}oz}, Ricardo R. and {Beaton}, Rachael L.},
        title = "{A Tidally Stripped Stellar Component of the Magellanic Bridge}",
      journal = {\apj},
         year = 2013,
        month = dec,
       volume = {779},
       number = {2},
          eid = {145},
        pages = {145},
          doi = {10.1088/0004-637X/779/2/145},
archivePrefix = {arXiv},
       eprint = {1310.4824},
 primaryClass = {astro-ph.GA},
       adsurl = {https://ui.adsabs.harvard.edu/abs/2013ApJ...779..145N}
}

@ARTICLE{Subramanian2012,
       author = {{Subramanian}, Smitha and {Subramaniam}, Annapurni},
        title = "{The Three-dimensional Structure of the Small Magellanic Cloud}",
      journal = {\apj},
         year = 2012,
        month = jan,
       volume = {744},
       number = {2},
          eid = {128},
        pages = {128},
          doi = {10.1088/0004-637X/744/2/128},
archivePrefix = {arXiv},
       eprint = {1109.3980},
 primaryClass = {astro-ph.CO},
       adsurl = {https://ui.adsabs.harvard.edu/abs/2012ApJ...744..128S}
}

@ARTICLE{Tatton2021,
       author = {{Tatton}, B.~L. and {van Loon}, J. Th and {Cioni}, M. -R.~L. and {Bekki}, K. and {Bell}, C.~P.~M. and {Choudhury}, S. and {de Grijs}, R. and {Groenewegen}, M.~A.~T. and {Ivanov}, V.~D. and {Marconi}, M. and {Oliveira}, J.~M. and {Ripepi}, V. and {Rubele}, S. and {Subramanian}, S. and {Sun}, N. -C.},
        title = "{The VMC Survey - XL. Three-dimensional structure of the Small Magellanic Cloud as derived from red clump stars}",
      journal = {\mnras},
         year = 2021,
        month = jun,
       volume = {504},
       number = {2},
        pages = {2983-2997},
          doi = {10.1093/mnras/staa3857},
archivePrefix = {arXiv},
       eprint = {2012.12288},
 primaryClass = {astro-ph.GA},
       adsurl = {https://ui.adsabs.harvard.edu/abs/2021MNRAS.504.2983T}
}

@ARTICLE{Muraveva2018,
       author = {{Muraveva}, T. and {Subramanian}, S. and {Clementini}, G. and {Cioni}, M. -R.~L. and {Palmer}, M. and {van Loon}, J. Th. and {Moretti}, M.~I. and {de Grijs}, R. and {Molinaro}, R. and {Ripepi}, V. and {Marconi}, M. and {Emerson}, J. and {Ivanov}, V.~D.},
        title = "{The VMC survey - XXVI. Structure of the Small Magellanic Cloud from RR Lyrae stars}",
      journal = {\mnras},
         year = 2018,
        month = jan,
       volume = {473},
       number = {3},
        pages = {3131-3146},
          doi = {10.1093/mnras/stx2514},
archivePrefix = {arXiv},
       eprint = {1709.09064},
 primaryClass = {astro-ph.SR},
       adsurl = {https://ui.adsabs.harvard.edu/abs/2018MNRAS.473.3131M}
}

@ARTICLE{Ripepi2017,
       author = {{Ripepi}, Vincenzo and {Cioni}, Maria-Rosa L. and {Moretti}, Maria Ida and {Marconi}, Marcella and {Bekki}, Kenji and {Clementini}, Gisella and {de Grijs}, Richard and {Emerson}, Jim and {Groenewegen}, Martin A.~T. and {Ivanov}, Valentin D. and {Molinaro}, Roberto and {Muraveva}, Tatiana and {Oliveira}, Joana M. and {Piatti}, Andr{\'e}s E. and {Subramanian}, Smitha and {van Loon}, Jacco Th.},
        title = "{The VMC survey - XXV. The 3D structure of the Small Magellanic Cloud from Classical Cepheids}",
      journal = {\mnras},
         year = 2017,
        month = nov,
       volume = {472},
       number = {1},
        pages = {808-827},
          doi = {10.1093/mnras/stx2096},
archivePrefix = {arXiv},
       eprint = {1707.04500},
 primaryClass = {astro-ph.GA},
       adsurl = {https://ui.adsabs.harvard.edu/abs/2017MNRAS.472..808R}
}

@ARTICLE{Subramanian2017,
       author = {{Subramanian}, Smitha and {Rubele}, Stefano and {Sun}, Ning-Chen and {Girardi}, L{\'e}o and {de Grijs}, Richard and {van Loon}, Jacco Th. and {Cioni}, Maria-Rosa L. and {Piatti}, Andr{\'e}s E. and {Bekki}, Kenji and {Emerson}, Jim and {Ivanov}, Valentin D. and {Kerber}, Leandro and {Marconi}, Marcella and {Ripepi}, Vincenzo and {Tatton}, Benjamin L.},
        title = "{The VMC Survey - XXIV. Signatures of tidally stripped stellar populations from the inner Small Magellanic Cloud}",
      journal = {\mnras},
         year = 2017,
        month = may,
       volume = {467},
       number = {3},
        pages = {2980-2995},
          doi = {10.1093/mnras/stx205},
archivePrefix = {arXiv},
       eprint = {1701.05722},
 primaryClass = {astro-ph.GA},
       adsurl = {https://ui.adsabs.harvard.edu/abs/2017MNRAS.467.2980S}
}

@ARTICLE{Johnson1961,
       author = {{Johnson}, Hugh M.},
        title = "{The Structure of the Small Magellanic Cloud}",
      journal = {\pasp},
         year = 1961,
        month = feb,
       volume = {73},
       number = {430},
        pages = {20},
          doi = {10.1086/127613},
       adsurl = {https://ui.adsabs.harvard.edu/abs/1961PASP...73...20J}
}

@ARTICLE{Patel2020,
       author = {{Patel}, Ekta and {Kallivayalil}, Nitya and {Garavito-Camargo}, Nicolas and {Besla}, Gurtina and {Weisz}, Daniel R. and {van der Marel}, Roeland P. and {Boylan-Kolchin}, Michael and {Pawlowski}, Marcel S. and {G{\'o}mez}, Facundo A.},
        title = "{The Orbital Histories of Magellanic Satellites Using Gaia DR2 Proper Motions}",
      journal = {\apj},
         year = 2020,
        month = apr,
       volume = {893},
       number = {2},
          eid = {121},
        pages = {121},
          doi = {10.3847/1538-4357/ab7b75},
archivePrefix = {arXiv},
       eprint = {2001.01746},
 primaryClass = {astro-ph.GA},
       adsurl = {https://ui.adsabs.harvard.edu/abs/2020ApJ...893..121P}
}

@ARTICLE{Kallivayalil2013,
       author = {{Kallivayalil}, Nitya and {van der Marel}, Roeland P. and {Besla}, Gurtina and {Anderson}, Jay and {Alcock}, Charles},
        title = "{Third-epoch Magellanic Cloud Proper Motions. I. Hubble Space Telescope/WFC3 Data and Orbit Implications}",
      journal = {\apj},
         year = 2013,
        month = feb,
       volume = {764},
       number = {2},
          eid = {161},
        pages = {161},
          doi = {10.1088/0004-637X/764/2/161},
archivePrefix = {arXiv},
       eprint = {1301.0832},
 primaryClass = {astro-ph.CO},
       adsurl = {https://ui.adsabs.harvard.edu/abs/2013ApJ...764..161K}
}

@ARTICLE{Kallivayalil2006,
       author = {{Kallivayalil}, Nitya and {van der Marel}, Roeland P. and {Alcock}, Charles},
        title = "{Is the SMC Bound to the LMC? The Hubble Space Telescope Proper Motion of the SMC}",
      journal = {\apj},
         year = 2006,
        month = dec,
       volume = {652},
       number = {2},
        pages = {1213-1229},
          doi = {10.1086/508014},
archivePrefix = {arXiv},
       eprint = {astro-ph/0606240},
 primaryClass = {astro-ph},
       adsurl = {https://ui.adsabs.harvard.edu/abs/2006ApJ...652.1213K}
}

@ARTICLE{Zivick2018,
       author = {{Zivick}, Paul and {Kallivayalil}, Nitya and {van der Marel}, Roeland P. and {Besla}, Gurtina and {Linden}, Sean T. and {Koz{\l}owski}, Szymon and {Fritz}, Tobias K. and {Kochanek}, C.~S. and {Anderson}, J. and {Sohn}, Sangmo Tony and {Geha}, Marla C. and {Alcock}, Charles R.},
        title = "{The Proper Motion Field of the Small Magellanic Cloud: Kinematic Evidence for Its Tidal Disruption}",
      journal = {\apj},
         year = 2018,
        month = sep,
       volume = {864},
       number = {1},
          eid = {55},
        pages = {55},
          doi = {10.3847/1538-4357/aad4b0},
archivePrefix = {arXiv},
       eprint = {1804.04110},
 primaryClass = {astro-ph.GA},
       adsurl = {https://ui.adsabs.harvard.edu/abs/2018ApJ...864...55Z}
}

@ARTICLE{Massana2022,
       author = {{Massana}, P. and {Ruiz-Lara}, T. and {No{\"e}l}, N.~E.~D. and {Gallart}, C. and {Nidever}, D.~L. and {Choi}, Y. and {Sakowska}, J.~D. and {Besla}, G. and {Olsen}, K.~A.~G. and {Monelli}, M. and {Dorta}, A. and {Stringfellow}, G.~S. and {Cassisi}, S. and {Bernard}, E.~J. and {Zaritsky}, D. and {Cioni}, M. -R.~L. and {Monachesi}, A. and {van der Marel}, R.~P. and {de Boer}, T.~J.~L. and {Walker}, A.~R.},
        title = "{The synchronized dance of the magellanic clouds' star formation history}",
      journal = {\mnras},
         year = 2022,
        month = jun,
       volume = {513},
       number = {1},
        pages = {L40-L45},
          doi = {10.1093/mnrasl/slac030},
archivePrefix = {arXiv},
       eprint = {2203.09523},
 primaryClass = {astro-ph.GA},
       adsurl = {https://ui.adsabs.harvard.edu/abs/2022MNRAS.513L..40M}
}

@ARTICLE{Choi2022,
       author = {{Choi}, Yumi and {Olsen}, Knut A.~G. and {Besla}, Gurtina and {van der Marel}, Roeland P. and {Zivick}, Paul and {Kallivayalil}, Nitya and {Nidever}, David L.},
        title = "{The Recent LMC-SMC Collision: Timing and Impact Parameter Constraints from Comparison of Gaia LMC Disk Kinematics and N-body Simulations}",
      journal = {\apj},
         year = 2022,
        month = mar,
       volume = {927},
       number = {2},
          eid = {153},
        pages = {153},
          doi = {10.3847/1538-4357/ac4e90},
archivePrefix = {arXiv},
       eprint = {2201.04648},
 primaryClass = {astro-ph.GA},
       adsurl = {https://ui.adsabs.harvard.edu/abs/2022ApJ...927..153C}
}

@ARTICLE{Besla2016,
       author = {{Besla}, Gurtina and {Mart{\'\i}nez-Delgado}, David and {van der Marel}, Roeland P. and {Beletsky}, Yuri and {Seibert}, Mark and {Schlafly}, Edward F. and {Grebel}, Eva K. and {Neyer}, Fabian},
        title = "{Low Surface Brightness Imaging of the Magellanic System: Imprints of Tidal Interactions between the Clouds in the Stellar Periphery}",
      journal = {\apj},
         year = 2016,
        month = jul,
       volume = {825},
       number = {1},
          eid = {20},
        pages = {20},
          doi = {10.3847/0004-637X/825/1/20},
archivePrefix = {arXiv},
       eprint = {1602.04222},
 primaryClass = {astro-ph.GA},
       adsurl = {https://ui.adsabs.harvard.edu/abs/2016ApJ...825...20B}
}

@ARTICLE{Zivick2019,
       author = {{Zivick}, Paul and {Kallivayalil}, Nitya and {Besla}, Gurtina and {Sohn}, Sangmo Tony and {van der Marel}, Roeland P. and {del Pino}, Andr{\'e}s and {Linden}, Sean T. and {Fritz}, Tobias K. and {Anderson}, J.},
        title = "{The Proper-motion Field along the Magellanic Bridge: A New Probe of the LMC-SMC Interaction}",
      journal = {\apj},
         year = 2019,
        month = mar,
       volume = {874},
       number = {1},
          eid = {78},
        pages = {78},
          doi = {10.3847/1538-4357/ab0554},
archivePrefix = {arXiv},
       eprint = {1811.09318},
 primaryClass = {astro-ph.GA},
       adsurl = {https://ui.adsabs.harvard.edu/abs/2019ApJ...874...78Z}
}

@ARTICLE{Dhanush2024,
       author = {{Dhanush}, S.~R. and {Subramaniam}, A. and {Subramanian}, S.},
        title = "{A Comprehensive Kinematic Model of the Large Magellanic Cloud Disk from Star Clusters and Field Stars using Gaia DR3: Tracing the Disk Characteristics, Rotation, Bar, and Outliers}",
      journal = {\apj},
         year = 2024,
        month = jun,
       volume = {968},
       number = {2},
          eid = {103},
        pages = {103},
          doi = {10.3847/1538-4357/ad4453},
archivePrefix = {arXiv},
       eprint = {2404.18658},
 primaryClass = {astro-ph.GA},
       adsurl = {https://ui.adsabs.harvard.edu/abs/2024ApJ...968..103D}
}

@ARTICLE{Hernquist1990,
       author = {{Hernquist}, Lars},
        title = "{An Analytical Model for Spherical Galaxies and Bulges}",
      journal = {\apj},
         year = 1990,
        month = jun,
       volume = {356},
        pages = {359},
          doi = {10.1086/168845},
       adsurl = {https://ui.adsabs.harvard.edu/abs/1990ApJ...356..359H}
}

@ARTICLE{NFW1997,
       author = {{Navarro}, Julio F. and {Frenk}, Carlos S. and {White}, Simon D.~M.},
        title = "{A Universal Density Profile from Hierarchical Clustering}",
      journal = {\apj},
         year = 1997,
        month = dec,
       volume = {490},
       number = {2},
        pages = {493-508},
          doi = {10.1086/304888},
archivePrefix = {arXiv},
       eprint = {astro-ph/9611107},
 primaryClass = {astro-ph},
       adsurl = {https://ui.adsabs.harvard.edu/abs/1997ApJ...490..493N}
}

@ARTICLE{Springel2005,
       author = {{Springel}, Volker},
        title = "{The cosmological simulation code GADGET-2}",
      journal = {\mnras},
         year = 2005,
        month = dec,
       volume = {364},
       number = {4},
        pages = {1105-1134},
          doi = {10.1111/j.1365-2966.2005.09655.x},
archivePrefix = {arXiv},
       eprint = {astro-ph/0505010},
 primaryClass = {astro-ph},
       adsurl = {https://ui.adsabs.harvard.edu/abs/2005MNRAS.364.1105S}
}

@ARTICLE{Springel2003,
       author = {{Springel}, Volker and {Hernquist}, Lars},
        title = "{Cosmological smoothed particle hydrodynamics simulations: a hybrid multiphase model for star formation}",
      journal = {\mnras},
         year = 2003,
        month = feb,
       volume = {339},
       number = {2},
        pages = {289-311},
          doi = {10.1046/j.1365-8711.2003.06206.x},
archivePrefix = {arXiv},
       eprint = {astro-ph/0206393},
 primaryClass = {astro-ph},
       adsurl = {https://ui.adsabs.harvard.edu/abs/2003MNRAS.339..289S}
}

@ARTICLE{Kallivayalil2006b,
       author = {{Kallivayalil}, Nitya and {van der Marel}, Roeland P. and {Alcock}, Charles and {Axelrod}, Tim and {Cook}, Kem H. and {Drake}, A.~J. and {Geha}, M.},
        title = "{The Proper Motion of the Large Magellanic Cloud Using HST}",
      journal = {\apj},
         year = 2006,
        month = feb,
       volume = {638},
       number = {2},
        pages = {772-785},
          doi = {10.1086/498972},
archivePrefix = {arXiv},
       eprint = {astro-ph/0508457},
 primaryClass = {astro-ph},
       adsurl = {https://ui.adsabs.harvard.edu/abs/2006ApJ...638..772K}
}

@ARTICLE{Mathewson1974,
       author = {{Mathewson}, D.~S. and {Cleary}, M.~N. and {Murray}, J.~D.},
        title = "{The Magellanic Stream.}",
      journal = {\apj},
         year = 1974,
        month = jun,
       volume = {190},
        pages = {291-296},
          doi = {10.1086/152875},
       adsurl = {https://ui.adsabs.harvard.edu/abs/1974ApJ...190..291M}
}

@ARTICLE{Nidever2010,
       author = {{Nidever}, David L. and {Majewski}, Steven R. and {Butler Burton}, W. and {Nigra}, Lou},
        title = "{The 200{\textdegree} Long Magellanic Stream System}",
      journal = {\apj},
         year = 2010,
        month = nov,
       volume = {723},
       number = {2},
        pages = {1618-1631},
          doi = {10.1088/0004-637X/723/2/1618},
archivePrefix = {arXiv},
       eprint = {1009.0001},
 primaryClass = {astro-ph.GA},
       adsurl = {https://ui.adsabs.harvard.edu/abs/2010ApJ...723.1618N}
}

@ARTICLE{Braun2004,
       author = {{Braun}, R. and {Thilker}, D.~A.},
        title = "{The WSRT wide-field H I survey. II. Local Group features}",
      journal = {\aap},
         year = 2004,
        month = apr,
       volume = {417},
        pages = {421-435},
          doi = {10.1051/0004-6361:20034423},
archivePrefix = {arXiv},
       eprint = {astro-ph/0312323},
 primaryClass = {astro-ph},
       adsurl = {https://ui.adsabs.harvard.edu/abs/2004A&A...417..421B}
}

@ARTICLE{Kerr1957,
       author = {{Kerr}, F.~J. and {Hindman}, J.~V.},
        title = "{Mass Distribution of Galactic Neutral Hydrogen}",
      journal = {\pasp},
         year = 1957,
        month = dec,
       volume = {69},
       number = {411},
        pages = {558},
          doi = {10.1086/127147},
       adsurl = {https://ui.adsabs.harvard.edu/abs/1957PASP...69..558K}
}

@ARTICLE{Putman2003,
       author = {{Putman}, Mary E. and {Staveley-Smith}, Lister and {Freeman}, Kenneth C. and {Gibson}, Brad K. and {Barnes}, David G.},
        title = "{The Magellanic Stream, High-Velocity Clouds, and the Sculptor Group}",
      journal = {\apj},
         year = 2003,
        month = mar,
       volume = {586},
       number = {1},
        pages = {170-194},
          doi = {10.1086/344477},
archivePrefix = {arXiv},
       eprint = {astro-ph/0209127},
 primaryClass = {astro-ph},
       adsurl = {https://ui.adsabs.harvard.edu/abs/2003ApJ...586..170P}
}

@ARTICLE{Bruns2005,
       author = {{Br{\"u}ns}, C. and {Kerp}, J. and {Staveley-Smith}, L. and {Mebold}, U. and {Putman}, M.~E. and {Haynes}, R.~F. and {Kalberla}, P.~M.~W. and {Muller}, E. and {Filipovic}, M.~D.},
        title = "{The Parkes H I Survey of the Magellanic System}",
      journal = {\aap},
         year = 2005,
        month = mar,
       volume = {432},
       number = {1},
        pages = {45-67},
          doi = {10.1051/0004-6361:20040321},
archivePrefix = {arXiv},
       eprint = {astro-ph/0411453},
 primaryClass = {astro-ph},
       adsurl = {https://ui.adsabs.harvard.edu/abs/2005A&A...432...45B}
}

@ARTICLE{Chandra2023,
       author = {{Chandra}, Vedant and {Naidu}, Rohan P. and {Conroy}, Charlie and {Bonaca}, Ana and {Zaritsky}, Dennis and {Cargile}, Phillip A. and {Caldwell}, Nelson and {Johnson}, Benjamin D. and {Han}, Jiwon Jesse and {Ting}, Yuan-Sen},
        title = "{Discovery of the Magellanic Stellar Stream Out to 100 kpc}",
      journal = {\apj},
         year = 2023,
        month = oct,
       volume = {956},
       number = {2},
          eid = {110},
        pages = {110},
          doi = {10.3847/1538-4357/acf7bf},
archivePrefix = {arXiv},
       eprint = {2306.15719},
 primaryClass = {astro-ph.GA},
       adsurl = {https://ui.adsabs.harvard.edu/abs/2023ApJ...956..110C}
}

@ARTICLE{Besla2013,
       author = {{Besla}, Gurtina and {Hernquist}, Lars and {Loeb}, Abraham},
        title = "{The origin of the microlensing events observed towards the LMC and the stellar counterpart of the Magellanic stream}",
      journal = {\mnras},
         year = 2013,
        month = jan,
       volume = {428},
       number = {3},
        pages = {2342-2365},
          doi = {10.1093/mnras/sts192},
archivePrefix = {arXiv},
       eprint = {1205.4724},
 primaryClass = {astro-ph.CO},
       adsurl = {https://ui.adsabs.harvard.edu/abs/2013MNRAS.428.2342B}
}

@ARTICLE{Paz2006,
       author = {{Paz}, D.~J. and {Lambas}, D.~G. and {Padilla}, N. and {Merch{\'a}n}, M.},
        title = "{Shapes of clusters and groups of galaxies: comparison of model predictions with observations}",
      journal = {\mnras},
         year = 2006,
        month = mar,
       volume = {366},
       number = {4},
        pages = {1503-1510},
          doi = {10.1111/j.1365-2966.2005.09934.x},
archivePrefix = {arXiv},
       eprint = {astro-ph/0509062},
 primaryClass = {astro-ph},
       adsurl = {https://ui.adsabs.harvard.edu/abs/2006MNRAS.366.1503P}
}

@ARTICLE{Pejcha2009,
       author = {{Pejcha}, Ond{\v{r}}ej and {Stanek}, K.~Z.},
        title = "{The Structure of the Large Magellanic Cloud Stellar Halo Derived using Ogle-III RR Lyr Stars}",
      journal = {\apj},
         year = 2009,
        month = oct,
       volume = {704},
       number = {2},
        pages = {1730-1734},
          doi = {10.1088/0004-637X/704/2/1730},
archivePrefix = {arXiv},
       eprint = {0905.3389},
 primaryClass = {astro-ph.GA},
       adsurl = {https://ui.adsabs.harvard.edu/abs/2009ApJ...704.1730P}
}

@ARTICLE{DB2012,
       author = {{Diaz}, Jonathan D. and {Bekki}, Kenji},
        title = "{The Tidal Origin of the Magellanic Stream and the Possibility of a Stellar Counterpart}",
      journal = {\apj},
         year = 2012,
        month = may,
       volume = {750},
       number = {1},
          eid = {36},
        pages = {36},
          doi = {10.1088/0004-637X/750/1/36},
archivePrefix = {arXiv},
       eprint = {1112.6191},
 primaryClass = {astro-ph.GA},
       adsurl = {https://ui.adsabs.harvard.edu/abs/2012ApJ...750...36D}
}

@ARTICLE{Toomre1972,
       author = {{Toomre}, Alar and {Toomre}, Juri},
        title = "{Galactic Bridges and Tails}",
      journal = {\apj},
         year = 1972,
        month = dec,
       volume = {178},
        pages = {623-666},
          doi = {10.1086/151823},
       adsurl = {https://ui.adsabs.harvard.edu/abs/1972ApJ...178..623T}
}

@ARTICLE{Mayer2001,
       author = {{Mayer}, Lucio and {Governato}, Fabio and {Colpi}, Monica and {Moore}, Ben and {Quinn}, Thomas and {Wadsley}, James and {Stadel}, Joachim and {Lake}, George},
        title = "{Tidal Stirring and the Origin of Dwarf Spheroidals in the Local Group}",
      journal = {\apjl},
         year = 2001,
        month = feb,
       volume = {547},
       number = {2},
        pages = {L123-L127},
          doi = {10.1086/318898},
archivePrefix = {arXiv},
       eprint = {astro-ph/0011041},
 primaryClass = {astro-ph},
       adsurl = {https://ui.adsabs.harvard.edu/abs/2001ApJ...547L.123M}
}

@ARTICLE{Semczuk2018,
       author = {{Semczuk}, Marcin and {{\L}okas}, Ewa L. and {Salomon}, Jean-Baptiste and {Athanassoula}, E. and {D'Onghia}, Elena},
        title = "{Tidally Induced Morphology of M33 in Hydrodynamical Simulations of Its Recent Interaction with M31}",
      journal = {\apj},
         year = 2018,
        month = sep,
       volume = {864},
       number = {1},
          eid = {34},
        pages = {34},
          doi = {10.3847/1538-4357/aad4ae},
archivePrefix = {arXiv},
       eprint = {1804.04536},
 primaryClass = {astro-ph.GA},
       adsurl = {https://ui.adsabs.harvard.edu/abs/2018ApJ...864...34S}
}

@ARTICLE{Lokas2015,
       author = {{{\L}okas}, Ewa L. and {Semczuk}, Marcin and {Gajda}, Grzegorz and {D'Onghia}, Elena},
        title = "{The Resonant Nature of Tidal Stirring of Disky Dwarf Galaxies Orbiting the Milky Way}",
      journal = {\apj},
         year = 2015,
        month = sep,
       volume = {810},
       number = {2},
          eid = {100},
        pages = {100},
          doi = {10.1088/0004-637X/810/2/100},
archivePrefix = {arXiv},
       eprint = {1505.00951},
 primaryClass = {astro-ph.GA},
       adsurl = {https://ui.adsabs.harvard.edu/abs/2015ApJ...810..100L}
}

@ARTICLE{SylosLabini2019,
       author = {{Sylos Labini}, Francesco and {Benhaiem}, David and {Comer{\'o}n}, S{\'e}bastien and {L{\'o}pez-Corredoira}, Mart{\'\i}n},
        title = "{Nonaxisymmetric models of galaxy velocity maps}",
      journal = {\aap},
         year = 2019,
        month = feb,
       volume = {622},
          eid = {A58},
        pages = {A58},
          doi = {10.1051/0004-6361/201833834},
archivePrefix = {arXiv},
       eprint = {1812.01447},
 primaryClass = {astro-ph.GA},
       adsurl = {https://ui.adsabs.harvard.edu/abs/2019A&A...622A..58S}
}

@ARTICLE{Murray2019,
       author = {{Murray}, Claire E. and {Peek}, J.~E.~G. and {Di Teodoro}, Enrico M. and {McClure-Griffiths}, N.~M. and {Dickey}, John M. and {D{\'e}nes}, Helga},
        title = "{The 3D Kinematics of Gas in the Small Magellanic Cloud}",
      journal = {\apj},
         year = 2019,
        month = dec,
       volume = {887},
       number = {2},
          eid = {267},
        pages = {267},
          doi = {10.3847/1538-4357/ab510f},
archivePrefix = {arXiv},
       eprint = {1910.11283},
 primaryClass = {astro-ph.GA},
       adsurl = {https://ui.adsabs.harvard.edu/abs/2019ApJ...887..267M}
}

@ARTICLE{Nakano2025a,
       author = {{Nakano}, Satoya and {Tachihara}, Kengo and {Tamashiro}, Mao},
        title = "{Evidence of Galactic Interaction in the Small Magellanic Cloud Probed by Gaia-selected Massive Star Candidates}",
      journal = {\apjs},
         year = 2025,
        month = apr,
       volume = {277},
       number = {2},
          eid = {62},
        pages = {62},
          doi = {10.3847/1538-4365/adb8de},
archivePrefix = {arXiv},
       eprint = {2502.12251},
 primaryClass = {astro-ph.GA},
       adsurl = {https://ui.adsabs.harvard.edu/abs/2025ApJS..277...62N}
}

@ARTICLE{Nakano2025b,
       author = {{Nakano}, Satoya and {Tachihara}, Kengo},
        title = "{Dual Directional Expansion of Classical Cepheids in the Small Magellanic Cloud Revealed by Gaia Data Release 3}",
      journal = {\apjl},
         year = 2025,
        month = may,
       volume = {985},
       number = {1},
          eid = {L5},
        pages = {L5},
          doi = {10.3847/2041-8213/adce0b},
archivePrefix = {arXiv},
       eprint = {2503.22866},
 primaryClass = {astro-ph.GA},
       adsurl = {https://ui.adsabs.harvard.edu/abs/2025ApJ...985L...5N}
}

@ARTICLE{Pardy2016,
       author = {{Pardy}, Stephen A. and {D'Onghia}, Elena and {Athanassoula}, E. and {Wilcots}, Eric M. and {Sheth}, Kartik},
        title = "{Tidally Induced Offset Disks in Magellanic Spiral Galaxies}",
      journal = {\apj},
         year = 2016,
        month = aug,
       volume = {827},
       number = {2},
          eid = {149},
        pages = {149},
          doi = {10.3847/0004-637X/827/2/149},
archivePrefix = {arXiv},
       eprint = {1602.07689},
 primaryClass = {astro-ph.GA},
       adsurl = {https://ui.adsabs.harvard.edu/abs/2016ApJ...827..149P}
}

@ARTICLE{Arranz2025,
       author = {{Jim{\'e}nez-Arranz}, {\'O}. and {Roca-F{\`a}brega}, S.},
        title = "{Tidal interaction can stop galactic bars: On the LMC non-rotating bar}",
      journal = {\aap},
         year = 2025,
        month = jun,
       volume = {698},
          eid = {L7},
        pages = {L7},
          doi = {10.1051/0004-6361/202555019},
archivePrefix = {arXiv},
       eprint = {2504.01870},
 primaryClass = {astro-ph.GA},
       adsurl = {https://ui.adsabs.harvard.edu/abs/2025A&A...698L...7J}
}

@ARTICLE{Wechsler2002,
       author = {{Wechsler}, Risa H. and {Bullock}, James S. and {Primack}, Joel R. and {Kravtsov}, Andrey V. and {Dekel}, Avishai},
        title = "{Concentrations of Dark Halos from Their Assembly Histories}",
      journal = {\apj},
         year = 2002,
        month = mar,
       volume = {568},
       number = {1},
        pages = {52-70},
          doi = {10.1086/338765},
archivePrefix = {arXiv},
       eprint = {astro-ph/0108151},
 primaryClass = {astro-ph},
       adsurl = {https://ui.adsabs.harvard.edu/abs/2002ApJ...568...52W}
}

@ARTICLE{Behroozi2019,
       author = {{Behroozi}, Peter and {Wechsler}, Risa H. and {Hearin}, Andrew P. and {Conroy}, Charlie},
        title = "{UNIVERSEMACHINE: The correlation between galaxy growth and dark matter halo assembly from z = 0-10}",
      journal = {\mnras},
         year = 2019,
        month = sep,
       volume = {488},
       number = {3},
        pages = {3143-3194},
          doi = {10.1093/mnras/stz1182},
archivePrefix = {arXiv},
       eprint = {1806.07893},
 primaryClass = {astro-ph.GA},
       adsurl = {https://ui.adsabs.harvard.edu/abs/2019MNRAS.488.3143B}
}

@ARTICLE{Bowden2023,
       author = {{Bowden}, Haley and {Behroozi}, Peter and {Hearin}, Andrew},
        title = "{Halo Properties from Observable Measures of Environment: I. Halo and Subhalo Masses}",
      journal = {The Open Journal of Astrophysics},
         year = 2023,
        month = oct,
       volume = {6},
          eid = {37},
        pages = {37},
          doi = {10.21105/astro.2307.07549},
archivePrefix = {arXiv},
       eprint = {2307.07549},
 primaryClass = {astro-ph.GA},
       adsurl = {https://ui.adsabs.harvard.edu/abs/2023OJAp....6E..37B}
}

@ARTICLE{Lynden-Bell1967,
       author = {{Lynden-Bell}, D.},
        title = "{Statistical mechanics of violent relaxation in stellar systems}",
      journal = {\mnras},
         year = 1967,
        month = jan,
       volume = {136},
        pages = {101},
          doi = {10.1093/mnras/136.1.101},
       adsurl = {https://ui.adsabs.harvard.edu/abs/1967MNRAS.136..101L}
}

@ARTICLE{Silva2017,
       author = {{Beraldo e Silva}, Leandro and {de Siqueira Pedra}, Walter and {Sodr{\'e}}, Laerte and {Perico}, Eder L.~D. and {Lima}, Marcos},
        title = "{The Arrow of Time in the Collapse of Collisionless Self-gravitating Systems: Non-validity of the Vlasov-Poisson Equation during Violent Relaxation}",
      journal = {\apj},
         year = 2017,
        month = sep,
       volume = {846},
       number = {2},
          eid = {125},
        pages = {125},
          doi = {10.3847/1538-4357/aa876e},
archivePrefix = {arXiv},
       eprint = {1703.07363},
 primaryClass = {astro-ph.GA},
       adsurl = {https://ui.adsabs.harvard.edu/abs/2017ApJ...846..125B}
}

@ARTICLE{Hamilton2024,
       author = {{Hamilton}, Chris and {Fouvry}, Jean-Baptiste},
        title = "{Kinetic theory of stellar systems: A tutorial}",
      journal = {Physics of Plasmas},
         year = 2024,
        month = dec,
       volume = {31},
       number = {12},
          eid = {120901},
        pages = {120901},
          doi = {10.1063/5.0204214},
archivePrefix = {arXiv},
       eprint = {2402.13322},
 primaryClass = {astro-ph.GA},
       adsurl = {https://ui.adsabs.harvard.edu/abs/2024PhPl...31l0901H}
}

@ARTICLE{Choi2018,
       author = {{Choi}, Yumi and {Nidever}, David L. and {Olsen}, Knut and {Blum}, Robert D. and {Besla}, Gurtina and {Zaritsky}, Dennis and {van der Marel}, Roeland P. and {Bell}, Eric F. and {Gallart}, Carme and {Cioni}, Maria-Rosa L. and {Johnson}, L. Clifton and {Vivas}, A. Katherina and {Saha}, Abhijit and {de Boer}, Thomas J.~L. and {No{\"e}l}, Noelia E.~D. and {Monachesi}, Antonela and {Massana}, Pol and {Conn}, Blair C. and {Martinez-Delgado}, David and {Mu{\~n}oz}, Ricardo R. and {Stringfellow}, Guy S.},
        title = "{SMASHing the LMC: A Tidally Induced Warp in the Outer LMC and a Large-scale Reddening Map}",
      journal = {\apj},
         year = 2018,
        month = oct,
       volume = {866},
       number = {2},
          eid = {90},
        pages = {90},
          doi = {10.3847/1538-4357/aae083},
archivePrefix = {arXiv},
       eprint = {1804.07765},
 primaryClass = {astro-ph.GA},
       adsurl = {https://ui.adsabs.harvard.edu/abs/2018ApJ...866...90C}
}

@ARTICLE{Arranz2025b,
       author = {{Jim{\'e}nez-Arranz}, {\'O}. and {Horta}, D. and {van der Marel}, R.~P. and {Nidever}, D. and {Laporte}, C.~F.~P. and {Patel}, E. and {Rix}, H. -W.},
        title = "{Vertical structure and kinematics of the LMC disc from SDSS/Gaia}",
      journal = {\aap},
         year = 2025,
        month = jun,
       volume = {698},
          eid = {A88},
        pages = {A88},
          doi = {10.1051/0004-6361/202553705},
archivePrefix = {arXiv},
       eprint = {2501.04616},
 primaryClass = {astro-ph.GA},
       adsurl = {https://ui.adsabs.harvard.edu/abs/2025A&A...698A..88J}
}

@ARTICLE{Gardiner-Noguchi1996,
       author = {{Gardiner}, Lance T. and {Noguchi}, Masafumi},
        title = "{N-body simulations of the Small Magellanic Cloud and the Magellanic Stream}",
      journal = {\mnras},
         year = 1996,
        month = jan,
       volume = {278},
       number = {1},
        pages = {191-208},
          doi = {10.1093/mnras/278.1.191},
archivePrefix = {arXiv},
       eprint = {astro-ph/9503095},
 primaryClass = {astro-ph},
       adsurl = {https://ui.adsabs.harvard.edu/abs/1996MNRAS.278..191G}
}

@ARTICLE{Connors2006,
       author = {{Connors}, Tim W. and {Kawata}, Daisuke and {Gibson}, Brad K.},
        title = "{N-body simulations of the Magellanic stream}",
      journal = {\mnras},
         year = 2006,
        month = sep,
       volume = {371},
       number = {1},
        pages = {108-120},
          doi = {10.1111/j.1365-2966.2006.10659.x},
archivePrefix = {arXiv},
       eprint = {astro-ph/0508390},
 primaryClass = {astro-ph},
       adsurl = {https://ui.adsabs.harvard.edu/abs/2006MNRAS.371..108C}
}

@ARTICLE{Arranz2024,
       author = {{Jim{\'e}nez-Arranz}, {\'O}. and {Roca-F{\`a}brega}, S. and {Romero-G{\'o}mez}, M. and {Luri}, X. and {Bernet}, M. and {McMillan}, P.~J. and {Chemin}, L.},
        title = "{KRATOS: A large suite of N-body simulations to interpret the stellar kinematics of LMC-like discs}",
      journal = {\aap},
         year = 2024,
        month = aug,
       volume = {688},
          eid = {A51},
        pages = {A51},
          doi = {10.1051/0004-6361/202349058},
archivePrefix = {arXiv},
       eprint = {2404.04061},
 primaryClass = {astro-ph.GA},
       adsurl = {https://ui.adsabs.harvard.edu/abs/2024A&A...688A..51J}
}

@ARTICLE{Besla2010,
       author = {{Besla}, G. and {Kallivayalil}, N. and {Hernquist}, L. and {van der Marel}, R.~P. and {Cox}, T.~J. and {Kere{\v{s}}}, D.},
        title = "{Simulations of the Magellanic Stream in a First Infall Scenario}",
      journal = {\apjl},
         year = 2010,
        month = oct,
       volume = {721},
       number = {2},
        pages = {L97-L101},
          doi = {10.1088/2041-8205/721/2/L97},
archivePrefix = {arXiv},
       eprint = {1008.2210},
 primaryClass = {astro-ph.GA},
       adsurl = {https://ui.adsabs.harvard.edu/abs/2010ApJ...721L..97B}
}

@ARTICLE{Pardy2018,
       author = {{Pardy}, Stephen A. and {D'Onghia}, Elena and {Fox}, Andrew J.},
        title = "{Models of Tidally Induced Gas Filaments in the Magellanic Stream}",
      journal = {\apj},
         year = 2018,
        month = apr,
       volume = {857},
       number = {2},
          eid = {101},
        pages = {101},
          doi = {10.3847/1538-4357/aab95b},
archivePrefix = {arXiv},
       eprint = {1802.01600},
 primaryClass = {astro-ph.GA},
       adsurl = {https://ui.adsabs.harvard.edu/abs/2018ApJ...857..101P}
}

@ARTICLE{Lucchini2024,
       author = {{Lucchini}, Scott and {D'Onghia}, Elena and {Fox}, Andrew J.},
        title = "{Properties of the Magellanic Corona}",
      journal = {\apj},
         year = 2024,
        month = may,
       volume = {967},
       number = {1},
          eid = {16},
        pages = {16},
          doi = {10.3847/1538-4357/ad3c3b},
archivePrefix = {arXiv},
       eprint = {2311.16221},
 primaryClass = {astro-ph.GA},
       adsurl = {https://ui.adsabs.harvard.edu/abs/2024ApJ...967...16L}
}

@ARTICLE{Cullinane2022,
       author = {{Cullinane}, L.~R. and {Mackey}, A.~D. and {Da Costa}, G.~S. and {Erkal}, D. and {Koposov}, S.~E. and {Belokurov}, V.},
        title = "{The Magellanic Edges Survey - III. Kinematics of the disturbed LMC outskirts}",
      journal = {\mnras},
         year = 2022,
        month = jun,
       volume = {512},
       number = {4},
        pages = {4798-4818},
          doi = {10.1093/mnras/stac733},
archivePrefix = {arXiv},
       eprint = {2203.05450},
 primaryClass = {astro-ph.GA},
       adsurl = {https://ui.adsabs.harvard.edu/abs/2022MNRAS.512.4798C}
}

@ARTICLE{Rubele2018,
       author = {{Rubele}, Stefano and {Pastorelli}, Giada and {Girardi}, L{\'e}o and {Cioni}, Maria-Rosa L. and {Zaggia}, Simone and {Marigo}, Paola and {Bekki}, Kenji and {Bressan}, Alessandro and {Clementini}, Gisella and {de Grijs}, Richard and {Emerson}, Jim and {Groenewegen}, Martin A.~T. and {Ivanov}, Valentin D. and {Muraveva}, Tatiana and {Nanni}, Ambra and {Oliveira}, Joana M. and {Ripepi}, Vincenzo and {Sun}, Ning-Chen and {van Loon}, Jacco Th},
        title = "{The VMC survey - XXXI: The spatially resolved star formation history of the main body of the Small Magellanic Cloud}",
      journal = {\mnras},
         year = 2018,
        month = aug,
       volume = {478},
       number = {4},
        pages = {5017-5036},
          doi = {10.1093/mnras/sty1279},
archivePrefix = {arXiv},
       eprint = {1805.04516},
 primaryClass = {astro-ph.GA},
       adsurl = {https://ui.adsabs.harvard.edu/abs/2018MNRAS.478.5017R}
}

@ARTICLE{vanderWalt2011,
       author = {{van der Walt}, St{\'e}fan and {Colbert}, S. Chris and {Varoquaux}, Ga{\"e}l},
        title = "{The NumPy Array: A Structure for Efficient Numerical Computation}",
      journal = {Computing in Science and Engineering},
         year = 2011,
        month = mar,
       volume = {13},
       number = {2},
        pages = {22-30},
          doi = {10.1109/MCSE.2011.37},
archivePrefix = {arXiv},
       eprint = {1102.1523},
 primaryClass = {cs.MS},
       adsurl = {https://ui.adsabs.harvard.edu/abs/2011CSE....13b..22V}
}

@ARTICLE{Harris2020,
       author = {{Harris}, Charles R. and {Millman}, K. Jarrod and {van der Walt}, St{\'e}fan J. and {Gommers}, Ralf and {Virtanen}, Pauli and {Cournapeau}, David and {Wieser}, Eric and {Taylor}, Julian and {Berg}, Sebastian and {Smith}, Nathaniel J. and {Kern}, Robert and {Picus}, Matti and {Hoyer}, Stephan and {van Kerkwijk}, Marten H. and {Brett}, Matthew and {Haldane}, Allan and {del R{\'\i}o}, Jaime Fern{\'a}ndez and {Wiebe}, Mark and {Peterson}, Pearu and {G{\'e}rard-Marchant}, Pierre and {Sheppard}, Kevin and {Reddy}, Tyler and {Weckesser}, Warren and {Abbasi}, Hameer and {Gohlke}, Christoph and {Oliphant}, Travis E.},
        title = "{Array programming with NumPy}",
      journal = {\nat},
         year = 2020,
        month = sep,
       volume = {585},
       number = {7825},
        pages = {357-362},
          doi = {10.1038/s41586-020-2649-2},
archivePrefix = {arXiv},
       eprint = {2006.10256},
 primaryClass = {cs.MS},
       adsurl = {https://ui.adsabs.harvard.edu/abs/2020Natur.585..357H}
}

@ARTICLE{Virtanen2020,
       author = {{Virtanen}, Pauli and {Gommers}, Ralf and {Oliphant}, Travis E. and {Haberland}, Matt and {Reddy}, Tyler and {Cournapeau}, David and {Burovski}, Evgeni and {Peterson}, Pearu and {Weckesser}, Warren and {Bright}, Jonathan and {van der Walt}, St{\'e}fan J. and {Brett}, Matthew and {Wilson}, Joshua and {Millman}, K. Jarrod and {Mayorov}, Nikolay and {Nelson}, Andrew R.~J. and {Jones}, Eric and {Kern}, Robert and {Larson}, Eric and {Carey}, C.~J. and {Polat}, {\.I}lhan and {Feng}, Yu and {Moore}, Eric W. and {VanderPlas}, Jake and {Laxalde}, Denis and {Perktold}, Josef and {Cimrman}, Robert and {Henriksen}, Ian and {Quintero}, E.~A. and {Harris}, Charles R. and {Archibald}, Anne M. and {Ribeiro}, Ant{\^o}nio H. and {Pedregosa}, Fabian and {van Mulbregt}, Paul and {SciPy 1. 0 Contributors}},
        title = "{SciPy 1.0: fundamental algorithms for scientific computing in Python}",
      journal = {Nature Methods},
         year = 2020,
        month = feb,
       volume = {17},
        pages = {261-272},
          doi = {10.1038/s41592-019-0686-2},
archivePrefix = {arXiv},
       eprint = {1907.10121},
 primaryClass = {cs.MS},
       adsurl = {https://ui.adsabs.harvard.edu/abs/2020NatMe..17..261V}
}

@ARTICLE{Hunter2007,
  author={Hunter, John D.},
  journal={Computing in Science \& Engineering}, 
  title={Matplotlib: A 2D Graphics Environment}, 
  year={2007},
  volume={9},
  number={3},
  pages={90-95},
  doi={10.1109/MCSE.2007.55}}

@ARTICLE{Gonidakis2009,
       author = {{Gonidakis}, I. and {Livanou}, E. and {Kontizas}, E. and {Klein}, U. and {Kontizas}, M. and {Belcheva}, M. and {Tsalmantza}, P. and {Karampelas}, A.},
        title = "{Structure of the SMC. Stellar component distribution from 2MASS data}",
      journal = {\aap},
         year = 2009,
        month = mar,
       volume = {496},
       number = {2},
        pages = {375-380},
          doi = {10.1051/0004-6361/200809828},
archivePrefix = {arXiv},
       eprint = {0812.0880},
 primaryClass = {astro-ph},
       adsurl = {https://ui.adsabs.harvard.edu/abs/2009A&A...496..375G}
}

@ARTICLE{Zaritsky2025,
       author = {{Zaritsky}, Dennis and {Chandra}, Vedant and {Conroy}, Charlie and {Bonaca}, Ana and {Cargile}, Phillip A. and {Naidu}, Rohan P.},
        title = "{Untangling Magellanic Streams}",
      journal = {The Open Journal of Astrophysics},
         year = 2025,
        month = feb,
       volume = {8},
          eid = {16},
        pages = {16},
          doi = {10.33232/001c.129885},
archivePrefix = {arXiv},
       eprint = {2411.15044},
 primaryClass = {astro-ph.GA},
       adsurl = {https://ui.adsabs.harvard.edu/abs/2025OJAp....8E..16Z}
}

@ARTICLE{Soszynski2010,
       author = {{Soszy{\'n}ski}, I. and {Udalski}, A. and {Szyma{\'n}ski}, M.~K. and {Kubiak}, J. and {Pietrzy{\'n}ski}, G. and {Wyrzykowski}, {\L}. and {Ulaczyk}, K. and {Poleski}, R.},
        title = "{The Optical Gravitational Lensing Experiment. The OGLE-III Catalog of Variable Stars. IX. RR Lyr Stars in the Small Magellanic Cloud}",
      journal = {\actaa},
         year = 2010,
        month = sep,
       volume = {60},
       number = {3},
        pages = {165-178},
          doi = {10.48550/arXiv.1009.0528},
archivePrefix = {arXiv},
       eprint = {1009.0528},
 primaryClass = {astro-ph.SR},
       adsurl = {https://ui.adsabs.harvard.edu/abs/2010AcA....60..165S}
}

@ARTICLE{Luri2021,
       author = {{Gaia Collaboration} and {Luri}, X. and {Chemin}, L. and {Clementini}, G. and {Delgado}, H.~E. and {McMillan}, P.~J. and {Romero-G{\'o}mez}, M. and {Balbinot}, E. and {Castro-Ginard}, A. and {Mor}, R. and {Ripepi}, V. and {Sarro}, L.~M. and {Cioni}, M. -R.~L. and {Fabricius}, C. and {Garofalo}, A. and {Helmi}, A. and {Muraveva}, T. and {Brown}, A.~G.~A. and {Vallenari}, A. and {Prusti}, T. and {de Bruijne}, J.~H.~J. and {Babusiaux}, C. and {Biermann}, M. and {Creevey}, O.~L. and {Evans}, D.~W. and {Eyer}, L. and {Hutton}, A. and {Jansen}, F. and {Jordi}, C. and {Klioner}, S.~A. and {Lammers}, U. and {Lindegren}, L. and {Mignard}, F. and {Panem}, C. and {Pourbaix}, D. and {Randich}, S. and {Sartoretti}, P. and {Soubiran}, C. and {Walton}, N.~A. and {Arenou}, F. and {Bailer-Jones}, C.~A.~L. and {Bastian}, U. and {Cropper}, M. and {Drimmel}, R. and {Katz}, D. and {Lattanzi}, M.~G. and {van Leeuwen}, F. and {Bakker}, J. and {Casta{\~n}eda}, J. and {De Angeli}, F. and {Ducourant}, C. and {Fouesneau}, M. and {Fr{\'e}mat}, Y. and {Guerra}, R. and {Guerrier}, A. and {Guiraud}, J. and {Jean-Antoine Piccolo}, A. and {Masana}, E. and {Messineo}, R. and {Mowlavi}, N. and {Nicolas}, C. and {Nienartowicz}, K. and {Pailler}, F. and {Panuzzo}, P. and {Riclet}, F. and {Roux}, W. and {Seabroke}, G.~M. and {Sordo}, R. and {Tanga}, P. and {Th{\'e}venin}, F. and {Gracia-Abril}, G. and {Portell}, J. and {Teyssier}, D. and {Altmann}, M. and {Andrae}, R. and {Bellas-Velidis}, I. and {Benson}, K. and {Berthier}, J. and {Blomme}, R. and {Brugaletta}, E. and {Burgess}, P.~W. and {Busso}, G. and {Carry}, B. and {Cellino}, A. and {Cheek}, N. and {Damerdji}, Y. and {Davidson}, M. and {Delchambre}, L. and {Dell'Oro}, A. and {Fern{\'a}ndez-Hern{\'a}ndez}, J. and {Galluccio}, L. and {Garc{\'\i}a-Lario}, P. and {Garcia-Reinaldos}, M. and {Gonz{\'a}lez-N{\'u}{\~n}ez}, J. and {Gosset}, E. and {Haigron}, R. and {Halbwachs}, J. -L. and {Hambly}, N.~C. and {Harrison}, D.~L. and {Hatzidimitriou}, D. and {Heiter}, U. and {Hern{\'a}ndez}, J. and {Hestroffer}, D. and {Hodgkin}, S.~T. and {Holl}, B. and {Jan{\ss}en}, K. and {Jevardat de Fombelle}, G. and {Jordan}, S. and {Krone-Martins}, A. and {Lanzafame}, A.~C. and {L{\"o}ffler}, W. and {Lorca}, A. and {Manteiga}, M. and {Marchal}, O. and {Marrese}, P.~M. and {Moitinho}, A. and {Mora}, A. and {Muinonen}, K. and {Osborne}, P. and {Pancino}, E. and {Pauwels}, T. and {Recio-Blanco}, A. and {Richards}, P.~J. and {Riello}, M. and {Rimoldini}, L. and {Robin}, A.~C. and {Roegiers}, T. and {Rybizki}, J. and {Siopis}, C. and {Smith}, M. and {Sozzetti}, A. and {Ulla}, A. and {Utrilla}, E. and {van Leeuwen}, M. and {van Reeven}, W. and {Abbas}, U. and {Abreu Aramburu}, A. and {Accart}, S. and {Aerts}, C. and {Aguado}, J.~J. and {Ajaj}, M. and {Altavilla}, G. and {{\'A}lvarez}, M.~A. and {{\'A}lvarez Cid-Fuentes}, J. and {Alves}, J. and {Anderson}, R.~I. and {Anglada Varela}, E. and {Antoja}, T. and {Audard}, M. and {Baines}, D. and {Baker}, S.~G. and {Balaguer-N{\'u}{\~n}ez}, L. and {Balog}, Z. and {Barache}, C. and {Barbato}, D. and {Barros}, M. and {Barstow}, M.~A. and {Bartolom{\'e}}, S. and {Bassilana}, J. -L. and {Bauchet}, N. and {Baudesson-Stella}, A. and {Becciani}, U. and {Bellazzini}, M. and {Bernet}, M. and {Bertone}, S. and {Bianchi}, L. and {Blanco-Cuaresma}, S. and {Boch}, T. and {Bombrun}, A. and {Bossini}, D. and {Bouquillon}, S. and {Bragaglia}, A. and {Bramante}, L. and {Breedt}, E. and {Bressan}, A. and {Brouillet}, N. and {Bucciarelli}, B. and {Burlacu}, A. and {Busonero}, D. and {Butkevich}, A.~G. and {Buzzi}, R. and {Caffau}, E. and {Cancelliere}, R. and {C{\'a}novas}, H. and {Cantat-Gaudin}, T. and {Carballo}, R. and {Carlucci}, T. and {Carnerero}, M.~I. and {Carrasco}, J.~M. and {Casamiquela}, L. and {Castellani}, M. and {Castro Sampol}, P. and {Chaoul}, L. and {Charlot}, P. and {Chiavassa}, A. and {Comoretto}, G. and {Cooper}, W.~J. and {Cornez}, T. and {Cowell}, S. and {Crifo}, F.},
        title = "{Gaia Early Data Release 3. Structure and properties of the Magellanic Clouds}",
      journal = {\aap},
         year = 2021,
        month = may,
       volume = {649},
          eid = {A7},
        pages = {A7},
          doi = {10.1051/0004-6361/202039588},
archivePrefix = {arXiv},
       eprint = {2012.01771},
 primaryClass = {astro-ph.GA},
       adsurl = {https://ui.adsabs.harvard.edu/abs/2021A&A...649A...7G}
}

@ARTICLE{Brown2018,
       author = {{Gaia Collaboration} and {Brown}, A.~G.~A. and {Vallenari}, A. and {Prusti}, T. and {de Bruijne}, J.~H.~J. and {Babusiaux}, C. and {Bailer-Jones}, C.~A.~L. and {Biermann}, M. and {Evans}, D.~W. and {Eyer}, L. and {Jansen}, F. and {Jordi}, C. and {Klioner}, S.~A. and {Lammers}, U. and {Lindegren}, L. and {Luri}, X. and {Mignard}, F. and {Panem}, C. and {Pourbaix}, D. and {Randich}, S. and {Sartoretti}, P. and {Siddiqui}, H.~I. and {Soubiran}, C. and {van Leeuwen}, F. and {Walton}, N.~A. and {Arenou}, F. and {Bastian}, U. and {Cropper}, M. and {Drimmel}, R. and {Katz}, D. and {Lattanzi}, M.~G. and {Bakker}, J. and {Cacciari}, C. and {Casta{\~n}eda}, J. and {Chaoul}, L. and {Cheek}, N. and {De Angeli}, F. and {Fabricius}, C. and {Guerra}, R. and {Holl}, B. and {Masana}, E. and {Messineo}, R. and {Mowlavi}, N. and {Nienartowicz}, K. and {Panuzzo}, P. and {Portell}, J. and {Riello}, M. and {Seabroke}, G.~M. and {Tanga}, P. and {Th{\'e}venin}, F. and {Gracia-Abril}, G. and {Comoretto}, G. and {Garcia-Reinaldos}, M. and {Teyssier}, D. and {Altmann}, M. and {Andrae}, R. and {Audard}, M. and {Bellas-Velidis}, I. and {Benson}, K. and {Berthier}, J. and {Blomme}, R. and {Burgess}, P. and {Busso}, G. and {Carry}, B. and {Cellino}, A. and {Clementini}, G. and {Clotet}, M. and {Creevey}, O. and {Davidson}, M. and {De Ridder}, J. and {Delchambre}, L. and {Dell'Oro}, A. and {Ducourant}, C. and {Fern{\'a}ndez-Hern{\'a}ndez}, J. and {Fouesneau}, M. and {Fr{\'e}mat}, Y. and {Galluccio}, L. and {Garc{\'\i}a-Torres}, M. and {Gonz{\'a}lez-N{\'u}{\~n}ez}, J. and {Gonz{\'a}lez-Vidal}, J.~J. and {Gosset}, E. and {Guy}, L.~P. and {Halbwachs}, J. -L. and {Hambly}, N.~C. and {Harrison}, D.~L. and {Hern{\'a}ndez}, J. and {Hestroffer}, D. and {Hodgkin}, S.~T. and {Hutton}, A. and {Jasniewicz}, G. and {Jean-Antoine-Piccolo}, A. and {Jordan}, S. and {Korn}, A.~J. and {Krone-Martins}, A. and {Lanzafame}, A.~C. and {Lebzelter}, T. and {L{\"o}ffler}, W. and {Manteiga}, M. and {Marrese}, P.~M. and {Mart{\'\i}n-Fleitas}, J.~M. and {Moitinho}, A. and {Mora}, A. and {Muinonen}, K. and {Osinde}, J. and {Pancino}, E. and {Pauwels}, T. and {Petit}, J. -M. and {Recio-Blanco}, A. and {Richards}, P.~J. and {Rimoldini}, L. and {Robin}, A.~C. and {Sarro}, L.~M. and {Siopis}, C. and {Smith}, M. and {Sozzetti}, A. and {S{\"u}veges}, M. and {Torra}, J. and {van Reeven}, W. and {Abbas}, U. and {Abreu Aramburu}, A. and {Accart}, S. and {Aerts}, C. and {Altavilla}, G. and {{\'A}lvarez}, M.~A. and {Alvarez}, R. and {Alves}, J. and {Anderson}, R.~I. and {Andrei}, A.~H. and {Anglada Varela}, E. and {Antiche}, E. and {Antoja}, T. and {Arcay}, B. and {Astraatmadja}, T.~L. and {Bach}, N. and {Baker}, S.~G. and {Balaguer-N{\'u}{\~n}ez}, L. and {Balm}, P. and {Barache}, C. and {Barata}, C. and {Barbato}, D. and {Barblan}, F. and {Barklem}, P.~S. and {Barrado}, D. and {Barros}, M. and {Barstow}, M.~A. and {Bartholom{\'e} Mu{\~n}oz}, S. and {Bassilana}, J. -L. and {Becciani}, U. and {Bellazzini}, M. and {Berihuete}, A. and {Bertone}, S. and {Bianchi}, L. and {Bienaym{\'e}}, O. and {Blanco-Cuaresma}, S. and {Boch}, T. and {Boeche}, C. and {Bombrun}, A. and {Borrachero}, R. and {Bossini}, D. and {Bouquillon}, S. and {Bourda}, G. and {Bragaglia}, A. and {Bramante}, L. and {Breddels}, M.~A. and {Bressan}, A. and {Brouillet}, N. and {Br{\"u}semeister}, T. and {Brugaletta}, E. and {Bucciarelli}, B. and {Burlacu}, A. and {Busonero}, D. and {Butkevich}, A.~G. and {Buzzi}, R. and {Caffau}, E. and {Cancelliere}, R. and {Cannizzaro}, G. and {Cantat-Gaudin}, T. and {Carballo}, R. and {Carlucci}, T. and {Carrasco}, J.~M. and {Casamiquela}, L. and {Castellani}, M. and {Castro-Ginard}, A. and {Charlot}, P. and {Chemin}, L. and {Chiavassa}, A. and {Cocozza}, G. and {Costigan}, G. and {Cowell}, S. and {Crifo}, F. and {Crosta}, M. and {Crowley}, C. and {Cuypers}, J. and {Dafonte}, C. and {Damerdji}, Y. and {Dapergolas}, A. and {David}, P. and {David}, M. and {de Laverny}, P. and {De Luise}, F.},
        title = "{Gaia Data Release 2. Summary of the contents and survey properties}",
      journal = {\aap},
         year = 2018,
        month = aug,
       volume = {616},
          eid = {A1},
        pages = {A1},
          doi = {10.1051/0004-6361/201833051},
archivePrefix = {arXiv},
       eprint = {1804.09365},
 primaryClass = {astro-ph.GA},
       adsurl = {https://ui.adsabs.harvard.edu/abs/2018A&A...616A...1G}
}

@ARTICLE{Majewski2017,
       author = {{Majewski}, Steven R. and {Schiavon}, Ricardo P. and {Frinchaboy}, Peter M. and {Allende Prieto}, Carlos and {Barkhouser}, Robert and {Bizyaev}, Dmitry and {Blank}, Basil and {Brunner}, Sophia and {Burton}, Adam and {Carrera}, Ricardo and {Chojnowski}, S. Drew and {Cunha}, K{\'a}tia and {Epstein}, Courtney and {Fitzgerald}, Greg and {Garc{\'\i}a P{\'e}rez}, Ana E. and {Hearty}, Fred R. and {Henderson}, Chuck and {Holtzman}, Jon A. and {Johnson}, Jennifer A. and {Lam}, Charles R. and {Lawler}, James E. and {Maseman}, Paul and {M{\'e}sz{\'a}ros}, Szabolcs and {Nelson}, Matthew and {Nguyen}, Duy Coung and {Nidever}, David L. and {Pinsonneault}, Marc and {Shetrone}, Matthew and {Smee}, Stephen and {Smith}, Verne V. and {Stolberg}, Todd and {Skrutskie}, Michael F. and {Walker}, Eric and {Wilson}, John C. and {Zasowski}, Gail and {Anders}, Friedrich and {Basu}, Sarbani and {Beland}, Stephane and {Blanton}, Michael R. and {Bovy}, Jo and {Brownstein}, Joel R. and {Carlberg}, Joleen and {Chaplin}, William and {Chiappini}, Cristina and {Eisenstein}, Daniel J. and {Elsworth}, Yvonne and {Feuillet}, Diane and {Fleming}, Scott W. and {Galbraith-Frew}, Jessica and {Garc{\'\i}a}, Rafael A. and {Garc{\'\i}a-Hern{\'a}ndez}, D. An{\'\i}bal and {Gillespie}, Bruce A. and {Girardi}, L{\'e}o and {Gunn}, James E. and {Hasselquist}, Sten and {Hayden}, Michael R. and {Hekker}, Saskia and {Ivans}, Inese and {Kinemuchi}, Karen and {Klaene}, Mark and {Mahadevan}, Suvrath and {Mathur}, Savita and {Mosser}, Beno{\^\i}t and {Muna}, Demitri and {Munn}, Jeffrey A. and {Nichol}, Robert C. and {O'Connell}, Robert W. and {Parejko}, John K. and {Robin}, A.~C. and {Rocha-Pinto}, Helio and {Schultheis}, Matthias and {Serenelli}, Aldo M. and {Shane}, Neville and {Silva Aguirre}, Victor and {Sobeck}, Jennifer S. and {Thompson}, Benjamin and {Troup}, Nicholas W. and {Weinberg}, David H. and {Zamora}, Olga},
        title = "{The Apache Point Observatory Galactic Evolution Experiment (APOGEE)}",
      journal = {\aj},
         year = 2017,
        month = sep,
       volume = {154},
       number = {3},
          eid = {94},
        pages = {94},
          doi = {10.3847/1538-3881/aa784d},
archivePrefix = {arXiv},
       eprint = {1509.05420},
 primaryClass = {astro-ph.IM},
       adsurl = {https://ui.adsabs.harvard.edu/abs/2017AJ....154...94M}
}

@ARTICLE{Marel2001,
       author = {{van der Marel}, Roeland P. and {Cioni}, Maria-Rosa L.},
        title = "{Magellanic Cloud Structure from Near-Infrared Surveys. I. The Viewing Angles of the Large Magellanic Cloud}",
      journal = {\aj},
         year = 2001,
        month = oct,
       volume = {122},
       number = {4},
        pages = {1807-1826},
          doi = {10.1086/323099},
archivePrefix = {arXiv},
       eprint = {astro-ph/0105339},
 primaryClass = {astro-ph},
       adsurl = {https://ui.adsabs.harvard.edu/abs/2001AJ....122.1807V}
}

@ARTICLE{Marel2002,
       author = {{van der Marel}, Roeland P. and {Alves}, David R. and {Hardy}, Eduardo and {Suntzeff}, Nicholas B.},
        title = "{New Understanding of Large Magellanic Cloud Structure, Dynamics, and Orbit from Carbon Star Kinematics}",
      journal = {\aj},
         year = 2002,
        month = nov,
       volume = {124},
       number = {5},
        pages = {2639-2663},
          doi = {10.1086/343775},
archivePrefix = {arXiv},
       eprint = {astro-ph/0205161},
 primaryClass = {astro-ph},
       adsurl = {https://ui.adsabs.harvard.edu/abs/2002AJ....124.2639V}
}

@ARTICLE{Gunn1972,
       author = {{Gunn}, James E. and {Gott}, III, J. Richard},
        title = "{On the Infall of Matter Into Clusters of Galaxies and Some Effects on Their Evolution}",
      journal = {\apj},
         year = 1972,
        month = aug,
       volume = {176},
        pages = {1},
          doi = {10.1086/151605},
       adsurl = {https://ui.adsabs.harvard.edu/abs/1972ApJ...176....1G}
}

@ARTICLE{Salem2015,
       author = {{Salem}, Munier and {Besla}, Gurtina and {Bryan}, Greg and {Putman}, Mary and {van der Marel}, Roeland P. and {Tonnesen}, Stephanie},
        title = "{Ram Pressure Stripping of the Large Magellanic Cloud's Disk as a Probe of the Milky Way's Circumgalactic Medium}",
      journal = {\apj},
         year = 2015,
        month = dec,
       volume = {815},
       number = {1},
          eid = {77},
        pages = {77},
          doi = {10.1088/0004-637X/815/1/77},
archivePrefix = {arXiv},
       eprint = {1507.07935},
 primaryClass = {astro-ph.GA},
       adsurl = {https://ui.adsabs.harvard.edu/abs/2015ApJ...815...77S}
}

@ARTICLE{Roediger2005,
       author = {{Roediger}, E. and {Hensler}, G.},
        title = "{Ram pressure stripping of disk galaxies. From high to low density environments}",
      journal = {\aap},
         year = 2005,
        month = apr,
       volume = {433},
       number = {3},
        pages = {875-895},
          doi = {10.1051/0004-6361:20042131},
       adsurl = {https://ui.adsabs.harvard.edu/abs/2005A&A...433..875R}
}

@ARTICLE{Roediger2006,
       author = {{Roediger}, Elke and {Br{\"u}ggen}, Marcus},
        title = "{Ram pressure stripping of disc galaxies: the role of the inclination angle}",
      journal = {\mnras},
         year = 2006,
        month = jun,
       volume = {369},
       number = {2},
        pages = {567-580},
          doi = {10.1111/j.1365-2966.2006.10335.x},
archivePrefix = {arXiv},
       eprint = {astro-ph/0512365},
 primaryClass = {astro-ph},
       adsurl = {https://ui.adsabs.harvard.edu/abs/2006MNRAS.369..567R}
}

@ARTICLE{Tonnesen2009,
       author = {{Tonnesen}, Stephanie and {Bryan}, Greg L.},
        title = "{Gas Stripping in Simulated Galaxies with a Multiphase Interstellar Medium}",
      journal = {\apj},
         year = 2009,
        month = apr,
       volume = {694},
       number = {2},
        pages = {789-804},
          doi = {10.1088/0004-637X/694/2/789},
archivePrefix = {arXiv},
       eprint = {0901.2115},
 primaryClass = {astro-ph.GA},
       adsurl = {https://ui.adsabs.harvard.edu/abs/2009ApJ...694..789T}
}

@ARTICLE{Agertz2007,
       author = {{Agertz}, Oscar and {Moore}, Ben and {Stadel}, Joachim and {Potter}, Doug and {Miniati}, Francesco and {Read}, Justin and {Mayer}, Lucio and {Gawryszczak}, Artur and {Kravtsov}, Andrey and {Nordlund}, {\r{A}}ke and {Pearce}, Frazer and {Quilis}, Vicent and {Rudd}, Douglas and {Springel}, Volker and {Stone}, James and {Tasker}, Elizabeth and {Teyssier}, Romain and {Wadsley}, James and {Walder}, Rolf},
        title = "{Fundamental differences between SPH and grid methods}",
      journal = {\mnras},
         year = 2007,
        month = sep,
       volume = {380},
       number = {3},
        pages = {963-978},
          doi = {10.1111/j.1365-2966.2007.12183.x},
archivePrefix = {arXiv},
       eprint = {astro-ph/0610051},
 primaryClass = {astro-ph},
       adsurl = {https://ui.adsabs.harvard.edu/abs/2007MNRAS.380..963A}
}

@ARTICLE{Hayward2014,
       author = {{Hayward}, Christopher C. and {Torrey}, Paul and {Springel}, Volker and {Hernquist}, Lars and {Vogelsberger}, Mark},
        title = "{Galaxy mergers on a moving mesh: a comparison with smoothed particle hydrodynamics}",
      journal = {\mnras},
         year = 2014,
        month = aug,
       volume = {442},
       number = {3},
        pages = {1992-2016},
          doi = {10.1093/mnras/stu957},
archivePrefix = {arXiv},
       eprint = {1309.2942},
 primaryClass = {astro-ph.CO},
       adsurl = {https://ui.adsabs.harvard.edu/abs/2014MNRAS.442.1992H}
}

@ARTICLE{Sijacki2012,
       author = {{Sijacki}, Debora and {Vogelsberger}, Mark and {Kere{\v{s}}}, Du{\v{s}}an and {Springel}, Volker and {Hernquist}, Lars},
        title = "{Moving mesh cosmology: the hydrodynamics of galaxy formation}",
      journal = {\mnras},
         year = 2012,
        month = aug,
       volume = {424},
       number = {4},
        pages = {2999-3027},
          doi = {10.1111/j.1365-2966.2012.21466.x},
archivePrefix = {arXiv},
       eprint = {1109.3468},
 primaryClass = {astro-ph.CO},
       adsurl = {https://ui.adsabs.harvard.edu/abs/2012MNRAS.424.2999S}
}

@ARTICLE{Xie2025,
       author = {{Xie}, Lizhi and {De Lucia}, Gabriella and {Fossati}, Matteo and {Fontanot}, Fabio and {Hirschmann}, Michaela},
        title = "{The impact of ram pressure on cluster galaxies, insights from GAEA and TNG}",
      journal = {\aap},
         year = 2025,
        month = jun,
       volume = {698},
          eid = {A73},
        pages = {A73},
          doi = {10.1051/0004-6361/202553915},
archivePrefix = {arXiv},
       eprint = {2504.12863},
 primaryClass = {astro-ph.GA},
       adsurl = {https://ui.adsabs.harvard.edu/abs/2025A&A...698A..73X}
}

@ARTICLE{Krishnarao2022,
       author = {{Krishnarao}, Dhanesh and {Fox}, Andrew J. and {D'Onghia}, Elena and {Wakker}, Bart P. and {Cashman}, Frances H. and {Howk}, J. Christopher and {Lucchini}, Scott and {French}, David M. and {Lehner}, Nicolas},
        title = "{Observations of a Magellanic Corona}",
      journal = {\nat},
         year = 2022,
        month = sep,
       volume = {609},
       number = {7929},
        pages = {915-918},
          doi = {10.1038/s41586-022-05090-5},
archivePrefix = {arXiv},
       eprint = {2209.15017},
 primaryClass = {astro-ph.GA},
       adsurl = {https://ui.adsabs.harvard.edu/abs/2022Natur.609..915K}
}

@ARTICLE{Emerick2016,
       author = {{Emerick}, Andrew and {Mac Low}, Mordecai-Mark and {Grcevich}, Jana and {Gatto}, Andrea},
        title = "{Gas Loss by Ram Pressure Stripping and Internal Feedback from Low-mass Milky Way Satellites}",
      journal = {\apj},
         year = 2016,
        month = aug,
       volume = {826},
       number = {2},
          eid = {148},
        pages = {148},
          doi = {10.3847/0004-637X/826/2/148},
archivePrefix = {arXiv},
       eprint = {1605.02746},
 primaryClass = {astro-ph.GA},
       adsurl = {https://ui.adsabs.harvard.edu/abs/2016ApJ...826..148E}
}

@ARTICLE{Gomez2017,
       author = {{G{\'o}mez}, Facundo A. and {White}, Simon D.~M. and {Grand}, Robert J.~J. and {Marinacci}, Federico and {Springel}, Volker and {Pakmor}, R{\"u}diger},
        title = "{Warps and waves in the stellar discs of the Auriga cosmological simulations}",
      journal = {\mnras},
         year = 2017,
        month = mar,
       volume = {465},
       number = {3},
        pages = {3446-3460},
          doi = {10.1093/mnras/stw2957},
archivePrefix = {arXiv},
       eprint = {1606.06295},
 primaryClass = {astro-ph.GA},
       adsurl = {https://ui.adsabs.harvard.edu/abs/2017MNRAS.465.3446G}
}

@ARTICLE{Gomez2016,
       author = {{G{\'o}mez}, Facundo A. and {White}, Simon D.~M. and {Marinacci}, Federico and {Slater}, Colin T. and {Grand}, Robert J.~J. and {Springel}, Volker and {Pakmor}, R{\"u}diger},
        title = "{A fully cosmological model of a Monoceros-like ring}",
      journal = {\mnras},
         year = 2016,
        month = mar,
       volume = {456},
       number = {3},
        pages = {2779-2793},
          doi = {10.1093/mnras/stv2786},
archivePrefix = {arXiv},
       eprint = {1509.08459},
 primaryClass = {astro-ph.GA},
       adsurl = {https://ui.adsabs.harvard.edu/abs/2016MNRAS.456.2779G}
}

@ARTICLE{RusselDopita1992,
       author = {{Russell}, Stephen C. and {Dopita}, Michael A.},
        title = "{Abundances of the Heavy Elements in the Magellanic Clouds. III. Interpretation of Results}",
      journal = {\apj},
         year = 1992,
        month = jan,
       volume = {384},
        pages = {508},
          doi = {10.1086/170893},
       adsurl = {https://ui.adsabs.harvard.edu/abs/1992ApJ...384..508R}
}

@ARTICLE{Gordon1998,
       author = {{Gordon}, Karl D. and {Clayton}, Geoffrey C.},
        title = "{Starburst-like Dust Extinction in the Small Magellanic Cloud}",
      journal = {\apj},
         year = 1998,
        month = jun,
       volume = {500},
       number = {2},
        pages = {816-824},
          doi = {10.1086/305774},
archivePrefix = {arXiv},
       eprint = {astro-ph/9802003},
 primaryClass = {astro-ph},
       adsurl = {https://ui.adsabs.harvard.edu/abs/1998ApJ...500..816G}
}

@ARTICLE{Kulkarni2021,
       author = {{Kulkarni}, S.~R. and {Harrison}, Fiona A. and {Grefenstette}, Brian W. and {Earnshaw}, Hannah P. and {Andreoni}, Igor and {Berg}, Danielle A. and {Bloom}, Joshua S. and {Cenko}, S. Bradley and {Chornock}, Ryan and {Christiansen}, Jessie L. and {Coughlin}, Michael W. and {Wuollet Criswell}, Alexander and {Darvish}, Behnam and {Das}, Kaustav K. and {De}, Kishalay and {Dessart}, Luc and {Dixon}, Don and {Dorsman}, Bas and {El-Badry}, Kareem and {Evans}, Christopher and {Ford}, K.~E. Saavik and {Fremling}, Christoffer and {Gansicke}, Boris T. and {Gezari}, Suvi and {Goetberg}, Y. and {Green}, Gregory M. and {Graham}, Matthew J. and {Heida}, Marianne and {Ho}, Anna Y.~Q. and {Jaodand}, Amruta D. and {Johns-Krull}, Christopher M. and {Kasliwal}, Mansi M. and {Lazzarini}, Margaret and {Lu}, Wenbin and {Margutti}, Raffaella and {Martin}, D. Christopher and {Masters}, Daniel Charles and {McKernan}, Barry and {Naze}, Yael and {Nissanke}, Samaya M. and {Parazin}, B. and {Perley}, Daniel A. and {Phinney}, E. Sterl and {Piro}, Anthony L. and {Raaijmakers}, G. and {Rauw}, Gregor and {Rodriguez}, Antonio C. and {Sana}, Hugues and {Senchyna}, Peter and {Singer}, Leo P. and {Spake}, Jessica J. and {Stassun}, Keivan G. and {Stern}, Daniel and {Teplitz}, Harry I. and {Weisz}, Daniel R. and {Yao}, Yuhan},
        title = "{Science with the Ultraviolet Explorer (UVEX)}",
      journal = {arXiv e-prints},
         year = 2021,
        month = nov,
          eid = {arXiv:2111.15608},
        pages = {arXiv:2111.15608},
          doi = {10.48550/arXiv.2111.15608},
archivePrefix = {arXiv},
       eprint = {2111.15608},
 primaryClass = {astro-ph.GA},
       adsurl = {https://ui.adsabs.harvard.edu/abs/2021arXiv211115608K}
}

@ARTICLE{Murray2024Scylla,
       author = {{Murray}, Claire E. and {Lindberg}, Christina W. and {Yanchulova Merica-Jones}, Petia and {Williams}, Benjamin F. and {Cohen}, Roger E. and {Gordon}, Karl D. and {McQuinn}, Kristen B.~W. and {Choi}, Yumi and {Burhenne}, Clare and {Sandstrom}, Karin M. and {Bot}, Caroline and {Johnson}, L. Clifton and {Goldman}, Steven R. and {Clark}, Christopher J.~R. and {Roman-Duval}, Julia C. and {Gilbert}, Karoline M. and {Peek}, J.~E.~G. and {Hirschauer}, Alec S. and {Boyer}, Martha L. and {Dolphin}, Andrew E.},
        title = "{Scylla. I. A Pure-parallel, Multiwavelength Imaging Survey of the ULLYSES Fields in the LMC and SMC}",
      journal = {\apjs},
         year = 2024,
        month = nov,
       volume = {275},
       number = {1},
          eid = {5},
        pages = {5},
          doi = {10.3847/1538-4365/ad6de2},
archivePrefix = {arXiv},
       eprint = {2410.11695},
 primaryClass = {astro-ph.GA},
       adsurl = {https://ui.adsabs.harvard.edu/abs/2024ApJS..275....5M}
}

@ARTICLE{Roman-Duval2025,
       author = {{Roman-Duval}, Julia and {Fischer}, William J. and {Fullerton}, Alexander W. and {Taylor}, Jo and {Plesha}, Rachel and {Proffitt}, Charles and {Monroe}, TalaWanda and {Fischer}, Travis C. and {Aloisi}, Alessandra and {Bouret}, Jean-Claude and {Britt}, Christopher and {Calvet}, Nuria and {Carlberg}, Joleen K. and {Crowther}, Paul A. and {De Rosa}, Gisella and {Dixon}, William V. and {Espaillat}, Catherine C. and {Evans}, Christopher J. and {Fox}, Andrew J. and {France}, Kevin and {Garcia}, Miriam and {Fleming}, Scott W. and {Frazer}, Elaine M. and {G{\'o}mez de Castro}, Ana I. and {Herczeg}, Gregory J. and {Hernandez}, Svea and {Hirschauer}, Alec S. and {James}, Bethan L. and {Johns-Krull}, Christopher M. and {Leitherer}, Claus and {Lockwood}, Sean and {Najita}, Joan and {Oey}, M.~S. and {Oliveira}, Cristina and {Pauly}, Tyler and {Reid}, I. Neill and {Riedel}, Adric and {Rodriguez}, David R. and {Sahnow}, David and {Sankrit}, Ravi and {Sembach}, Kenneth R. and {Shaw}, Richard and {Smith}, Linda J. and {Sohn}, S. Tony and {Som}, Debopam and {{\'U}beda}, Leonardo and {Welty}, Daniel E.},
        title = "{The UV Legacy Library of Young Stars as Essential Standards (ULLYSES) Large Director's Discretionary Program with Hubble. I. Goals, Design, and Initial Results}",
      journal = {\apj},
         year = 2025,
        month = may,
       volume = {985},
       number = {1},
          eid = {109},
        pages = {109},
          doi = {10.3847/1538-4357/adc45b},
archivePrefix = {arXiv},
       eprint = {2504.05446},
 primaryClass = {astro-ph.SR},
       adsurl = {https://ui.adsabs.harvard.edu/abs/2025ApJ...985..109R}
}

@ARTICLE{Williams2010,
       author = {{Williams}, Michael J. and {Bureau}, Martin and {Cappellari}, Michele},
        title = "{The Tully-Fisher relations of early-type spiral and S0 galaxies}",
      journal = {\mnras},
         year = 2010,
        month = dec,
       volume = {409},
       number = {4},
        pages = {1330-1346},
          doi = {10.1111/j.1365-2966.2010.17406.x},
archivePrefix = {arXiv},
       eprint = {1007.4072},
 primaryClass = {astro-ph.GA},
       adsurl = {https://ui.adsabs.harvard.edu/abs/2010MNRAS.409.1330W}
}

@ARTICLE{Hinz2003,
       author = {{Hinz}, J.~L. and {Rieke}, G.~H. and {Caldwell}, N.},
        title = "{The Tully-Fisher Relation in Coma and Virgo Cluster S0 Galaxies}",
      journal = {\aj},
         year = 2003,
        month = dec,
       volume = {126},
       number = {6},
        pages = {2622-2634},
          doi = {10.1086/379555},
       adsurl = {https://ui.adsabs.harvard.edu/abs/2003AJ....126.2622H}
}

@ARTICLE{Bolatto2011,
       author = {{Bolatto}, Alberto D. and {Leroy}, Adam K. and {Jameson}, Katherine and {Ostriker}, Eve and {Gordon}, Karl and {Lawton}, Brandon and {Stanimirovi{\'c}}, Sne{\v{z}}ana and {Israel}, Frank P. and {Madden}, Suzanne C. and {Hony}, Sacha and {Sandstrom}, Karin M. and {Bot}, Caroline and {Rubio}, M{\'o}nica and {Winkler}, P. Frank and {Roman-Duval}, Julia and {van Loon}, Jacco Th. and {Oliveira}, Joana M. and {Indebetouw}, R{\'e}my},
        title = "{The State of the Gas and the Relation between Gas and Star Formation at Low Metallicity: The Small Magellanic Cloud}",
      journal = {\apj},
         year = 2011,
        month = nov,
       volume = {741},
       number = {1},
          eid = {12},
        pages = {12},
          doi = {10.1088/0004-637X/741/1/12},
archivePrefix = {arXiv},
       eprint = {1107.1717},
 primaryClass = {astro-ph.CO},
       adsurl = {https://ui.adsabs.harvard.edu/abs/2011ApJ...741...12B}
}

@ARTICLE{Jameson2016,
       author = {{Jameson}, Katherine E. and {Bolatto}, Alberto D. and {Leroy}, Adam K. and {Meixner}, Margaret and {Roman-Duval}, Julia and {Gordon}, Karl and {Hughes}, Annie and {Israel}, Frank P. and {Rubio}, Monica and {Indebetouw}, Remy and {Madden}, Suzanne C. and {Bot}, Caroline and {Hony}, Sacha and {Cormier}, Diane and {Pellegrini}, Eric W. and {Galametz}, Maud and {Sonneborn}, George},
        title = "{The Relationship Between Molecular Gas, H I, and Star Formation in the Low-mass, Low-metallicity Magellanic Clouds}",
      journal = {\apj},
         year = 2016,
        month = jul,
       volume = {825},
       number = {1},
          eid = {12},
        pages = {12},
          doi = {10.3847/0004-637X/825/1/12},
archivePrefix = {arXiv},
       eprint = {1510.08084},
 primaryClass = {astro-ph.GA},
       adsurl = {https://ui.adsabs.harvard.edu/abs/2016ApJ...825...12J}
}

@ARTICLE{Harris2009,
       author = {{Harris}, Jason and {Zaritsky}, Dennis},
        title = "{The Star Formation History of the Large Magellanic Cloud}",
      journal = {\aj},
         year = 2009,
        month = nov,
       volume = {138},
       number = {5},
        pages = {1243-1260},
          doi = {10.1088/0004-6256/138/5/1243},
archivePrefix = {arXiv},
       eprint = {0908.1422},
 primaryClass = {astro-ph.CO},
       adsurl = {https://ui.adsabs.harvard.edu/abs/2009AJ....138.1243H}
}

@ARTICLE{Weisz2013,
       author = {{Weisz}, Daniel R. and {Dolphin}, Andrew E. and {Skillman}, Evan D. and {Holtzman}, Jon and {Dalcanton}, Julianne J. and {Cole}, Andrew A. and {Neary}, Kyle},
        title = "{Comparing the ancient star formation histories of the Magellanic Clouds}",
      journal = {\mnras},
         year = 2013,
        month = may,
       volume = {431},
       number = {1},
        pages = {364-371},
          doi = {10.1093/mnras/stt165},
archivePrefix = {arXiv},
       eprint = {1301.7422},
 primaryClass = {astro-ph.CO},
       adsurl = {https://ui.adsabs.harvard.edu/abs/2013MNRAS.431..364W}
}

@ARTICLE{Lucchini2021,
       author = {{Lucchini}, Scott and {D'Onghia}, Elena and {Fox}, Andrew J.},
        title = "{The Magellanic Stream at 20 kpc: A New Orbital History for the Magellanic Clouds}",
      journal = {\apjl},
         year = 2021,
        month = nov,
       volume = {921},
       number = {2},
          eid = {L36},
        pages = {L36},
          doi = {10.3847/2041-8213/ac3338},
archivePrefix = {arXiv},
       eprint = {2110.11355},
 primaryClass = {astro-ph.GA},
       adsurl = {https://ui.adsabs.harvard.edu/abs/2021ApJ...921L..36L}
}

@ARTICLE{Yoshizawa2003,
       author = {{Yoshizawa}, Akira M. and {Noguchi}, Masafumi},
        title = "{The dynamical evolution and star formation history of the Small Magellanic Cloud: effects of interactions with the Galaxy and the Large Magellanic Cloud}",
      journal = {\mnras},
         year = 2003,
        month = mar,
       volume = {339},
       number = {4},
        pages = {1135-1154},
          doi = {10.1046/j.1365-8711.2003.06263.x},
       adsurl = {https://ui.adsabs.harvard.edu/abs/2003MNRAS.339.1135Y}
}

@ARTICLE{Foote2023,
       author = {{Foote}, Hayden R. and {Besla}, Gurtina and {Mocz}, Philip and {Garavito-Camargo}, Nicol{\'a}s and {Lancaster}, Lachlan and {Sparre}, Martin and {Cunningham}, Emily C. and {Vogelsberger}, Mark and {G{\'o}mez}, Facundo A. and {Laporte}, Chervin F.~P.},
        title = "{Structure, Kinematics, and Observability of the Large Magellanic Cloud's Dynamical Friction Wake in Cold versus Fuzzy Dark Matter}",
      journal = {\apj},
         year = 2023,
        month = sep,
       volume = {954},
       number = {2},
          eid = {163},
        pages = {163},
          doi = {10.3847/1538-4357/ace533},
archivePrefix = {arXiv},
       eprint = {2307.00053},
 primaryClass = {astro-ph.GA},
       adsurl = {https://ui.adsabs.harvard.edu/abs/2023ApJ...954..163F}
}

@ARTICLE{Bekki2009,
       author = {{Bekki}, Kenji and {Stanimirovi{\'c}}, Sne{\v{z}}ana},
        title = "{The total mass and dark halo properties of the Small Magellanic Cloud}",
      journal = {\mnras},
         year = 2009,
        month = may,
       volume = {395},
       number = {1},
        pages = {342-350},
          doi = {10.1111/j.1365-2966.2009.14514.x},
archivePrefix = {arXiv},
       eprint = {0807.2102},
 primaryClass = {astro-ph},
       adsurl = {https://ui.adsabs.harvard.edu/abs/2009MNRAS.395..342B}
}

@ARTICLE{BekkiChiba2009,
       author = {{Bekki}, Kenji and {Chiba}, Masashi},
        title = "{Origin of Structural and Kinematic Properties of the Small Magellanic Cloud}",
      journal = {\pasa},
         year = 2009,
        month = apr,
       volume = {26},
       number = {1},
        pages = {37-57},
          doi = {10.1071/AS08020},
archivePrefix = {arXiv},
       eprint = {0806.4657},
 primaryClass = {astro-ph},
       adsurl = {https://ui.adsabs.harvard.edu/abs/2009PASA...26...37B}
}

@ARTICLE{Dopita1985,
       author = {{Dopita}, M.~A. and {Ford}, H.~C. and {Lawrence}, C.~J. and {Webster}, B.~L.},
        title = "{The kinematics and internal dynamics of planetary nebulae in the Small Magellanic Cloud.}",
      journal = {\apj},
         year = 1985,
        month = sep,
       volume = {296},
        pages = {390-398},
          doi = {10.1086/163457},
       adsurl = {https://ui.adsabs.harvard.edu/abs/1985ApJ...296..390D}
}

@ARTICLE{BekkiChiba2008,
       author = {{Bekki}, Kenji and {Chiba}, Masashi},
        title = "{Formation of the Small Magellanic Cloud: An Ancient Major Merger as a Solution to the Kinematical Differences between Old Stars and H I Gas}",
      journal = {\apjl},
         year = 2008,
        month = jun,
       volume = {679},
       number = {2},
        pages = {L89},
          doi = {10.1086/589441},
archivePrefix = {arXiv},
       eprint = {0804.4563},
 primaryClass = {astro-ph},
       adsurl = {https://ui.adsabs.harvard.edu/abs/2008ApJ...679L..89B}
}

@ARTICLE{Hatzidimitriou1993,
       author = {{Hatzidimitriou}, D. and {Cannon}, R.~D. and {Hawkins}, M.~R.~S.},
        title = "{Kinematics in the outer parts of the SMC.}",
      journal = {\mnras},
         year = 1993,
        month = apr,
       volume = {261},
        pages = {873-882},
          doi = {10.1093/mnras/261.4.873},
       adsurl = {https://ui.adsabs.harvard.edu/abs/1993MNRAS.261..873H}
}

@ARTICLE{Hardy1989,
       author = {{Hardy}, Eduardo and {Suntzeff}, Nicholas B. and {Azzopardi}, Marc},
        title = "{The Kinematics of the Small Magellanic Cloud from Its Field Carbon Stars}",
      journal = {\apj},
         year = 1989,
        month = sep,
       volume = {344},
        pages = {210},
          doi = {10.1086/167790},
       adsurl = {https://ui.adsabs.harvard.edu/abs/1989ApJ...344..210H}
}

@ARTICLE{Mathewson1986,
       author = {{Mathewson}, D.~S. and {Ford}, V.~L. and {Visvanathan}, N.},
        title = "{The Structure of the Small Magellanic Cloud}",
      journal = {\apj},
         year = 1986,
        month = feb,
       volume = {301},
        pages = {664},
          doi = {10.1086/163932},
       adsurl = {https://ui.adsabs.harvard.edu/abs/1986ApJ...301..664M}
}

@ARTICLE{Murai1980,
       author = {{Murai}, Tadayuki and {Fujimoto}, Mitsuaki},
        title = "{The Magellanic Stream and the Galaxy with a Massive Halo}",
      journal = {\pasj},
         year = 1980,
        month = dec,
       volume = {32},
       number = {4},
        pages = {581-603},
          doi = {10.1093/pasj/32.4.581},
       adsurl = {https://ui.adsabs.harvard.edu/abs/1980PASJ...32..581M}
}

@ARTICLE{Torres1987,
       author = {{Torres}, G. and {Carranza}, G.~J.},
        title = "{The H II radial velocity field and kinematics of the Small MagellanicCloud.}",
      journal = {\mnras},
         year = 1987,
        month = jun,
       volume = {226},
        pages = {513-529},
          doi = {10.1093/mnras/226.3.513},
       adsurl = {https://ui.adsabs.harvard.edu/abs/1987MNRAS.226..513T}
}

@ARTICLE{Ocvirk2025,
       author = {{Ocvirk}, Pierre and {Lewis}, Joseph S.~W. and {Conaboy}, Luke and {Dubois}, Yohan and {Bethermin}, Matthieu and {Sorce}, Jenny G. and {Aubert}, Dominique and {Shapiro}, Paul R. and {Dawoodbhoy}, Taha and {Lee}, Joohyun and {Teyssier}, Romain and {Yepes}, Gustavo and {Gottl{\"o}ber}, Stefan and {Iliev}, Ilian T. and {Ahn}, Kyungjin and {Park}, Hyunbae and {Palanque}, Mei},
        title = "{Dust-UV offsets in high-redshift galaxies in the Cosmic Dawn III simulation}",
      journal = {\aap},
         year = 2025,
        month = nov,
       volume = {703},
          eid = {A98},
        pages = {A98},
          doi = {10.1051/0004-6361/202452098},
archivePrefix = {arXiv},
       eprint = {2409.05946},
 primaryClass = {astro-ph.GA},
       adsurl = {https://ui.adsabs.harvard.edu/abs/2025A&A...703A..98O}
}

@ARTICLE{Joseph2023,
       author = {{Lewis}, Joseph S.~W. and {Ocvirk}, Pierre and {Dubois}, Yohan and {Aubert}, Dominique and {Chardin}, Jonathan and {Gillet}, Nicolas and {Th{\'e}lie}, {\'E}milie},
        title = "{DUSTiER (DUST in the Epoch of Reionization): dusty galaxies in cosmological radiation-hydrodynamical simulations of the Epoch of Reionization with RAMSES-CUDATON}",
      journal = {\mnras},
         year = 2023,
        month = mar,
       volume = {519},
       number = {4},
        pages = {5987-6007},
          doi = {10.1093/mnras/stad081},
archivePrefix = {arXiv},
       eprint = {2204.03949},
 primaryClass = {astro-ph.GA},
       adsurl = {https://ui.adsabs.harvard.edu/abs/2023MNRAS.519.5987L}
}

@ARTICLE{Markov2023,
       author = {{Markov}, V. and {Gallerani}, S. and {Pallottini}, A. and {Sommovigo}, L. and {Carniani}, S. and {Ferrara}, A. and {Parlanti}, E. and {Di Mascia}, F.},
        title = "{Dust attenuation law in JWST galaxies at z {\ensuremath{\sim}} 7-8}",
      journal = {\aap},
         year = 2023,
        month = nov,
       volume = {679},
          eid = {A12},
        pages = {A12},
          doi = {10.1051/0004-6361/202346723},
archivePrefix = {arXiv},
       eprint = {2304.11178},
 primaryClass = {astro-ph.GA},
       adsurl = {https://ui.adsabs.harvard.edu/abs/2023A&A...679A..12M}
}

@ARTICLE{Schouws2022,
       author = {{Schouws}, Sander and {Stefanon}, Mauro and {Bouwens}, Rychard and {Smit}, Renske and {Hodge}, Jacqueline and {Labb{\'e}}, Ivo and {Algera}, Hiddo and {Boogaard}, Leindert and {Carniani}, Stefano and {Fudamoto}, Yoshinobu and {Holwerda}, Benne W. and {Illingworth}, Garth D. and {Maiolino}, Roberto and {Maseda}, Michael and {Oesch}, Pascal and {van der Werf}, Paul},
        title = "{Significant Dust-obscured Star Formation in Luminous Lyman-break Galaxies at z   7-8}",
      journal = {\apj},
         year = 2022,
        month = mar,
       volume = {928},
       number = {1},
          eid = {31},
        pages = {31},
          doi = {10.3847/1538-4357/ac4605},
archivePrefix = {arXiv},
       eprint = {2105.12133},
 primaryClass = {astro-ph.GA},
       adsurl = {https://ui.adsabs.harvard.edu/abs/2022ApJ...928...31S}
}

@ARTICLE{Olsen2011,
       author = {{Olsen}, Knut A.~G. and {Zaritsky}, Dennis and {Blum}, Robert D. and {Boyer}, Martha L. and {Gordon}, Karl D.},
        title = "{A Population of Accreted Small Magellanic Cloud Stars in the Large Magellanic Cloud}",
      journal = {\apj},
         year = 2011,
        month = aug,
       volume = {737},
       number = {1},
          eid = {29},
        pages = {29},
          doi = {10.1088/0004-637X/737/1/29},
archivePrefix = {arXiv},
       eprint = {1106.0044},
 primaryClass = {astro-ph.GA},
       adsurl = {https://ui.adsabs.harvard.edu/abs/2011ApJ...737...29O}
}

@ARTICLE{Noel2007,
       author = {{No{\"e}l}, Noelia E.~D. and {Gallart}, Carme and {Costa}, Edgardo and {M{\'e}ndez}, Ren{\'e} A.},
        title = "{Old Main-Sequence Turnoff Photometry in the Small Magellanic Cloud. I. Constraints on the Star Formation History in Different Fields}",
      journal = {\aj},
         year = 2007,
        month = may,
       volume = {133},
       number = {5},
        pages = {2037-2052},
          doi = {10.1086/512668},
archivePrefix = {arXiv},
       eprint = {0704.1151},
 primaryClass = {astro-ph},
       adsurl = {https://ui.adsabs.harvard.edu/abs/2007AJ....133.2037N}
}

@ARTICLE{Noel2009,
       author = {{No{\"e}l}, Noelia E.~D. and {Aparicio}, Antonio and {Gallart}, Carme and {Hidalgo}, Sebasti{\'a}n L. and {Costa}, Edgardo and {M{\'e}ndez}, Ren{\'e} A.},
        title = "{Old Main-Sequence Turnoff Photometry in the Small Magellanic Cloud. II. Star Formation History and its Spatial Gradients}",
      journal = {\apj},
         year = 2009,
        month = nov,
       volume = {705},
       number = {2},
        pages = {1260-1274},
          doi = {10.1088/0004-637X/705/2/1260},
archivePrefix = {arXiv},
       eprint = {0909.2873},
 primaryClass = {astro-ph.CO},
       adsurl = {https://ui.adsabs.harvard.edu/abs/2009ApJ...705.1260N}
}

@ARTICLE{DeLeo2020,
       author = {{De Leo}, Michele and {Carrera}, Ricardo and {No{\"e}l}, Noelia E.~D. and {Read}, Justin I. and {Erkal}, Denis and {Gallart}, Carme},
        title = "{Revealing the tidal scars of the Small Magellanic Cloud}",
      journal = {\mnras},
         year = 2020,
        month = jun,
       volume = {495},
       number = {1},
        pages = {98-113},
          doi = {10.1093/mnras/staa1122},
archivePrefix = {arXiv},
       eprint = {2002.11138},
 primaryClass = {astro-ph.GA},
       adsurl = {https://ui.adsabs.harvard.edu/abs/2020MNRAS.495...98D}
}

@ARTICLE{DeLeo2024,
       author = {{De Leo}, Michele and {Read}, Justin I. and {No{\"e}l}, Noelia E.~D. and {Erkal}, Denis and {Massana}, Pol and {Carrera}, Ricardo},
        title = "{Surviving the waves: evidence for a dark matter cusp in the tidally disrupting Small Magellanic Cloud}",
      journal = {\mnras},
         year = 2024,
        month = nov,
       volume = {535},
       number = {1},
        pages = {1015-1034},
          doi = {10.1093/mnras/stae2428},
archivePrefix = {arXiv},
       eprint = {2303.08838},
 primaryClass = {astro-ph.GA},
       adsurl = {https://ui.adsabs.harvard.edu/abs/2024MNRAS.535.1015D}
}

@BOOK{BT2008,
       author = {{Binney}, James and {Tremaine}, Scott},
        title = "{Galactic Dynamics: Second Edition}",
         year = 2008,
       adsurl = {https://ui.adsabs.harvard.edu/abs/2008gady.book.....B}
}

@ARTICLE{Kreckel2011,
       author = {{Kreckel}, K. and {Platen}, E. and {Arag{\'o}n-Calvo}, M.~A. and {van Gorkom}, J.~H. and {van de Weygaert}, R. and {van der Hulst}, J.~M. and {Kova{\v{c}}}, K. and {Yip}, C.-W. and {Peebles}, P.~J.~E.},
        title = "{Only the Lonely: H I Imaging of Void Galaxies}",
      journal = {\aj},
         year = 2011,
        month = jan,
       volume = {141},
       number = {1},
          eid = {4},
        pages = {4},
          doi = {10.1088/0004-6256/141/1/4},
archivePrefix = {arXiv},
       eprint = {1008.4616},
 primaryClass = {astro-ph.CO},
       adsurl = {https://ui.adsabs.harvard.edu/abs/2011AJ....141....4K}
}

@ARTICLE{Swaters2002,
       author = {{Swaters}, R.~A. and {van Albada}, T.~S. and {van der Hulst}, J.~M. and {Sancisi}, R.},
        title = "{The Westerbork HI survey of spiral and irregular galaxies. I. HI imaging of late-type dwarf galaxies}",
      journal = {\aap},
         year = 2002,
        month = aug,
       volume = {390},
        pages = {829-861},
          doi = {10.1051/0004-6361:20011755},
archivePrefix = {arXiv},
       eprint = {astro-ph/0204525},
 primaryClass = {astro-ph},
       adsurl = {https://ui.adsabs.harvard.edu/abs/2002A&A...390..829S}
}

@ARTICLE{Connors2004,
       author = {{Connors}, Tim W. and {Kawata}, Daisuke and {Maddison}, Sarah T. and {Gibson}, Brad K.},
        title = "{High-Resolution N-body Simulations of Galactic Cannibalism: The Magellanic Stream}",
      journal = {\pasa},
         year = 2004,
        month = jan,
       volume = {21},
       number = {2},
        pages = {222-227},
          doi = {10.1071/AS04014},
archivePrefix = {arXiv},
       eprint = {astro-ph/0402187},
 primaryClass = {astro-ph},
       adsurl = {https://ui.adsabs.harvard.edu/abs/2004PASA...21..222C}
}

@ARTICLE{Lokas2012,
       author = {{{\L}okas}, Ewa L. and {Majewski}, Steven R. and {Kazantzidis}, Stelios and {Mayer}, Lucio and {Carlin}, Jeffrey L. and {Nidever}, David L. and {Moustakas}, Leonidas A.},
        title = "{The Shapes of Milky Way Satellites: Looking for Signatures of Tidal Stirring}",
      journal = {\apj},
         year = 2012,
        month = may,
       volume = {751},
       number = {1},
          eid = {61},
        pages = {61},
          doi = {10.1088/0004-637X/751/1/61},
archivePrefix = {arXiv},
       eprint = {1112.5336},
 primaryClass = {astro-ph.CO},
       adsurl = {https://ui.adsabs.harvard.edu/abs/2012ApJ...751...61L}
}

@ARTICLE{Lokas2014,
       author = {{{\L}okas}, E.~L. and {Athanassoula}, E. and {Debattista}, V.~P. and {Valluri}, M. and {Pino}, A. del and {Semczuk}, M. and {Gajda}, G. and {Kowalczyk}, K.},
        title = "{Adventures of a tidally induced bar}",
      journal = {\mnras},
         year = 2014,
        month = dec,
       volume = {445},
       number = {2},
        pages = {1339-1350},
          doi = {10.1093/mnras/stu1846},
archivePrefix = {arXiv},
       eprint = {1404.1211},
 primaryClass = {astro-ph.GA},
       adsurl = {https://ui.adsabs.harvard.edu/abs/2014MNRAS.445.1339L}
}

@ARTICLE{Kazantzidis2011,
       author = {{Kazantzidis}, Stelios and {{\L}okas}, Ewa L. and {Callegari}, Simone and {Mayer}, Lucio and {Moustakas}, Leonidas A.},
        title = "{On the Efficiency of the Tidal Stirring Mechanism for the Origin of Dwarf Spheroidals: Dependence on the Orbital and Structural Parameters of the Progenitor Disky Dwarfs}",
      journal = {\apj},
         year = 2011,
        month = jan,
       volume = {726},
       number = {2},
          eid = {98},
        pages = {98},
          doi = {10.1088/0004-637X/726/2/98},
archivePrefix = {arXiv},
       eprint = {1009.2499},
 primaryClass = {astro-ph.CO},
       adsurl = {https://ui.adsabs.harvard.edu/abs/2011ApJ...726...98K}
}

@ARTICLE{Kazantzidis2013,
       author = {{Kazantzidis}, Stelios and {{\L}okas}, Ewa L. and {Mayer}, Lucio},
        title = "{Tidal Stirring of Disky Dwarfs with Shallow Dark Matter Density Profiles: Enhanced Transformation into Dwarf Spheroidals}",
      journal = {\apjl},
         year = 2013,
        month = feb,
       volume = {764},
       number = {2},
          eid = {L29},
        pages = {L29},
          doi = {10.1088/2041-8205/764/2/L29},
archivePrefix = {arXiv},
       eprint = {1302.0008},
 primaryClass = {astro-ph.CO},
       adsurl = {https://ui.adsabs.harvard.edu/abs/2013ApJ...764L..29K}
}

@ARTICLE{Kazantzidis2017,
       author = {{Kazantzidis}, Stelios and {Mayer}, Lucio and {Callegari}, Simone and {Dotti}, Massimo and {Moustakas}, Leonidas A.},
        title = "{The Effects of Ram-pressure Stripping and Supernova Winds on the Tidal Stirring of Disky Dwarfs: Enhanced Transformation into Dwarf Spheroidals}",
      journal = {\apjl},
         year = 2017,
        month = feb,
       volume = {836},
       number = {1},
          eid = {L13},
        pages = {L13},
          doi = {10.3847/2041-8213/aa5b8f},
archivePrefix = {arXiv},
       eprint = {1703.08381},
 primaryClass = {astro-ph.GA},
       adsurl = {https://ui.adsabs.harvard.edu/abs/2017ApJ...836L..13K}
}

@ARTICLE{Gatto2013,
       author = {{Gatto}, A. and {Fraternali}, F. and {Read}, J.~I. and {Marinacci}, F. and {Lux}, H. and {Walch}, S.},
        title = "{Unveiling the corona of the Milky Way via ram-pressure stripping of dwarf satellites}",
      journal = {\mnras},
         year = 2013,
        month = aug,
       volume = {433},
       number = {4},
        pages = {2749-2763},
          doi = {10.1093/mnras/stt896},
archivePrefix = {arXiv},
       eprint = {1305.4176},
 primaryClass = {astro-ph.GA},
       adsurl = {https://ui.adsabs.harvard.edu/abs/2013MNRAS.433.2749G}
}

@ARTICLE{Mayer2006,
       author = {{Mayer}, Lucio and {Mastropietro}, Chiara and {Wadsley}, James and {Stadel}, Joachim and {Moore}, Ben},
        title = "{Simultaneous ram pressure and tidal stripping; how dwarf spheroidals lost their gas}",
      journal = {\mnras},
         year = 2006,
        month = jul,
       volume = {369},
       number = {3},
        pages = {1021-1038},
          doi = {10.1111/j.1365-2966.2006.10403.x},
archivePrefix = {arXiv},
       eprint = {astro-ph/0504277},
 primaryClass = {astro-ph},
       adsurl = {https://ui.adsabs.harvard.edu/abs/2006MNRAS.369.1021M}
}

@ARTICLE{Zhu2025,
       author = {{Zhu}, Jingyao and {Asali}, Yasmeen and {Putman}, Mary and {Westmeier}, Tobias and {de Blok}, W.~J. G and {Catinella}, Barbara and {Deg}, Nathan and {For}, Bi-Qing and {Kleiner}, Dane and {Lee-Waddell}, Karen and et al.},
        title = "{Baryonic Masses and Properties of Gaseous Satellite Galaxies}",
      journal = {arXiv e-prints},
         year = 2025,
        month = oct,
          eid = {arXiv:2510.27019},
        pages = {arXiv:2510.27019},
          doi = {10.48550/arXiv.2510.27019},
archivePrefix = {arXiv},
       eprint = {2510.27019},
 primaryClass = {astro-ph.GA},
       adsurl = {https://ui.adsabs.harvard.edu/abs/2025arXiv251027019Z}
}

@ARTICLE{Geha2024,
       author = {{Geha}, Marla and {Mao}, Yao-Yuan and {Wechsler}, Risa H. and {Asali}, Yasmeen and {Kado-Fong}, Erin and {Kallivayalil}, Nitya and {Nadler}, Ethan O. and {Tollerud}, Erik J. and {Weiner}, Benjamin and {de los Reyes}, Mithi A.~C. and et al.},
        title = "{The SAGA Survey. IV. The Star Formation Properties of 101 Satellite Systems around Milky Way{\textendash}mass Galaxies}",
      journal = {\apj},
         year = 2024,
        month = nov,
       volume = {976},
       number = {1},
          eid = {118},
        pages = {118},
          doi = {10.3847/1538-4357/ad61e7},
archivePrefix = {arXiv},
       eprint = {2404.14499},
 primaryClass = {astro-ph.GA},
       adsurl = {https://ui.adsabs.harvard.edu/abs/2024ApJ...976..118G}
}

@ARTICLE{Ostriker2010,
       author = {{Ostriker}, Eve C. and {McKee}, Christopher F. and {Leroy}, Adam K.},
        title = "{Regulation of Star Formation Rates in Multiphase Galactic Disks: A Thermal/Dynamical Equilibrium Model}",
      journal = {\apj},
         year = 2010,
        month = oct,
       volume = {721},
       number = {2},
        pages = {975-994},
          doi = {10.1088/0004-637X/721/2/975},
archivePrefix = {arXiv},
       eprint = {1008.0410},
 primaryClass = {astro-ph.CO},
       adsurl = {https://ui.adsabs.harvard.edu/abs/2010ApJ...721..975O}
}

@ARTICLE{Mercia-Jones2021,
       author = {{Yanchulova Merica-Jones}, Petia and {Sandstrom}, Karin M. and {Johnson}, L. Clifton and {Dolphin}, Andrew E. and {Dalcanton}, Julianne J. and {Gordon}, Karl and {Roman-Duval}, Julia and {Weisz}, Daniel R. and {Williams}, Benjamin F.},
        title = "{Three-dimensional Structure and Dust Extinction in the Small Magellanic Cloud}",
      journal = {\apj},
         year = 2021,
        month = jan,
       volume = {907},
       number = {1},
          eid = {50},
        pages = {50},
          doi = {10.3847/1538-4357/abc48b},
archivePrefix = {arXiv},
       eprint = {2010.11181},
 primaryClass = {astro-ph.GA},
       adsurl = {https://ui.adsabs.harvard.edu/abs/2021ApJ...907...50Y}
}

@ARTICLE{Burhenne2026,
       author = {{Burhenne}, Clare and {McQuinn}, Kristen B.~W. and {Cohen}, Roger E. and {Murray}, Claire E. and {Patel}, Ekta and {Williams}, Benjamin F. and {Lindberg}, Christina W. and {Merica-Jones}, Petia Yanchulova and {Gordon}, Karl D. and {Choi}, Yumi and {Dolphin}, Andrew E. and {Roman-Duval}, Julia C.},
        title = "{Scylla. V. Constraints on the Spatial and Temporal Distribution of Bursts and the Interaction History of the Magellanic Clouds from Their Resolved Stellar Populations}",
      journal = {\apj},
         year = 2026,
        month = jan,
       volume = {997},
       number = {1},
          eid = {23},
        pages = {23},
          doi = {10.3847/1538-4357/ae13af},
archivePrefix = {arXiv},
       eprint = {2511.02947},
 primaryClass = {astro-ph.GA},
       adsurl = {https://ui.adsabs.harvard.edu/abs/2026ApJ...997...23B}
}

@ARTICLE{Almeida2024,
       author = {{Almeida}, Andres and {Majewski}, Steven R. and {Nidever}, David L. and {Olsen}, Knut A.~G. and {Monachesi}, Antonela and {Kallivayalil}, Nitya and {Hasselquist}, Sten and {Choi}, Yumi and {Povick}, Joshua T. and {Wilson}, John C. and {Geisler}, Doug and {Lane}, Richard R. and {Nitschelm}, Christian and {Sobeck}, Jennifer S. and {Stringfellow}, Guy S.},
        title = "{Exploring the origin of the distance bimodality of stars in the periphery of the Small Magellanic Cloud with APOGEE and Gaia}",
      journal = {\mnras},
         year = 2024,
        month = apr,
       volume = {529},
       number = {4},
        pages = {3858-3876},
          doi = {10.1093/mnras/stae373},
archivePrefix = {arXiv},
       eprint = {2308.13631},
 primaryClass = {astro-ph.GA},
       adsurl = {https://ui.adsabs.harvard.edu/abs/2024MNRAS.529.3858A}
}

@ARTICLE{Dias2021,
       author = {{Dias}, B. and {Angelo}, M.~S. and {Oliveira}, R.~A.~P. and {Maia}, F. and {Parisi}, M.~C. and {De Bortoli}, B. and {Souza}, S.~O. and {Katime Santrich}, O.~J. and {Bassino}, L.~P. and {Barbuy}, B. and {Bica}, E. and {Geisler}, D. and {Kerber}, L. and {P{\'e}rez-Villegas}, A. and {Quint}, B. and {Sanmartim}, D. and {Santos}, J.~F.~C. and {Westera}, P.},
        title = "{The VISCACHA survey. III. Star clusters counterpart of the Magellanic Bridge and Counter-Bridge in 8D}",
      journal = {\aap},
         year = 2021,
        month = mar,
       volume = {647},
          eid = {L9},
        pages = {L9},
          doi = {10.1051/0004-6361/202040015},
archivePrefix = {arXiv},
       eprint = {2103.02600},
 primaryClass = {astro-ph.GA},
       adsurl = {https://ui.adsabs.harvard.edu/abs/2021A&A...647L...9D}
}

@ARTICLE{Dias2022,
       author = {{Dias}, B. and {Parisi}, M.~C. and {Angelo}, M. and {Maia}, F. and {Oliveira}, R.~A.~P. and {Souza}, S.~O. and {Kerber}, L.~O. and {Santos}, J.~F.~C. and {P{\'e}rez-Villegas}, A. and {Sanmartim}, D. and {Quint}, B. and {Fraga}, L. and {Barbuy}, B. and {Bica}, E. and {Santrich}, O.~J. Katime and {Hernandez-Jimenez}, J.~A. and {Geisler}, D. and {Minniti}, D. and {De B{\'o}rtoli}, B.~J. and {Bassino}, L.~P. and {Rocha}, J.~P.},
        title = "{The VISCACHA survey - IV. The SMC West Halo in 8D}",
      journal = {\mnras},
         year = 2022,
        month = may,
       volume = {512},
       number = {3},
        pages = {4334-4351},
          doi = {10.1093/mnras/stac259},
archivePrefix = {arXiv},
       eprint = {2201.11119},
 primaryClass = {astro-ph.GA},
       adsurl = {https://ui.adsabs.harvard.edu/abs/2022MNRAS.512.4334D}
}

@ARTICLE{Parisi2024,
       author = {{Parisi}, M.~C. and {Oliveira}, R.~A.~P. and {Angelo}, M.~S. and {Dias}, B. and {Maia}, F.~F.~S. and {Saroon}, S. and {Feinstein}, C. and {Santos}, J.~F.~C. and {Bica}, E. and {Ferreira}, B. Pereira Lima and {Fern{\'a}ndez-Trincado}, J.~G. and {Westera}, P. and {Minniti}, D. and {Garro}, E.~R. and {Santrich}, O.~J. Katime and {De Bortoli}, B.~J. and {Souza}, S.~O. and {Kerber}, L. and {P{\'e}rez-Villegas}, A.},
        title = "{The VISCACHA survey - IX. The SMC Southern Bridge in 8D}",
      journal = {\mnras},
         year = 2024,
        month = feb,
       volume = {527},
       number = {4},
        pages = {10632-10648},
          doi = {10.1093/mnras/stad3871},
archivePrefix = {arXiv},
       eprint = {2312.09756},
 primaryClass = {astro-ph.GA},
       adsurl = {https://ui.adsabs.harvard.edu/abs/2024MNRAS.52710632P}
}

@ARTICLE{Matijevic2025,
       author = {{Matijevi{\'c}}, Luka and {Tomi{\v{c}}i{\'c}}, Neven and {Marasco}, Antonino and {Ignesti}, Alessandro and {Lassen}, Augusto E. and {Smith}, Rory and {Sell}, Paul and {Roberts}, Ian D. and {Zezas}, Andreas and {Anastasopoulou}, Konstantina and {Kotoulas}, Panagiotis and {Ba{\v{s}}i{\'c}}, Roko},
        title = "{The Competing Influence of Ram Pressure and Tidal Interaction in NGC 2276}",
      journal = {arXiv e-prints},
         year = 2025,
        month = dec,
          eid = {arXiv:2512.17486},
        pages = {arXiv:2512.17486},
          doi = {10.48550/arXiv.2512.17486},
archivePrefix = {arXiv},
       eprint = {2512.17486},
 primaryClass = {astro-ph.GA},
       adsurl = {https://ui.adsabs.harvard.edu/abs/2025arXiv251217486M}
}

@ARTICLE{Oden2025,
       author = {{Oden}, Slater J. and {Nidever}, David L. and {Povick}, Joshua and {Massana}, Pol and {Choi}, Yumi and {van der Marel}, Roeland P. and {Cioni}, Maria-Rosa L. and {Sakowska}, Joanna and {Olsen}, Knut A.~G. and {Cullinane}, Lara and {Carballo-Bello}, J.~A. and {Crnojevi{\'c}}, D. and {Ferguson}, P.~S. and {Mart{\'\i}nez-V{\'a}zquez}, C.~E. and {Medina}, G.~E. and {Mutlu-Pakdil}, B. and {Navabi}, M. and {Pace}, A.~B. and {Riley}, A.~H. and {Stringfellow}, Guy S. and {Vivas}, A.~K.},
        title = "{Warped \& Hooked: Mapping the Magellanic Clouds in 3D using Red Clump stars}",
      journal = {arXiv e-prints},
         year = 2025,
        month = dec,
          eid = {arXiv:2512.04200},
        pages = {arXiv:2512.04200},
          doi = {10.48550/arXiv.2512.04200},
archivePrefix = {arXiv},
       eprint = {2512.04200},
 primaryClass = {astro-ph.GA},
       adsurl = {https://ui.adsabs.harvard.edu/abs/2025arXiv251204200O}
}
\bibliographystyle{aasjournal}

\end{document}